\documentclass[floatfix,aps,pre,preprint,groupedaddress]{revtex4}

\usepackage{graphicx} 
\usepackage{epstopdf}
\usepackage{enumerate}
\usepackage{paralist}
\usepackage{caption}
\usepackage[title]{appendix}
\usepackage{stackengine}

\usepackage{booktabs}
\usepackage{subfigure}
\usepackage{color}
\usepackage{amsfonts}
\usepackage{amssymb}
\usepackage{amsmath}
\usepackage[hidelinks]{hyperref}
\usepackage{tabularx}
\graphicspath{{/}}

\renewcommand{\vec}[1]{\ensuremath\boldsymbol{#1}}

\newcommand{\Pe}{\ensuremath \textnormal{Pe}}	
\newcommand{\Cn}{\ensuremath \textnormal{Cn}}	

\renewcommand{\Re}{\ensuremath \textnormal{Re}}	
\newcommand{\Ca}{\ensuremath \textnormal{Ca}}	
\usepackage{multirow}

\begin{document}

\begin{abstract}
In this work, the nonlinear dynamics of a fully three-dimensional multicomponent vesicle in shear flow are explored. 
Using a volume- and area-conserving projection method coupled to a gradient-augmented level set and surface phase method,
the dynamics are systematically studied as a function of the membrane bending rigidity difference between the components, 
the speed of diffusion compared to the underlying shear flow, and the strength of the phase domain energy compared to the bending energy.
Using a pre-segregated vesicle, three dynamics are observed: stationary phase, phase-treading, and a new dynamic called vertical banding.
These regimes are very sensitive to the strength of the domain line energy, as the vertical banding regime is not observed when 
line energy is larger than the bending energy. These findings demonstrate that a complete understanding of multicomponent vesicle
dynamics require that the full three-dimensional system be modeled, and show the complexity obtained when considering 
heterogeneous material properties.

\end{abstract}

\title{Three-Dimensional Multicomponent Vesicles: Dynamics \& Influence of Material Properties}
\author{Prerna Gera$^\dag$}
\affiliation{Department of Mechanical and Aerospace Engineering, University at Buffalo,Buffalo, New York 14260-4400, USA}
\altaffiliation{Present Address: Department of Mathematics, University of Wisconsin--Madison, Van Vleck Hall, 480 Lincoln Drive, Madison, WI  53706, USA}
\author{David Salac}%
\email[Corresponding author: ]{davidsal@buffalo.edu}
\affiliation{Department of Mechanical and Aerospace Engineering, University at Buffalo,Buffalo, New York 14260-4400, USA}
\date{\today}%
\maketitle

\section{Introduction}

Biological cells have a bilayer membrane which protects the enclosed material and acts as a medium of
communication between the intra- and extra-cellular environments. 
Variation in the composition of the membrane have shown to impact
fundamental cellular processes such as signal
transduction~\cite{simons2000lipid,sackmann2014physics}, membrane trafficking~\cite{simons2004model}, and membrane
sorting~\cite{mukherjee2004membrane}.
The inhomogeneous membranes of living cells have a complex and a
dynamic structure and thus the simplified membrane model system of lipid
vesicles is of major significance~\cite{Lipowsky1995}.

A multicomponent membrane of a vesicle is typically
composed of a mixture of saturated lipids, unsaturated lipids, and cholesterol.
Due to the molecular structure, saturated lipids tend to combine with cholesterol
to form energetically stable and relatively ordered domains, also known as an
ordered phase~\cite{baumgart2003imaging,veatch2003separation}.
These are surrounded by
the unsaturated lipids, known as the disordered phase, and the process of phase segregation
occurs to achieve a lower state of energy~\cite{veatch2005seeing}. In general, the phases have differing material
properties, which causes morphological changes to the underlying surface of the vesicle,
as observed in experiments~\cite{braggart2005membrane,baumgart2003imaging,McMahon2005}.
While not considered here, variation of the domains between the inner and outer leaflet
have also been shown to influence the overall material properties and dynamics~\cite{C6SM01349J}.

Extensive work in the literature is present for a single component
membrane~\cite{Deschamps2009,biben2003tumbling, aranda2008morphological, Staykova2008, vlahovska2009electrohydrodynamic, kolahdouz2015electrohydrodynamics,Salac2012,vlahovska2007dynamics,kantsler2008critical}, 
but there is limited work for
inhomogeneous vesicles~\cite{elliott2013computation,elliott2010modeling,barrett2017finite2}.
In cases where multicomponent vesicles are studied,
material properties, such as the
bending rigidity, have been shown to dramatically influence the dynamics~\cite{Allain2004,Fournier2006}.
Du et al. demonstrated
interesting and exotic dynamic patterns in 3D multicomponent vesicles~\cite{Wang2008}.
Funkhouser et al. examined the dynamics of vesicles with non-uniform
mechanical properties~\cite{funkhouser2014dynamics}. These works are however done in absence of an aqueous
medium. Other works which include the influence of the fluid
are limited to two-dimensions, which cannot include all aspects of multicomponent
vesicle such as the line energy associated with domain boundaries~\cite{C6SM02452A}.

In this work, the hydrodynamics of a three-dimensional multicomponent
vesicle in shear flow is systematically explored. 
The goal of this work is to investigate and understand the
influence of material properties on the dynamics of multicomponent vesicles
in the presence of shear flow. 
To the best of the authors' knowledge, this is the first work
to carry out such an investigation in three-dimensional space.
The parameters considered are the membrane bending
rigidity, the rate of surface diffusion, and the influence of the domain line energy.
Using the model presented here, three major dynamics are observed: 
stationary phases, vertical stationary band, and the treading of
the surface phases. Using two characteristic values of domain line energy,
the phase diagrams of these dynamics as a function of surface diffusion rate and membrane bending rigidity
is presented. An extensive investigation is performed and
sample dynamics, energy curves, treading period, and time for
domain merging are presented.

The remainder of this work describes the models and numerical methods
used. This is followed by a demonstration of the major observed dynamics.
Considering two characteristic domain line energy values, 
a systematic investigation of the dynamics as a function of the surface diffusion
rate and bending rigidity difference is performed. This is followed
by further discussion and conclusions.

\section{Model and Methods}
\begin{figure}[!h]
	\begin{center}
		\includegraphics[height=4cm]{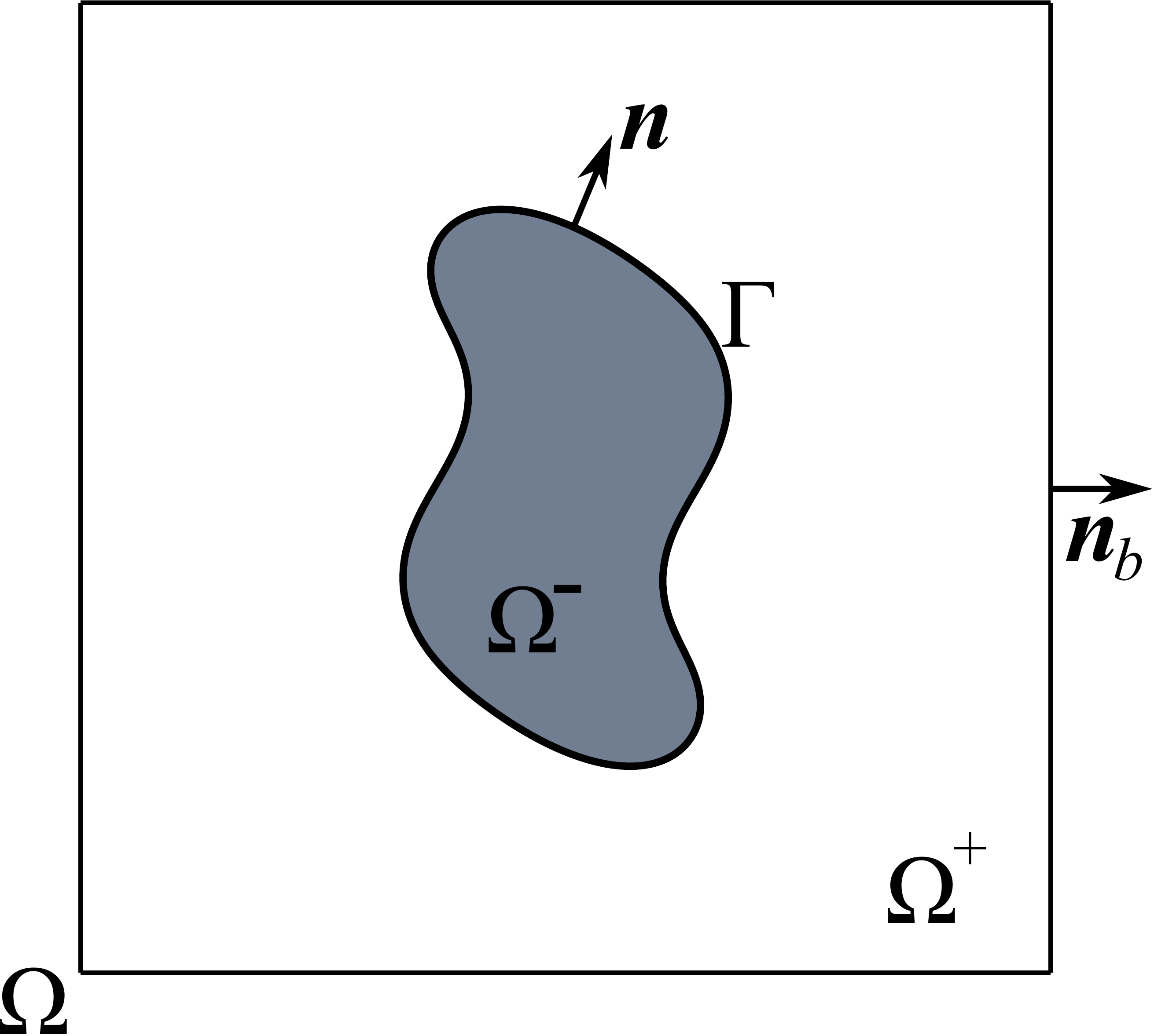}
	\end{center}
  \caption{A vesicle membrane $\Gamma$ separating the fluid outside $\Omega^+$
  from the fluid inside $\Omega^-$.}
  \label{fig:compDomain}
\end{figure}

Consider a multicomponent vesicle suspended in an aqueous fluid that differs from the fluid encapsulated inside the membrane. 
The membrane $\Gamma$ separates the fluid outside $\Omega^+$ from the fluid
inside $\Omega^-$ as shown in Fig.~\ref{fig:compDomain}. 
The vesicle is characterized using a reduced volume parameter $\nu$, which is defined as the ratio of the vesicle volume 
$V$ to the volume of a sphere with the same surface area $A$: $\nu=3V/4\pi a^3$, where $a=\sqrt{A/4\pi}$. 
The vesicles considered here have a radius $\sim10 \mu$m and a membrane thickness of $\sim5$ nm, 
and therefore the membrane is considered as an infinitesimally thin interface. 
Additionally, the membrane is impermeable to fluids and the number of lipid molecules on the surface of the membrane 
does not change over time, which results in an inextensible membrane. Therefore, such systems are both volume and surface
area conserving.

\subsection{Fluid Field}
In general, such systems are governed by the Navier-Stokes equation and a volume-incompressibility constraint,
\begin{align*}
	\rho\dfrac{D\vec{u}^{\pm}}{Dt} = \nabla\cdot\vec{T}_{hd}^{\pm} \text{\hspace{3mm} and \hspace{3mm}}
	\nabla\cdot\vec{u}^{\pm}=0 \hspace{3mm} \text{in \hspace{3mm}}  \Omega^{\pm},
\end{align*}
where $\rho$ is the density, $\vec{u}$ is the fluid velocity vector, and $\vec{T}_{hd}$ is the bulk
hydrodynamic stress tensor. This tensor is given by
\begin{align}
	\vec{T}_{hd}=-p^{\pm}\vec{I}+\mu^{\pm}(\nabla
	\vec{u}^{\pm}+\nabla^T\vec{u}^{\pm}) \hspace{3 mm} \text{in 
	\hspace{3mm}} \Omega^{\pm},
\end{align}
where $p$ is the pressure and $\mu$ is the fluid viscosity. 

The inextensible membrane introduces three additional conditions.
First, the velocity on the surface of the membrane is assumed to be continuous
$[\vec{u}]=0$, where $[\cdot]$ represents the jump of a parameter across the interface.
Second, the hydrodynamic stress tensor undergoes a jump across the interface 
which is balanced by the forces exerted by the membrane,
\begin{align}
	\vec{n}\cdot[\vec{T}_{hd}]=\vec{f},
\end{align}
where $\vec{f}$ is the total membrane force.
Finally, since the membrane area is conserved, the local area is conserved via a surface-incompressibility constraint on the fluid field,
\begin{equation*}
	\nabla_s\cdot\vec{u}=0 \hspace{3mm} \text{on} \hspace{3mm}  \Gamma.
	\label{eq:surf_div_free}
\end{equation*}

\subsection{Surface Material Field}
The membrane surface of the vesicle is composed of saturated lipids that combine with
cholesterol to form energetically stable domains also known as the ordered phase. The
ordered phase is surrounded by unsaturated lipids, also called the disordered phase. To model 
the multicomponent surface dynamics, a two phase Cahn-Hilliard system is used.
Consider one surface phase $B$ with surface concentration $q(\vec{x},t)$ 
while $1-q(\vec{x},t)$ indicates the amount of the second, $A$ phase. 
There is no mass transfer from the
bulk to the interface or vice versa, and therefore the mass of the
surface concentration is conserved,
\begin{equation}
	M_q(t)=\int_{\Gamma(t)}q(\vec{x},t)=M_q(0).
\end{equation} 
This surface concentration evolves on the interface via a mass-conserving convection-diffusion equation, which in Eulerian form
is written as
\begin{equation}
	\dfrac{\partial q}{\partial t} +\vec{u}\cdot \nabla q =\nabla_s \cdot \vec{J}_s,
	\label{eq:phaseEvolution}
\end{equation}
where $J_s$ is the surface flux and is defined in the next section. Note that
the advection as written accounts for motion both in the tangential 
direction and in the normal direction, which accounts for movement of the interface.

\subsection{Constitutive Equations}
The total energy of the system, $E$, consists of three contributions:
\begin{equation}	
	E=E_b+E_\gamma+E_q,
\end{equation}
where
\begin{align}
	E_b&=\int_\Gamma\dfrac{\kappa_c(q)}{2}H^2\;\textrm{dA},\label{eq:bendingEnergy}\\	
	E_\gamma&=\int_\Gamma \gamma \;\textrm{dA},\\
	E_q&=\int_\Gamma \left(g(q)+\dfrac{k_f^2}{2}\|\nabla_s q\|^2\right)\;\textrm{dA}\label{eq:phaseEnergy}.
\end{align}
The first energy functional, $E_b$, is the total bending energy of the interface where
$\kappa_c(q)$ and $H$ are the bending rigidity and total curvature, respectively.
The total curvature $H$ is defined as
$H=c_1+c_2$, where $c_1$ and $c_2$ are the principal curvatures on the surface. 
Note that this bending energy form assumes zero spontaneous curvature and constant Gaussian bending rigidity.
Due to the Gauss-Bonnet theorem the Gaussian bending energy term is constant for 
membranes which do no undergo splitting or merging events~\cite{guillemin2010differential}.

The energy due to surface tension is given by $E_\gamma$,
where $\gamma$ is the surface tension. 
The phase field energy $E_q$ has two components. The first term $g(q)$ is the mixing energy of a phase
and is typically taken as a double well potential. In this work it is defined as $g(q)=q^2(1-q)^2$, 
with two minimas at $q=0$ and $q=1$. The second component of the surface phase field free energy is
associated with surface domain boundaries, where $k_f$ is a constant associated
with the surface domain energy.

The forces applied by the membrane can be computed by taking the variation of the energy
with respect to the interface position,
\begin{equation}
	\vec{f}=-\dfrac{\partial E}{\partial \Gamma}=-\left(\dfrac{\partial E_b}{\partial \Gamma}+ \dfrac{\partial E_\gamma}{\partial \Gamma}
	+ \dfrac{\partial E_q}{\partial \Gamma}\right).
\end{equation}
Each component is given by
\begin{align}
	\label{eq:dBendingG}
	\dfrac{\partial E_b}{\partial\Gamma}			&= -\kappa_c\left(\dfrac{1}{2}H^3-2HK+\Delta_s H\right)\vec{n}-\dfrac{1}{2}H^2\nabla_s\kappa_c-\vec{n}H\Delta_s\kappa_c,\\	
	\label{eq:dTensionG}
	\dfrac{\partial E_\gamma}{\partial \Gamma}	&= -\nabla_s\gamma + \gamma H\vec{n},\\
	\label{eq:dSPFG}
	\dfrac{\partial E_q}{\partial \Gamma}		&= -k_f\left(\nabla_s q\cdot\vec{L}\nabla_s q\right)\vec{n} + \dfrac{k_f}{2}\|\nabla_s q\|^2 H \vec{n}			
													+k_f \left(\nabla_s q\right)\Delta_s q,
\end{align}
where $\vec{L}=\nabla_s\vec{n}$ is the surface curvature tensor.
Full details
of the derivation for the above expressions can be found in Gera and Salac~\cite{GERA2018}. As the tension will be determined to enforce
surface incompressibility, all terms which have a similar form as Eq.~\ref{eq:dTensionG} are neglected~\cite{GERA2018}.

The surface flux in the phase evolution system is given by Fick's law,
\begin{align}
	\vec{J}_s=\nu \nabla_s \beta,
	\label{eq:surfaceFlux}
\end{align}
where $\nu$ is the mobility and $\beta$ is the chemical potential. 
In this work, it is assumed that mobility is constant, $\nu=\nu_0$.
This chemical
potential is computed by variation of the total energy in the system with respect to the
surface concentration~\cite{GERA2018},
\begin{align}
	\beta=\dfrac{\partial E}{\partial q}= \dfrac{\partial E_b}{\partial q}	
	+ \dfrac{\partial E_q}{\partial q},
	\label{eq:ChemPot}
\end{align}
with
\begin{align}
	\dfrac{\partial E_b}{\partial q}			&= \dfrac{1}{2}\dfrac{d \kappa_c}{dq}\;H^2, \\	
	\dfrac{\partial E_q}{\partial q}			&= \dfrac{d g}{dq} - k_f\Delta_s q.
\end{align}

\subsection{Nondimensional Model}

All properties in the system are made dimensionless using the properties of the outer fluid, 
lipid phase $A$, a characteristic length given by $r_0$ and a characteristic time of $t_0$.
The Reynolds number relates the strength of fluid advection to viscosity 
and is taken to be $\Re=\rho^+ u_0 r_0/\mu^+$, where the characteristic velocity is $u_0=r_0/t_0$. 
The capillary bending number is defined as the strength of the membrane bending
compared to viscous effects, $\Ca=\mu^+r_0^3/(\kappa_c^A t_0)$. 
The rate of diffusion of lipid phases compared to the
characteristic time is given by the surface Peclet number, $\Pe=r_0^2/(t_0\beta_0\nu_0)$,
where $\nu_0$ is the characteristic and constant mobility and $\beta_0$ is the characteristic surface chemical
potential. The strength of the bending forces to the domain tension 
force is characterized by $\alpha= \kappa_c^A/k_f$, 
while the Cahn number relates the strength of domain line tension to the chemical potential,
$\Cn^2=k_f/\beta_0 r_0^2$.

Using the dimensionless parameters and the Continuum-Surface-Force Method~\cite{Chang1996}, a single 
equation describes the dynamics of the fluid over the entire domain, 
		\begin{equation}
			\begin{split}
				\rho\frac{D \vec{u}}{Dt}=&-\nabla p	+\frac{1}{\Re}\nabla \cdot \left[\mu(\phi)\left(\nabla\vec{u}+\nabla^T \vec{u}\right)\right]  \\
					&+\delta(\phi)||\nabla\phi||\left(\nabla_s \gamma - \gamma H \vec{n}\right) \\
					&+\dfrac{\delta(\phi)}{\Re}\|\nabla\phi\|\left(\frac{1}{\Ca}\vec{f}_{b}
                    + \frac{1}{\alpha\Ca }\vec{f}_{spf}\right),\\
                \nabla\cdot\vec{u}=&0,
			\end{split}
			\label{eq:nonDimNS}
		\end{equation}
        where
        \begin{align}
            \vec{f}_{b}&= \kappa_c\left(\dfrac{1}{2}H^3-2HK+\Delta_s H\right)\vec{n}+\dfrac{1}{2}H^2\nabla_s\kappa_c+\vec{n}H\Delta_s\kappa_c,\\
            \vec{f}_{spf}&= k_f\left(\nabla_s q\cdot\vec{L}\nabla_s
					q\right)\vec{n} - \dfrac{k_f}{2}\|\nabla_s q\|^2 H \vec{n}
                    -k_f \left(\nabla_sq\right)\Delta_s q, 
        \end{align}
and $\delta(\phi)$ is a smoothed Dirac-delta function~\cite{Towers2008} and $\phi$ is an implicit representation of the interface.

Finally, the Cahn-Hilliard system defining the dynamics on the surface, Eqs. \eqref{eq:phaseEvolution}, \eqref{eq:surfaceFlux}, and \eqref{eq:ChemPot} is written as
a pair of coupled partial differential equation,
		\begin{align}
			\dfrac{Dq}{Dt}=&\dfrac{1}{\Pe}\nabla_s\cdot\left(\nu\nabla_s\beta\right),\label{eq:nonDimPhaseEvol}
			\\\bigskip
			\beta=&\dfrac{d g}{dq}-\Cn^2\nabla_s^2 q 
				+  \alpha \dfrac{\Cn^2}{2}\dfrac{d \kappa_c}{dq}H^2.\label{eq:nonDimChemPotential}
		\end{align}

\subsection{Numerical Methods}

The vesicle surface is modeled using a level-set Jet scheme where the membrane
$\Gamma$ is represented using the zero of a mathematical function $\phi$, which called the
level set function~\cite{nave2010gradient,seibold2012jet},
\begin{equation} 
	\Gamma(\vec{x},t) = \{\vec{x}:\phi(\vec{x},t)=0\}.
\end{equation}
In a given flow-field, the membrane motion is captured
using standard advection. Written in Lagrangian form this is
\begin{equation}
	\dfrac{\mathrm{D} \phi}{\mathrm{D} t}=0,
\end{equation}
which indicates that the level set function behaves as if it was a material
property being advected by the underlying fluid field. 

The above equation is 
discretized using a second-order semi-implicit, semi-Lagrangian scheme as
follows,
\begin{equation}
	\dfrac{3 \phi^{n+1}-4\phi_d^n+\phi_d^{n-1}}{2\Delta t}+\dfrac{1}{2}\Delta \phi^{n+1}=\dfrac{1}{2}\Delta \phi^n,
\end{equation}
where $\phi_d^n$ and $\phi_d^{n-1}$ are the departure level set values at the two prior time steps
$t^n$ and $t^{n-1}$ where $\Delta t=t^{n}-t^{n-1}$ is a constant time step. 
The inclusion of the $\Delta \phi$ results in better stability properties than fully explicit schemes.
When applied to a level set jet, this scheme is known as the SemiJet
level-set method. Details, including convergence results, can be found in
Velmurugan et al.~\cite{Velmurugan2015}.

The coupled surface Cahn-Hilliard equation Eq.~(\ref{eq:nonDimPhaseEvol}) and Eq.~(\ref{eq:nonDimChemPotential}) is discretized using a second-order
backward-finite-difference scheme~\cite{Fornberg1988},
\begin{equation}
	\begin{bmatrix}
		\vec{I} & \Cn^2\vec{L}_s\\
		-\dfrac{2\Delta t}{3\Pe}\vec{L}_s & \vec{I}
	\end{bmatrix}
	\begin{bmatrix}
		\vec{\beta}^{n+1}\\
		\vec{q}^{n+1}
	\end{bmatrix}
	=
	\begin{bmatrix}
		2\vec{\beta}^n_{rhs}-\vec{\beta}^{n-1}_{rhs}\\
	    \dfrac{4}{3} \vec{q}^n -\dfrac{1}{3}
        \vec{q}^{n-1}
	\end{bmatrix},
	\label{eq:CH_block}
\end{equation}
where $\vec{q}^n$ and $\vec{q}^{n-1}$ are the solutions at times
$t^n$ and $t^{n-1}$, respectively, $\vec{\beta}_{rhs}=g'+0.5\Cn^2\alpha\kappa'_cH^2$, and
$\vec{I}$ is the identity matrix, while the
surface Laplacian is given by $\vec{L}_s\approx\Delta_s$.
The surface partial differential equation is approximated using a closest point
method, where the solution of a surface partial differential equation is
extended such that it is constant in normal direction. This enables the solution 
to the surface partial differential equation using discretizations
in the embedding space. For more details on this method, readers are referred to 
Chen et al.~\cite{doi:10.1137/130929497}.

To compute the velocity, pressure, and tension, a projection method is employed.
A semi-implicit and semi-Lagrangian update is performed to obtain a tentative velocity field,
\begin{align}
	\dfrac{3\vec{u}^{\ast}-4\vec{u}_d^n+\vec{u}_d^{n-1}}{2\Delta t}=&-\nabla p^n+\delta(\phi)\|\nabla\phi\|\left(\nabla_s \gamma^n-\gamma^n H\|\nabla \phi\|\vec{n}\right)\nonumber \\
		&+\dfrac{1}{\Re}\nabla\cdot\left(\mu\left(\nabla
        \vec{u}^{\ast}+\left(\nabla\hat{\vec{u}}\right)^T\right)\right)\nonumber \\
        &+\dfrac{\delta(\phi)}{\Re\Ca}\left(\vec{f}_{b}+\dfrac{1}{\alpha}\vec{f}_{spf}\right),
	\label{eq:tentativeVelocity}
\end{align}
where the material derivative is described using a Lagrangian approach with
$\vec{u}_d^n$ being the departure velocity at time $t^n$ and $\vec{u}_d^{n-1}$ the 
departure velocity at time $t^{n-1}$.
The tentative velocity field is then projected onto the volume- and surface-divergence free velocity space,
\begin{equation}
	\dfrac{3\left(\vec{u}^{n+1}-\vec{u}^{\ast}\right)}{2\Delta t}=-\nabla r +\delta(\phi)\|\nabla\phi\|\left(\nabla_s \xi-\xi H\nabla\phi \right),
	\label{eq:projection}
\end{equation}
where $r$ and $\xi$ are the corrections needed for the pressure and tension, respectively. 
Finally, the pressure and tension are updated by including the corrections,
\begin{align}
	p^{n+1}&=p^n+r,\label{eq:pressureCorrection}\\
	\gamma^{n+1}&=\gamma^n+\xi\label{eq:tensionCorrection}.
\end{align}
Complete details of the method including convergence study can be found in
Kolahdouz et al.~\cite{kolahdouz2015electrohydrodynamics}. For more details on the
algorithm used to couple the system, readers are referred to Gera et al.~\cite{GERA2018}.

\begin{figure*}[!ht]
	\begin{center}
		\subfigure[]{
			\begin{tabular}{
				>{\centering}m{1.0cm}>{\centering}m{1.7cm}>{\centering}m{1.7cm}>{\centering}m{1.7cm}>{\centering}m{1.7cm}>{\centering}m{1.7cm}>{\centering}m{1.7cm}>{\centering}m{1.7cm}>{\centering}m{1.7cm}}
				& \multicolumn{8}{c}{Time} \tabularnewline
				\cline{2-9}
				View & $0.0$ & $0.5$ & $2.5$ & $5.0$ & $15.0$ & $18.5$ & $20.0$ & $22.5$ \tabularnewline
				Iso & 
					\includegraphics[height=1.7cm]{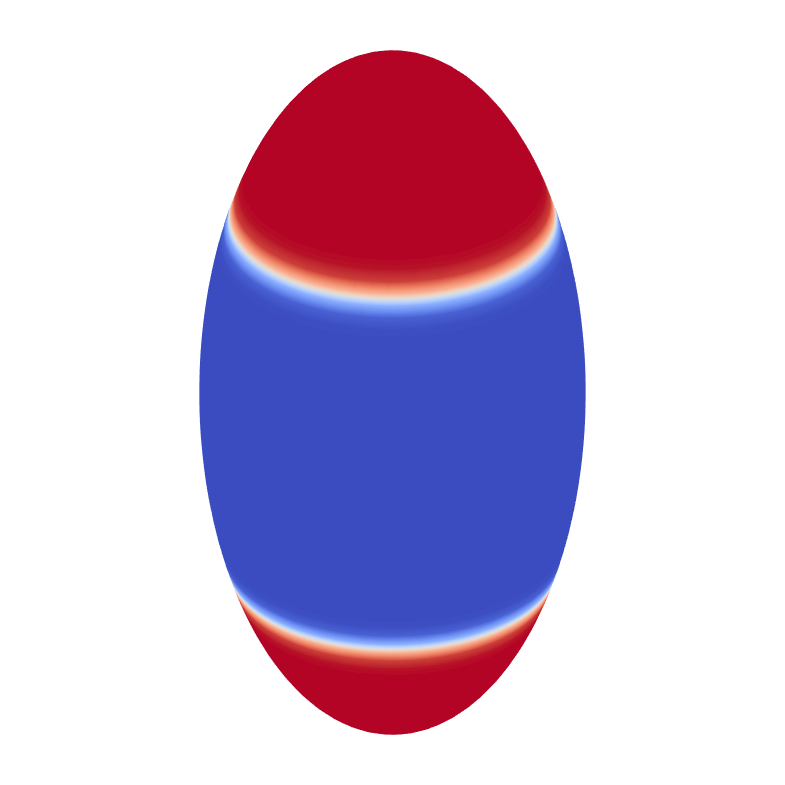} &
					\includegraphics[height=1.7cm]{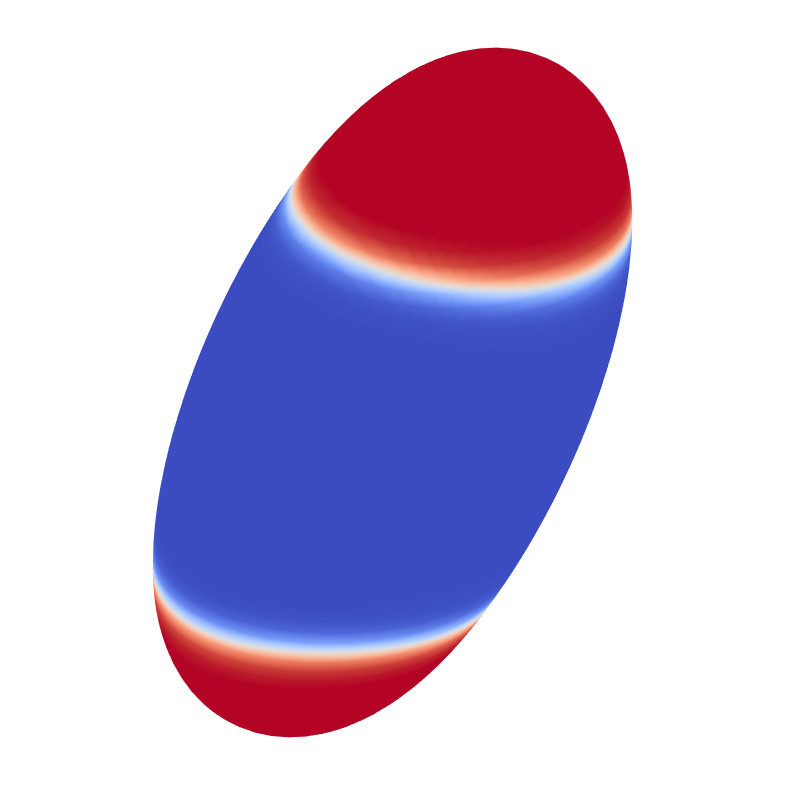} &
					\includegraphics[height=1.7cm]{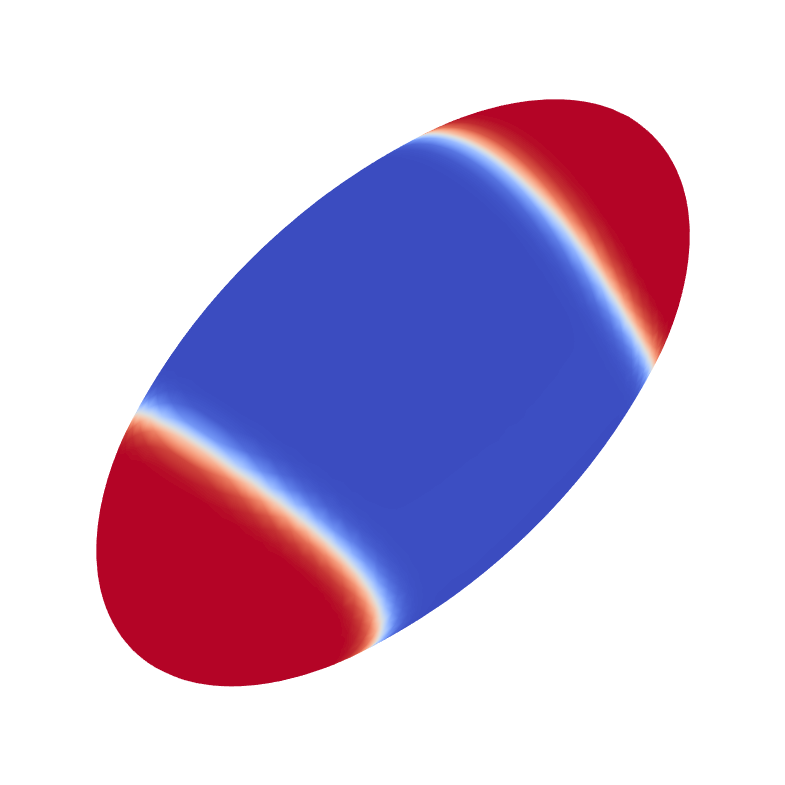} &
					\includegraphics[height=1.7cm]{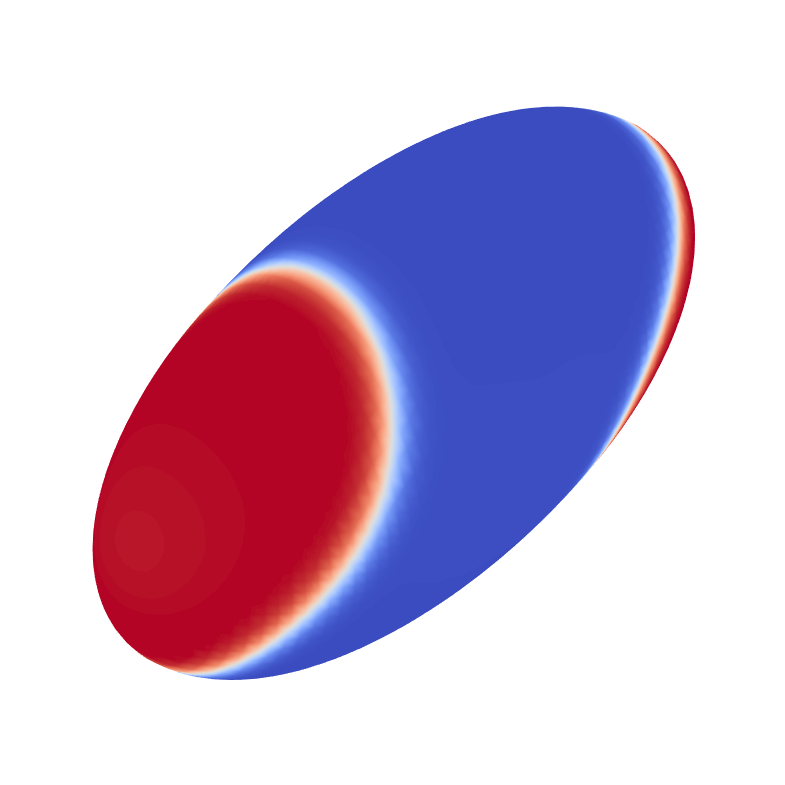} &
					\includegraphics[height=1.7cm]{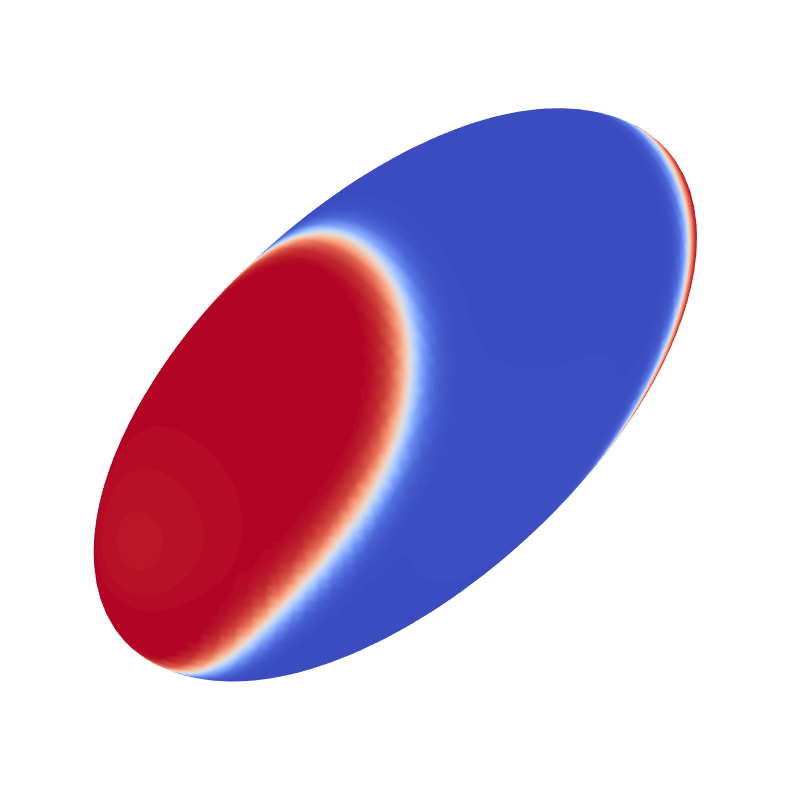} &
					\includegraphics[height=1.7cm]{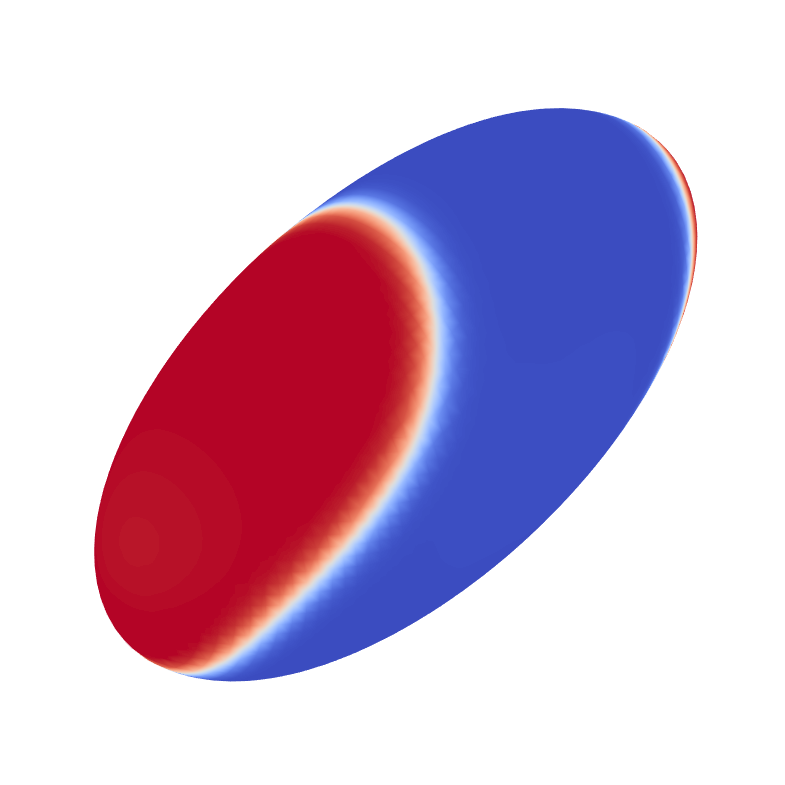} &
					\includegraphics[height=1.7cm]{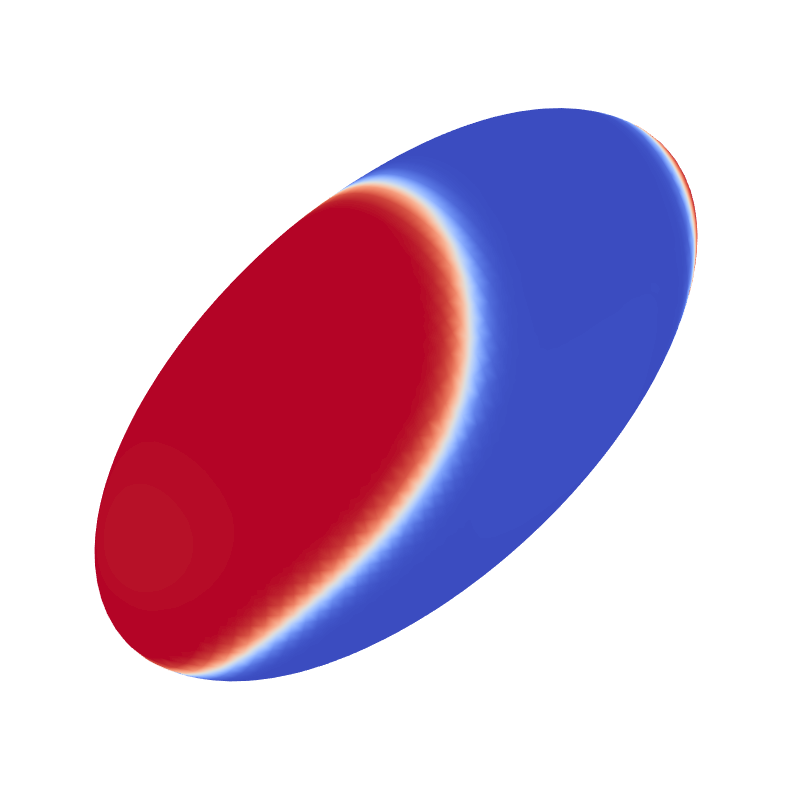} &
					\includegraphics[height=1.7cm]{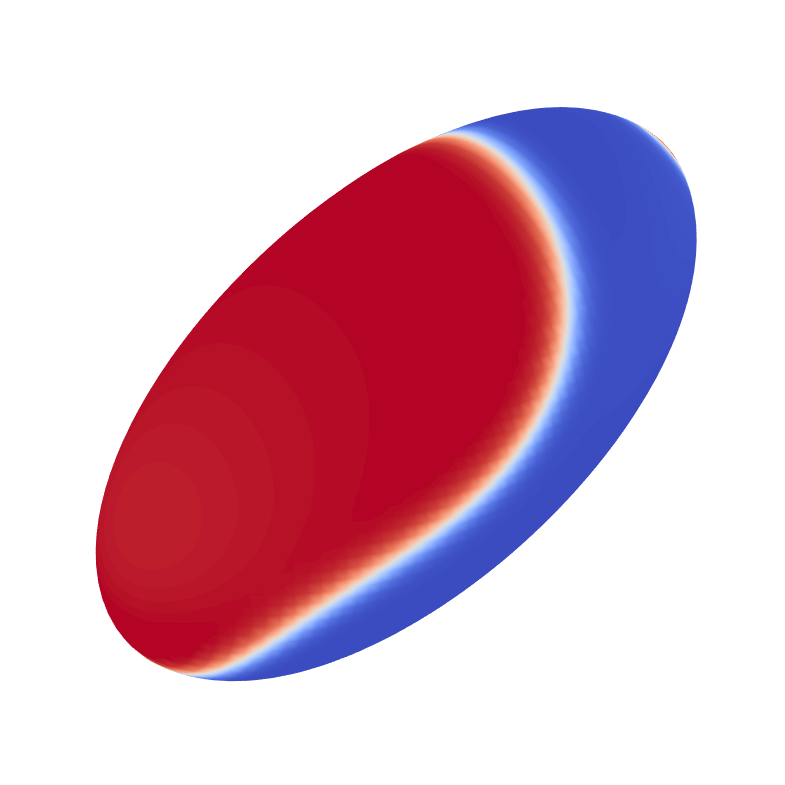} \tabularnewline
				X-Y & 
					\includegraphics[height=1.7cm]{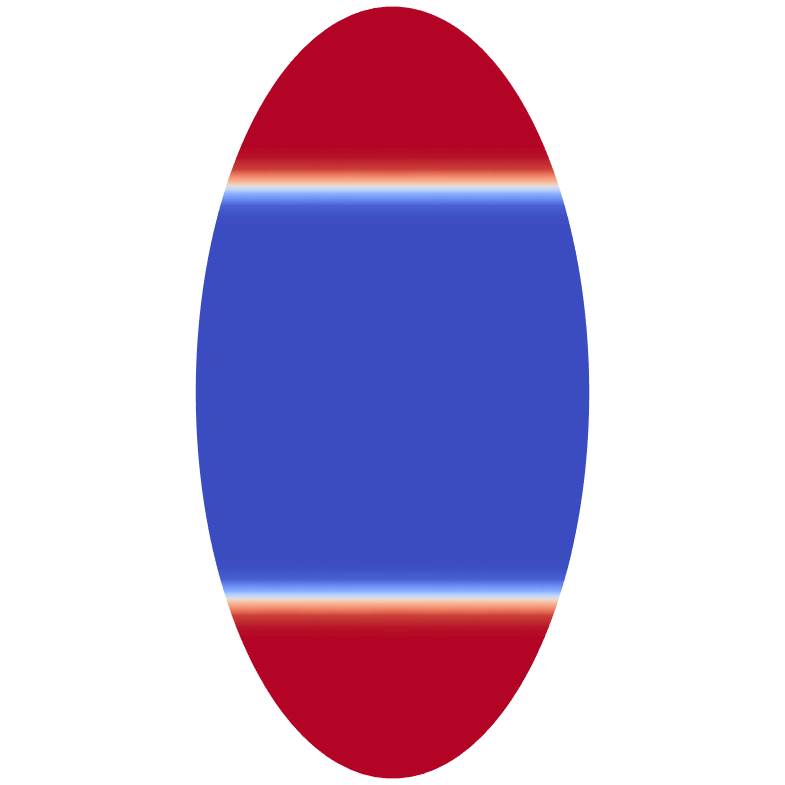} &
					\includegraphics[height=1.7cm]{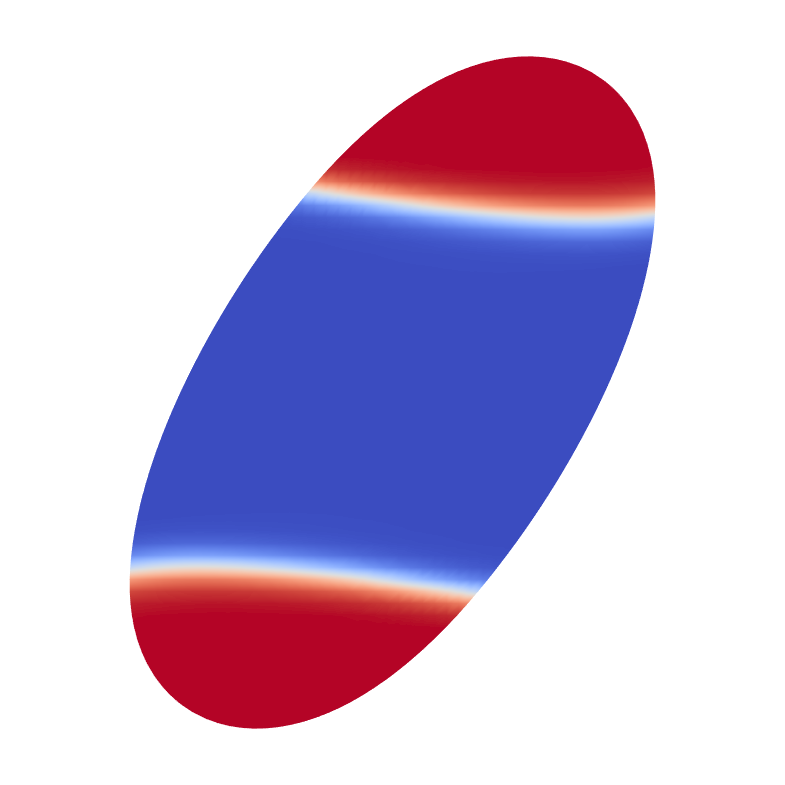} &
					\includegraphics[height=1.7cm]{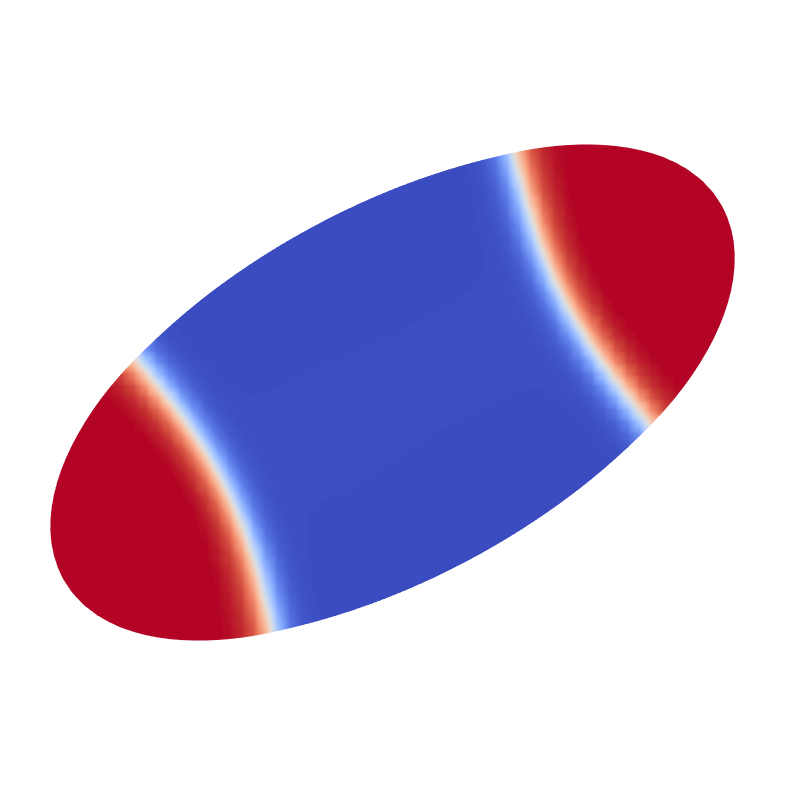} &
					\includegraphics[height=1.7cm]{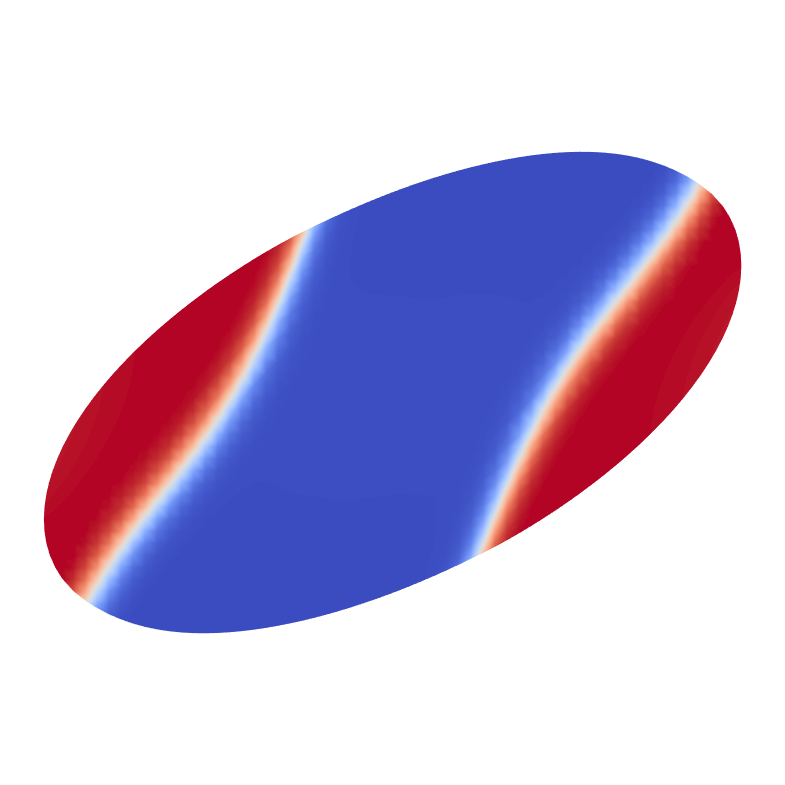} &
					\includegraphics[height=1.7cm]{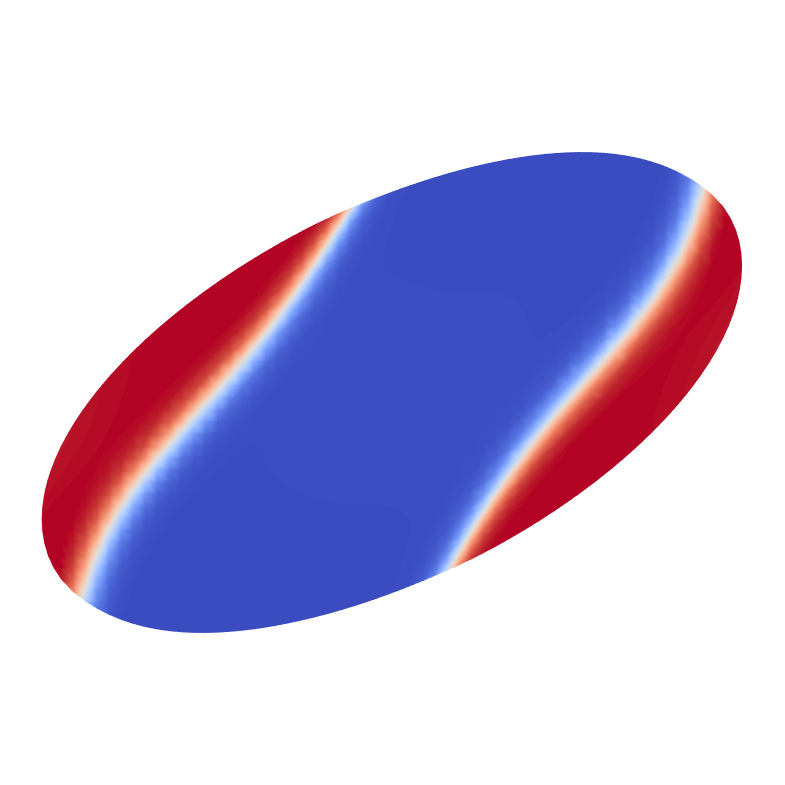} &
					\includegraphics[height=1.7cm]{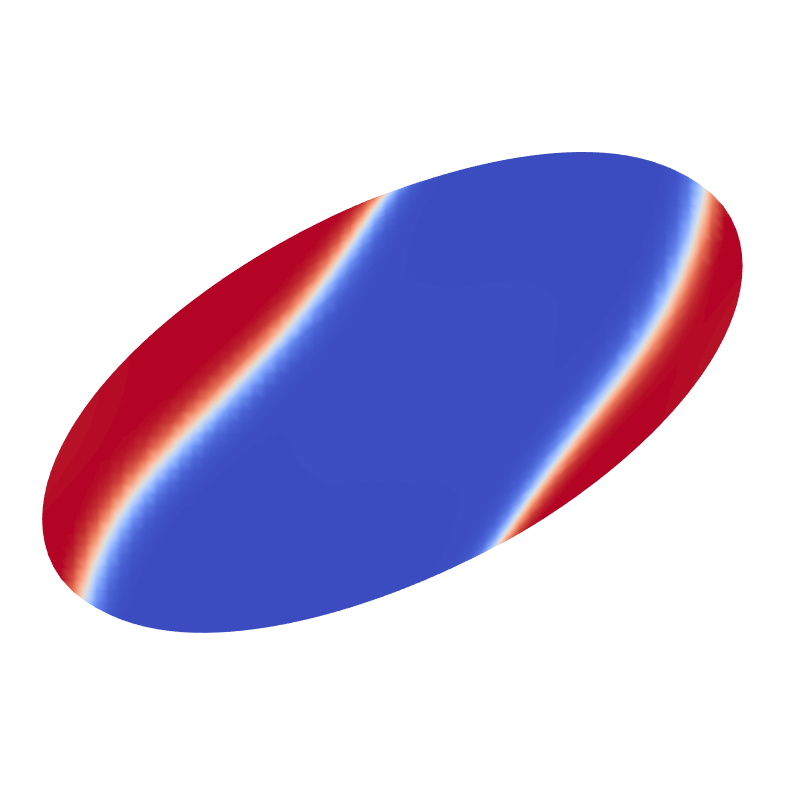} &
					\includegraphics[height=1.7cm]{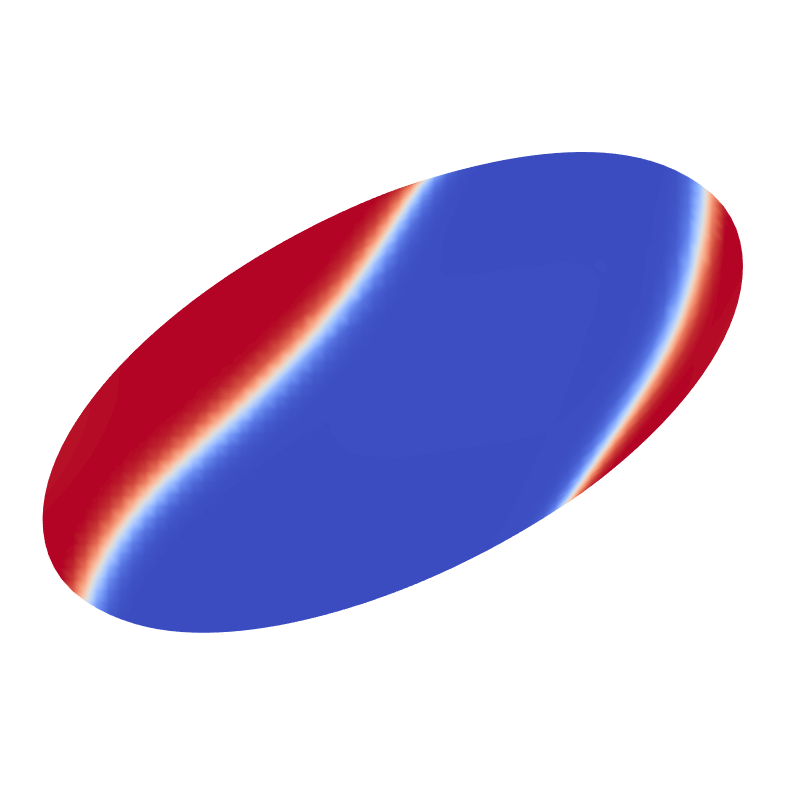} &
					\includegraphics[height=1.7cm]{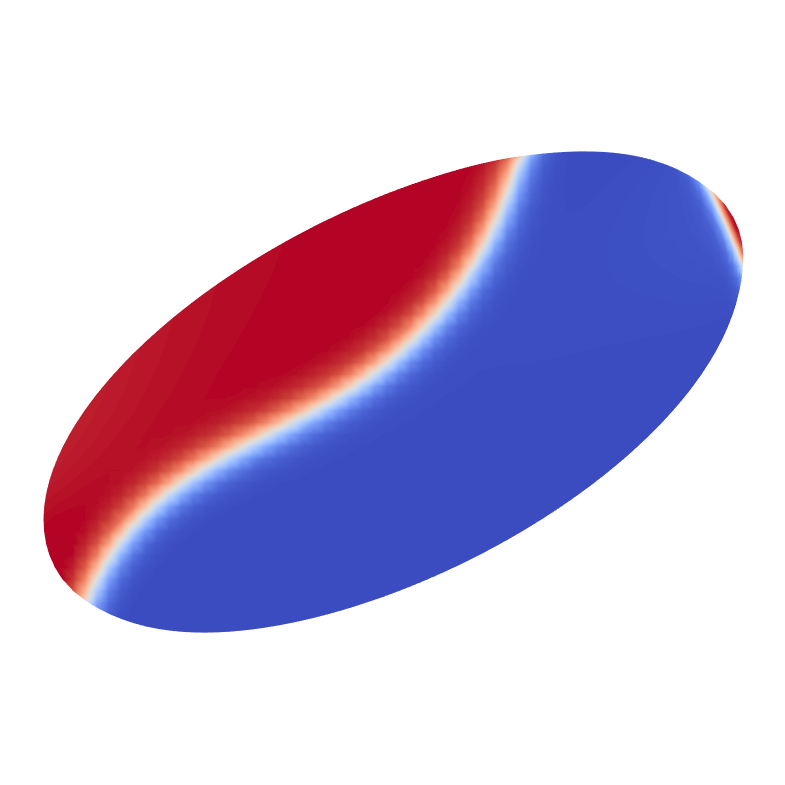} \tabularnewline			
			\end{tabular}
		}
		\subfigure[Bending Energy]{
			\includegraphics[height=5.5cm]{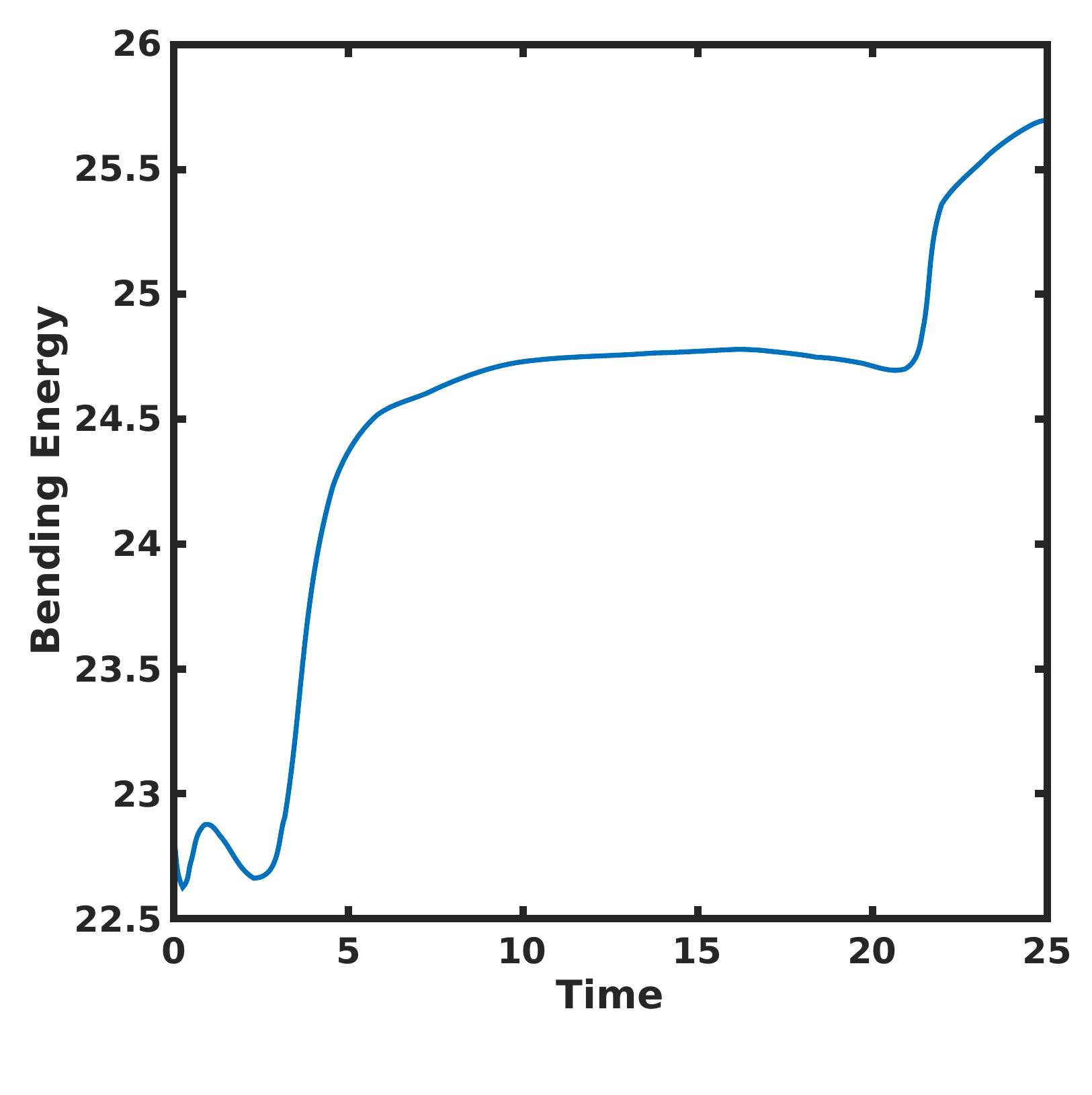}
		}
		\qquad
		\subfigure[Domain Boundary Energy]{			
			\includegraphics[height=5.5cm]{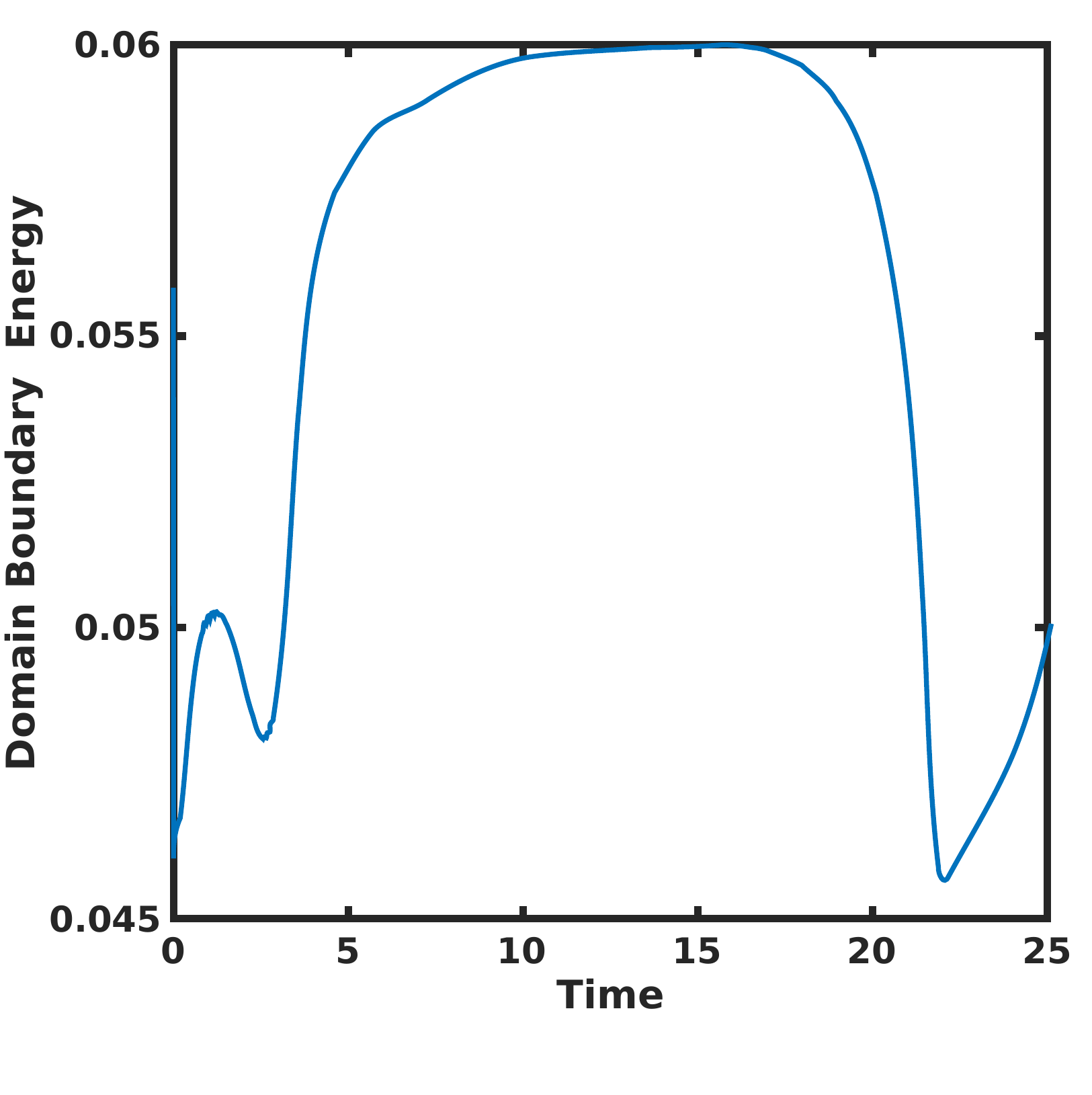}	
		}
	\end{center}
	\caption{Stationary dynamics of a vesicle with
	 $\bar{c}=0.4$ and $\alpha=20$. The soft
	phase has a bending rigidity of $\kappa_c^B=0.6$ and the Peclet number
	is $\Pe=0.2$. The domains remain stationary until one domain diffuses
	into the other.}
	\label{fig:DD}
\end{figure*}

\section{Results}
In this section, the dynamics of a multicomponent vesicle in the presence of shear flow is examined.
Unless otherwise noted, the initial shape is a prolate-ellipsoid
with half-axes lengths of $0.77$, $1.51$, $0.77$ in the x-, y-, and z-directions, 
respectively. This results in a reduced volume of $0.9$,
which measures the deviation of the volume from a perfect sphere with the same surface
area. The computational domain spans $[-3,3]^3$ with
a mesh size of $128^3$, while a constant time step of $5\times 10^{-3}$ is used. 
In a prior work the authors demonstrate qualitative convergence with these
parameters~\cite{GERA2018}. The computational domain is periodic in the x- and z- directions,
with wall boundary conditions in the y-direction. 
Shear flow is applied by imposing a velocity of $\vec{u}_{bc}=(\chi\textnormal{y},0,0)$, where $\chi=1$ is the normalized shear
rate, on the wall boundaries.

To remove dependence on the initial phase distribution, 
the initial condition for the lipid phases are assumed to be pre-segregated
domains covering the tips of the vesicles. Specifically,
the initial field is given by
$q_0=((2 + \textnormal{tanh}(20(y-y_0))) + \textnormal{tanh}(-20(y+y_0)))/2$,
where $y_0=0.682$, which results in an average concentration of 
$\bar{q}=0.4$.

In all cases the Cahn number is taken to be $\Cn=0.05$ while the capillary
bending number is fixed at $\Ca=20$ and the Reynolds number is $\Re=10^{-3}$. 
For simplicity, matched viscosity and density between the inner and outer fluids is assumed.
As stated previously, the spontaneous curvature is zero and the two lipid domains have matched Gaussian bending rigidity.
The normalized bending rigidity of the $q=0$ (hard-)phase, shown in blue below, is taken to be one, $\kappa_c^A=1$,
while the bending rigidity of the $q=1$ (soft-)phase, shown in red, has a value less than one, $\kappa_c^B<1$.
In most cases the surface Peclet number will vary between $\Pe=0.01$ and $\Pe=1$. 
Two ratios between the bending rigidity and the domain line energy
are considered: $\alpha=0.5$ and $\alpha=20$. The value of $\alpha$ will be clearly stated.

The results begin with a characterization of the different dynamics observed while using the present model.
The bending energy and the interfacial
energy curves are shown to describe and explain the distinguishing features observed.
Following this, the influence of varying bending rigidity, surface Peclet number, and domain line
energy is explored.

\subsection{Sample Dynamics}

This section describes the spectrum of dynamics observed with the variation of
the bending rigidities and surface Peclet number when a multicomponent vesicle
is subjected to an external shear flow.
For a single-component vesicle, matching the inner and outer fluid
viscosities results in the tank-treading regime~\cite{Kantsler2006}.
When a multicomponent vesicle is exposed to shear flow, this may no longer be true
as two primary competing forces exist. 
First, the surrounding fluid attempts to advect the phase along the vesicle membrane. As will be demonstrated,
this movement results in changes in the overall energy of the vesicle. This results in the second,
restorative, force: the surface diffusion of the domains to reduce the energy of the system.
Varying the soft phase bending rigidity and surface Peclet number results 
in three different types of observed dynamics: 1) Stationary/Diffusion Dominated, 
2) Vertical Banding, and 3) Phase Treading.

\begin{figure*}[!ht]
	\begin{center}
		\subfigure[]{
			\begin{tabular}{
				>{\centering}m{1.0cm}>{\centering}m{1.7cm}>{\centering}m{1.7cm}>{\centering}m{1.7cm}>{\centering}m{1.7cm}>{\centering}m{1.7cm}>{\centering}m{1.7cm}>{\centering}m{1.7cm}}
				& \multicolumn{7}{c}{Time} \tabularnewline
				\cline{2-8}
				View & $0.0$ & $2.0$ & $5.0$ & $10.0$ & $12.5$ & $15.0$ & $20.0$ \tabularnewline
				Iso & 
					\includegraphics[height=1.7cm]{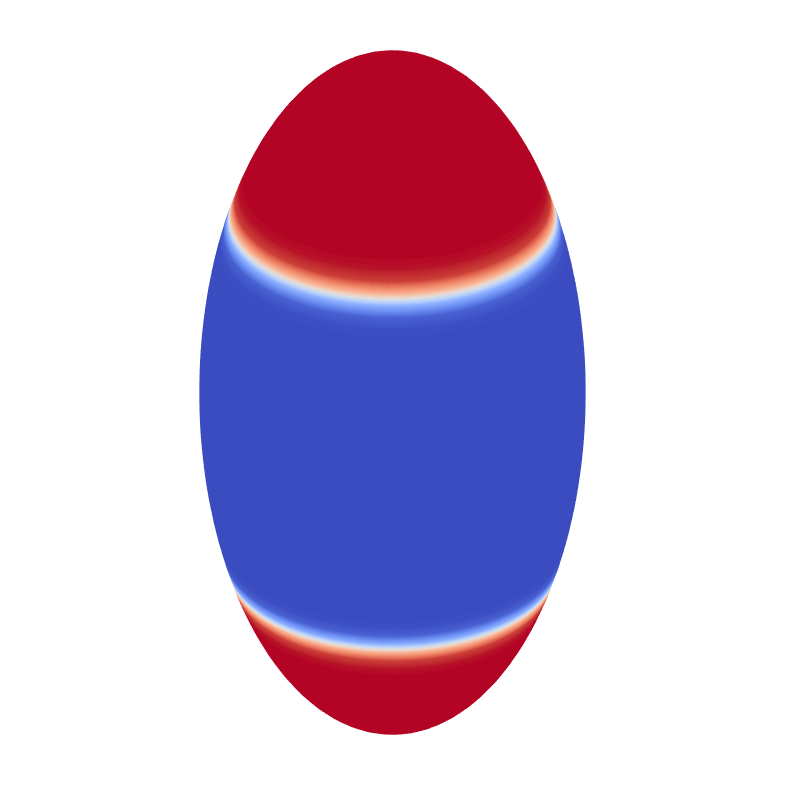} &
					\includegraphics[height=1.7cm]{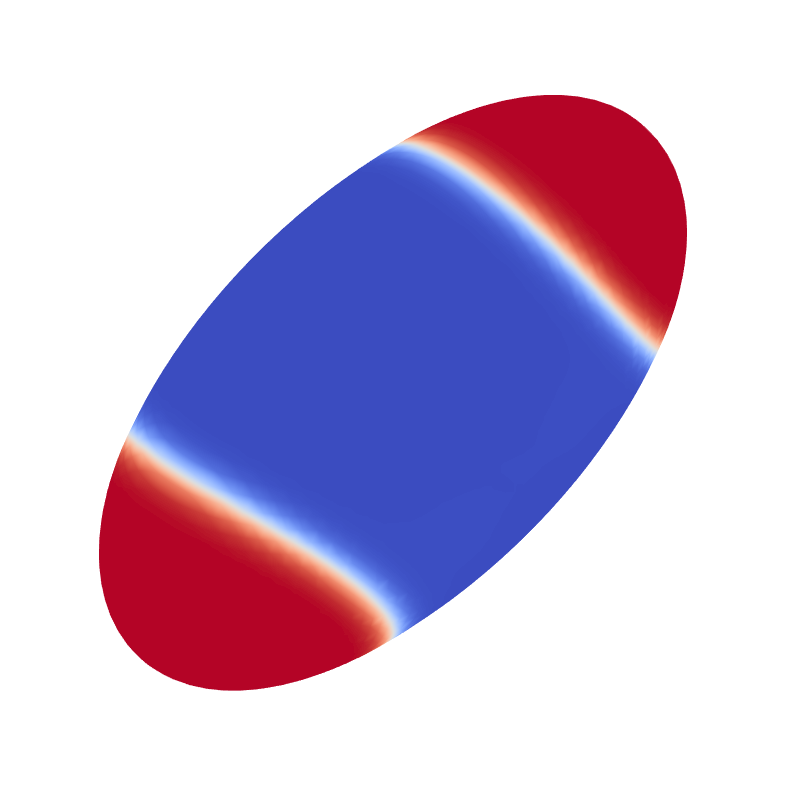} &
					\includegraphics[height=1.7cm]{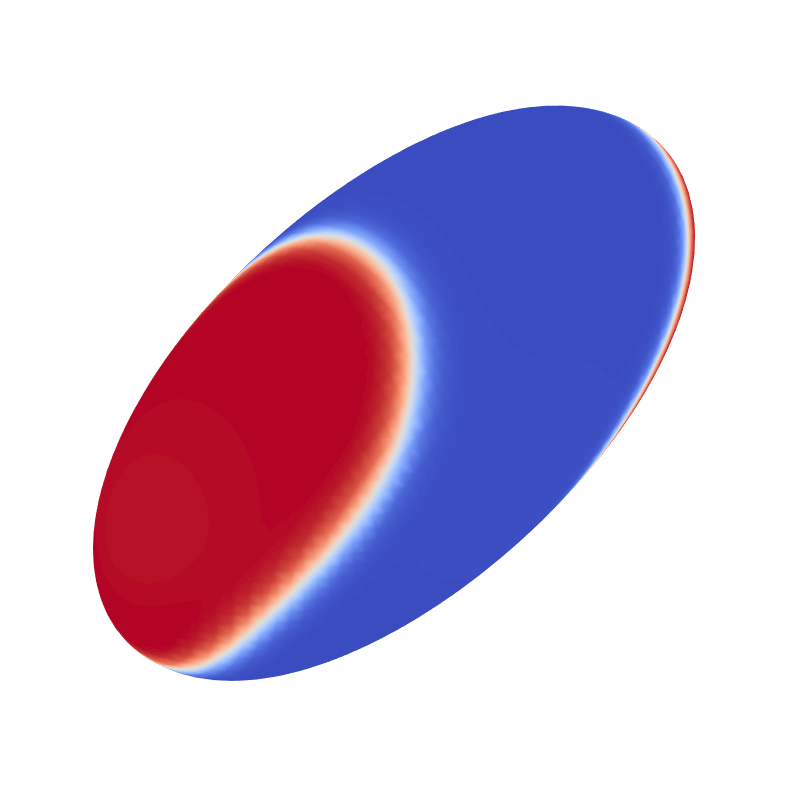} &
					\includegraphics[height=1.7cm]{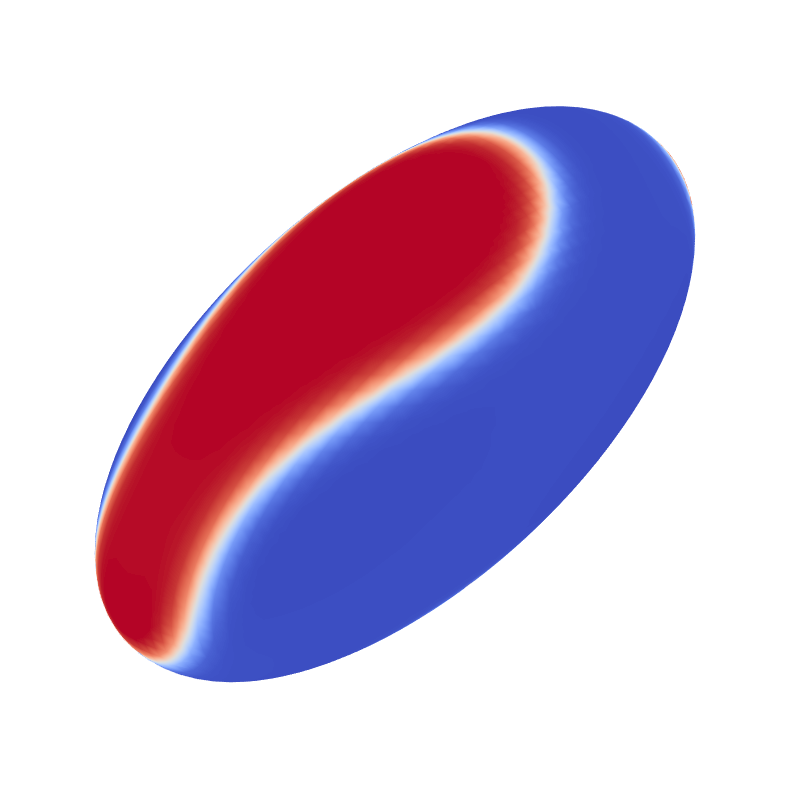} &
					\includegraphics[height=1.7cm]{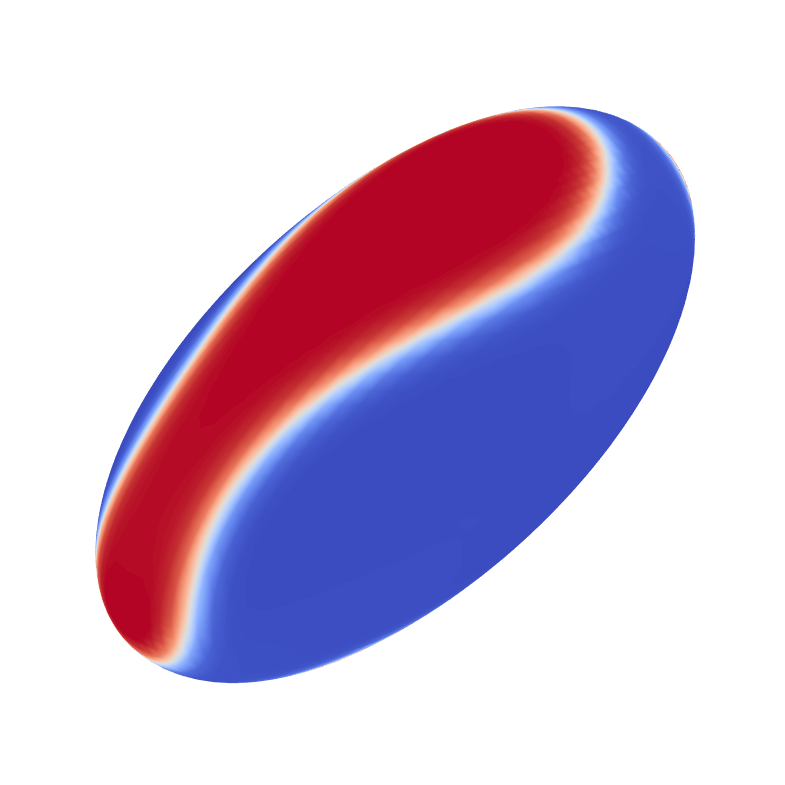} &
					\includegraphics[height=1.7cm]{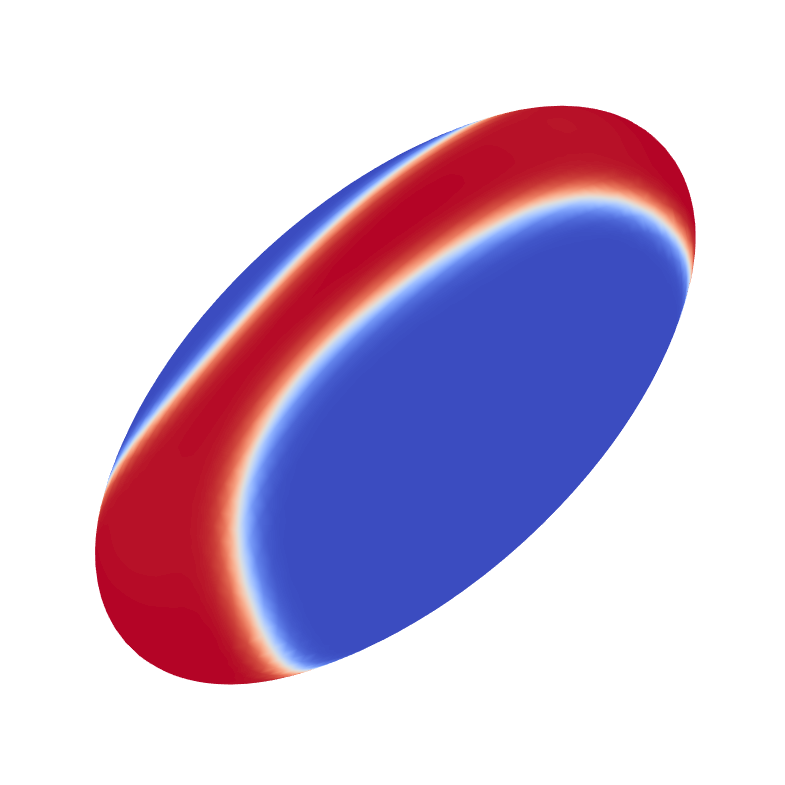} &
					\includegraphics[height=1.7cm]{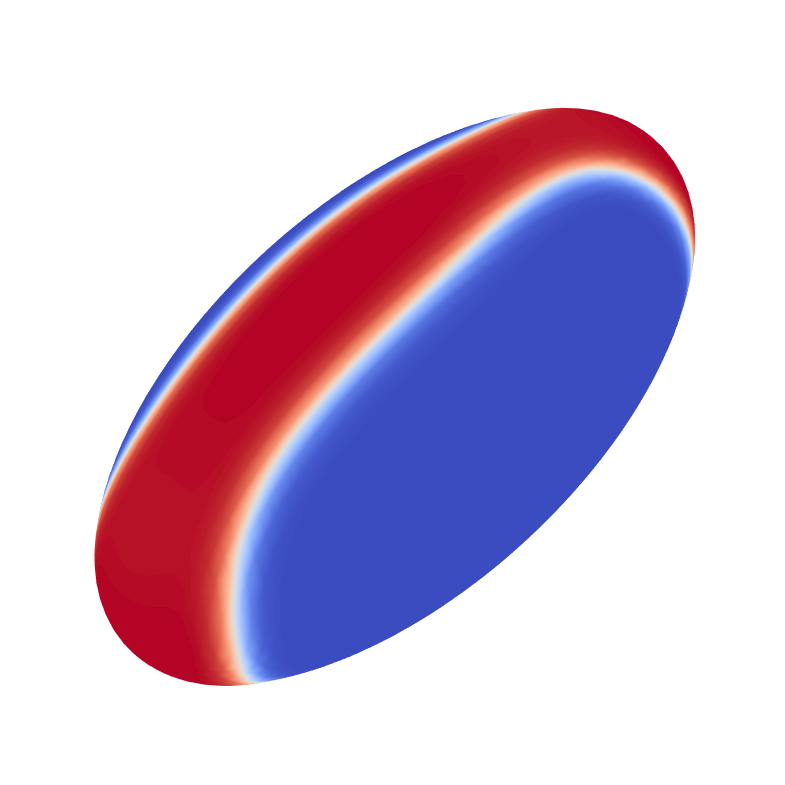} \tabularnewline
				X-Y & 
					\includegraphics[height=1.7cm]{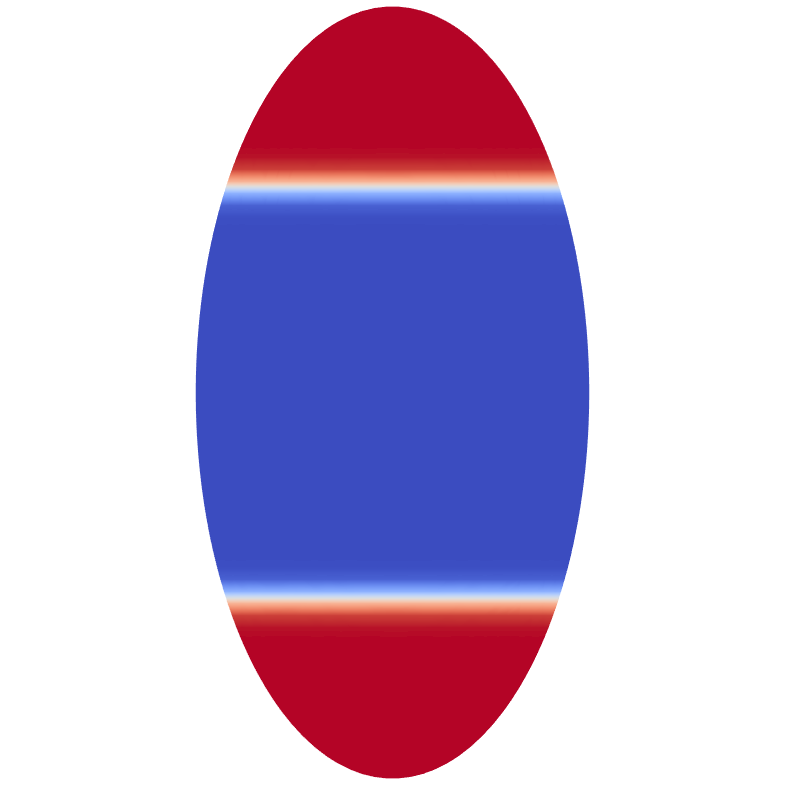} &
					\includegraphics[height=1.7cm]{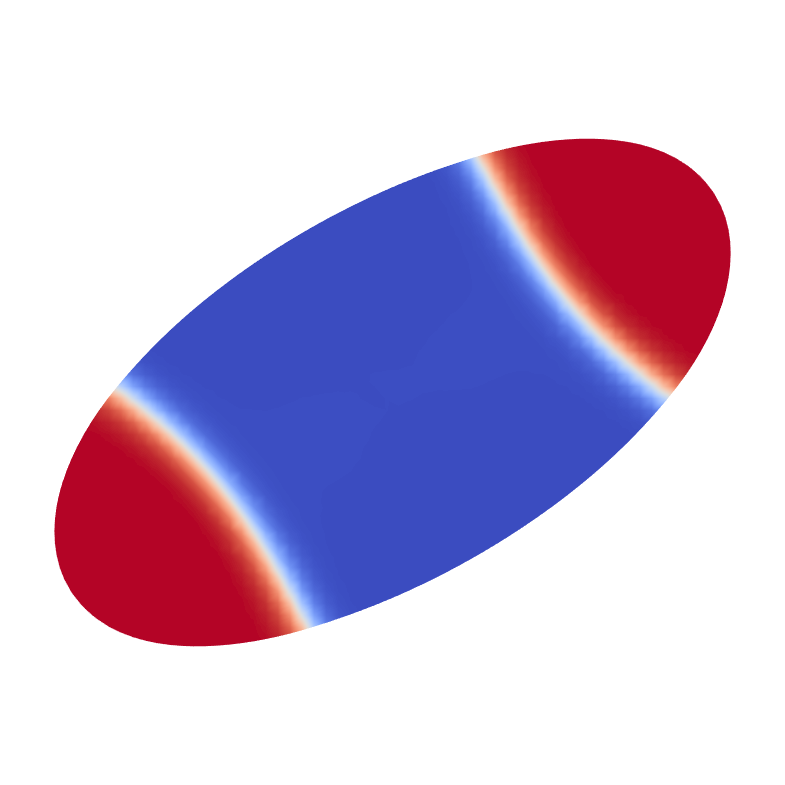} &
					\includegraphics[height=1.7cm]{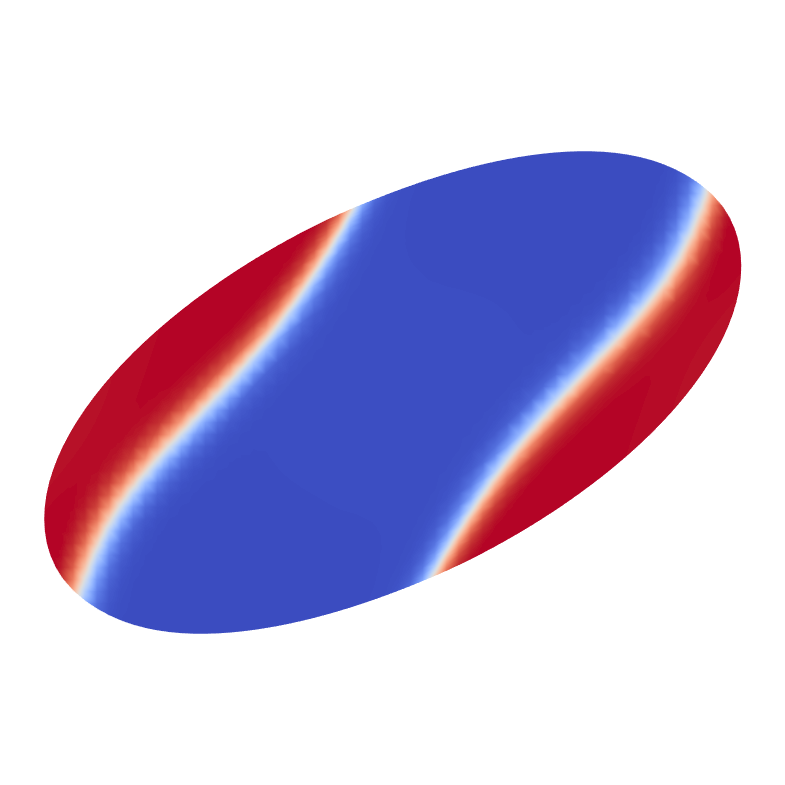} &
					\includegraphics[height=1.7cm]{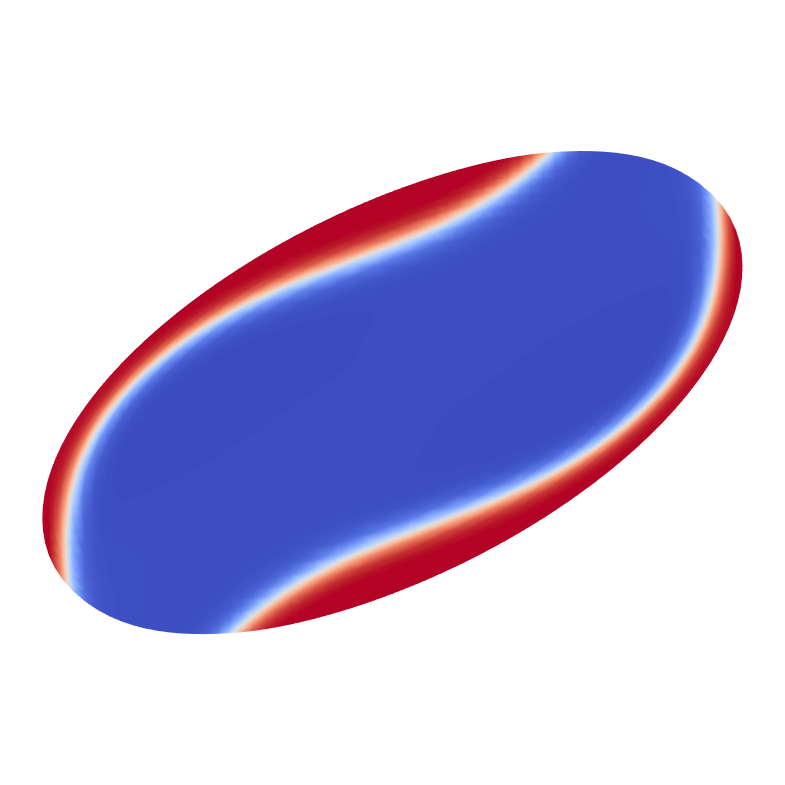} &
					\includegraphics[height=1.7cm]{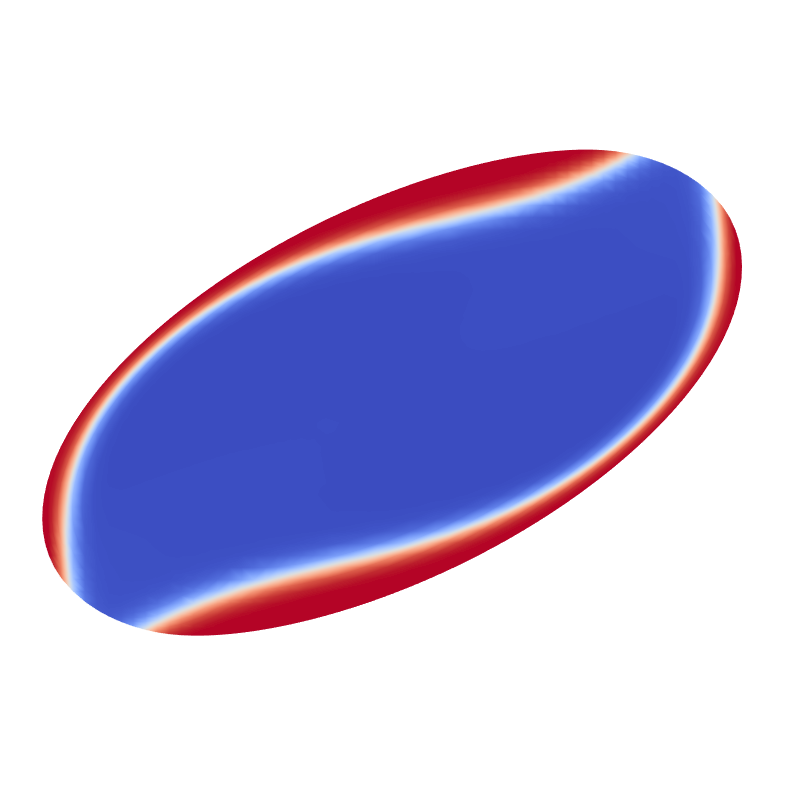} &
					\includegraphics[height=1.7cm]{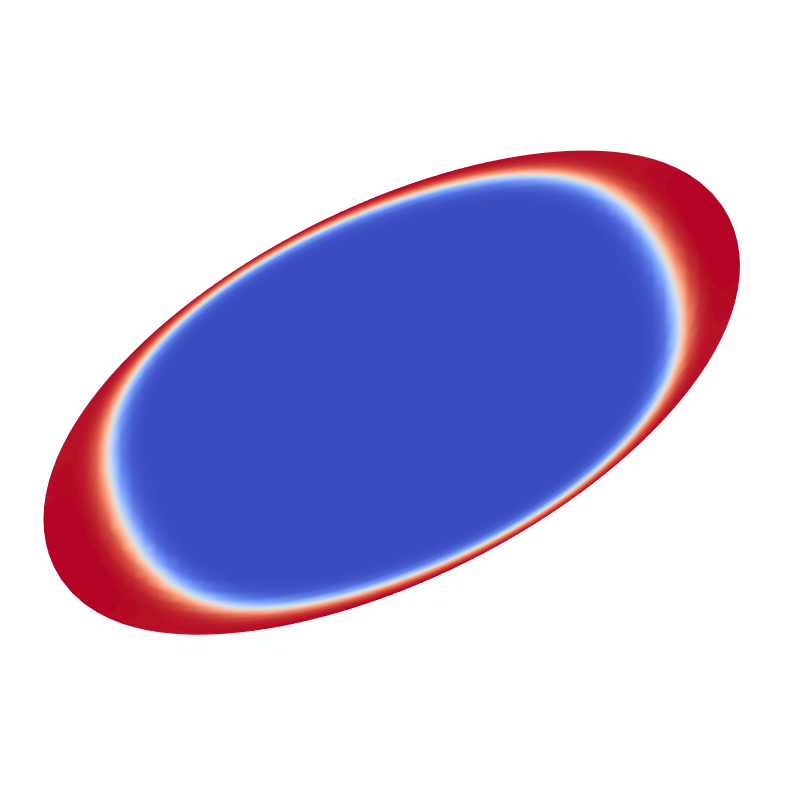} &
					\includegraphics[height=1.7cm]{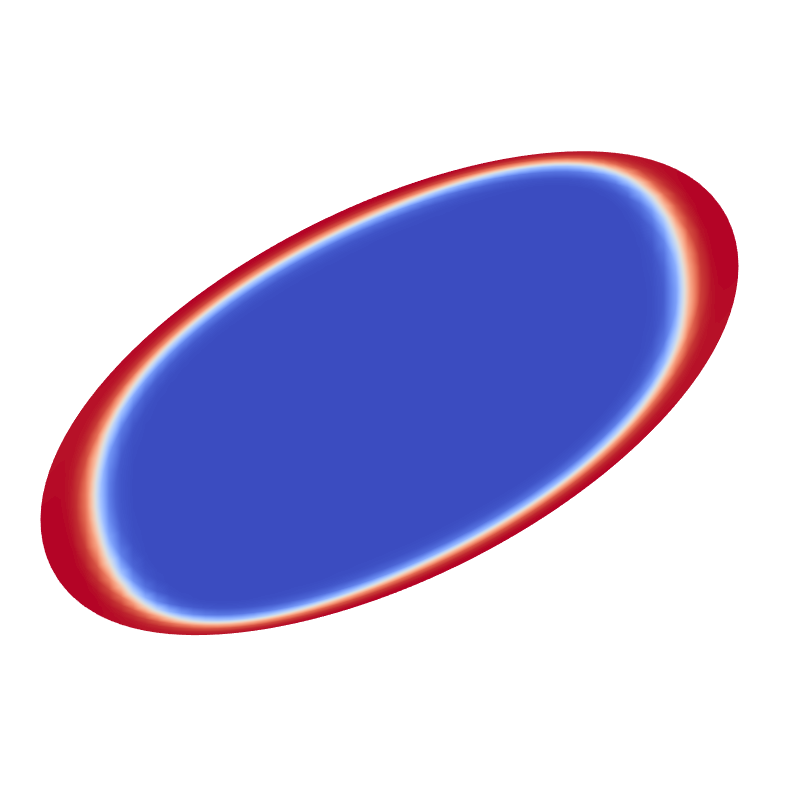} \tabularnewline
				X-Z & 
					\includegraphics[height=1.7cm]{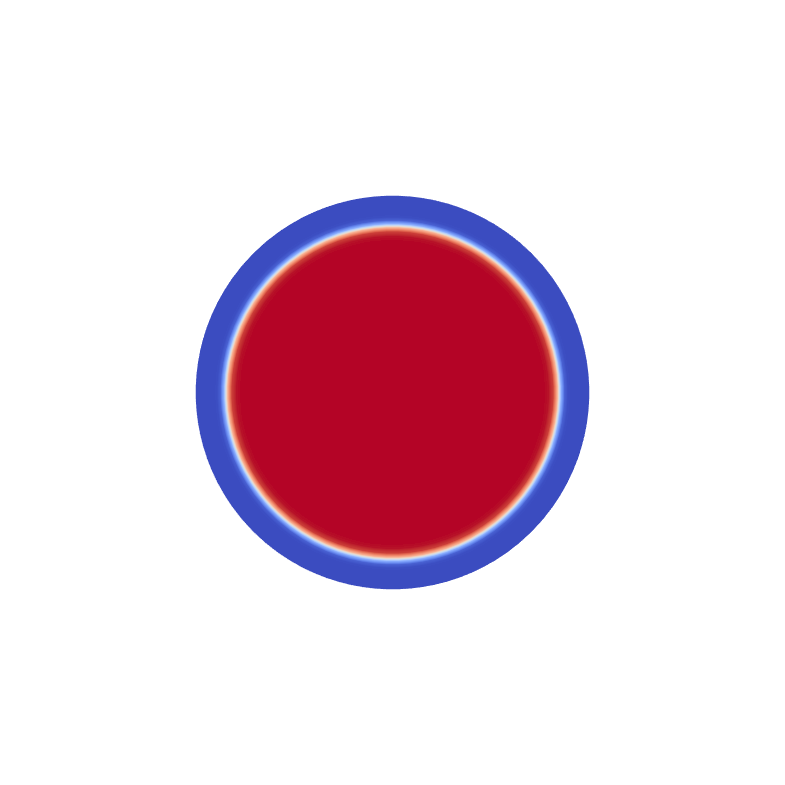} &
					\includegraphics[height=1.7cm]{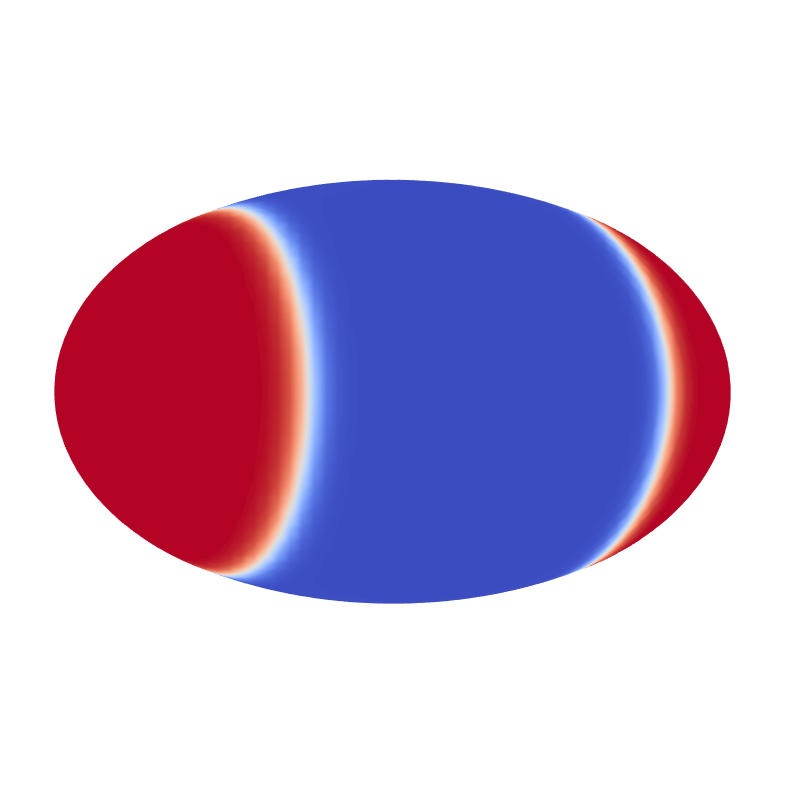} &
					\includegraphics[height=1.7cm]{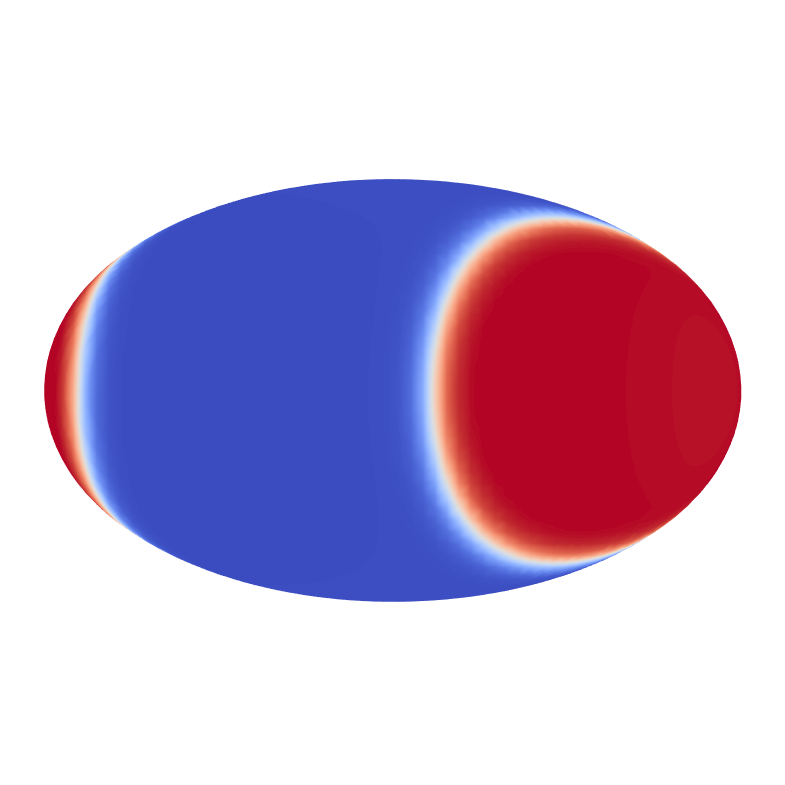} &
					\includegraphics[height=1.7cm]{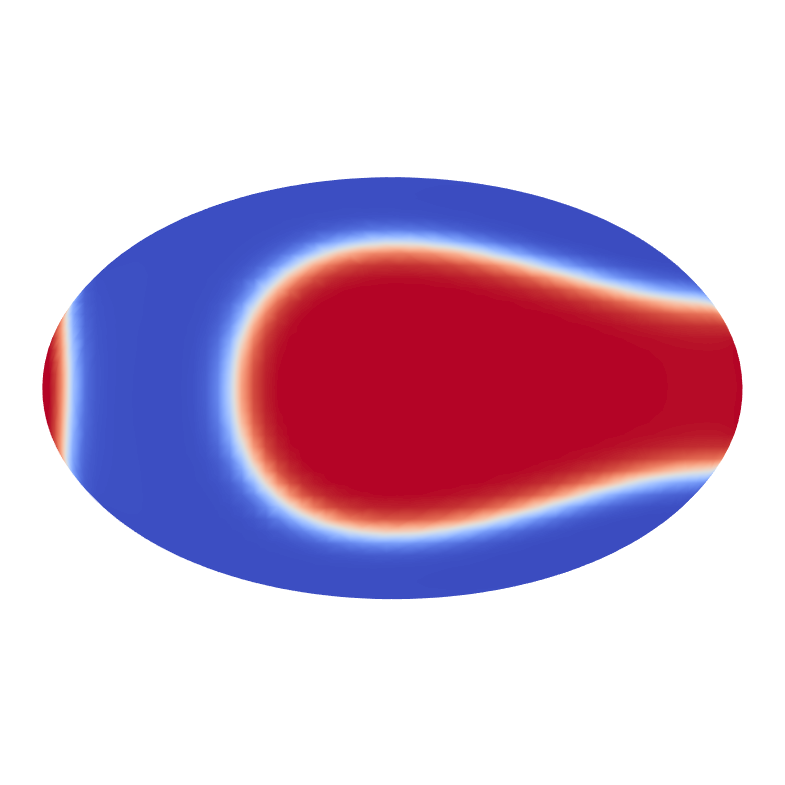} &
					\includegraphics[height=1.7cm]{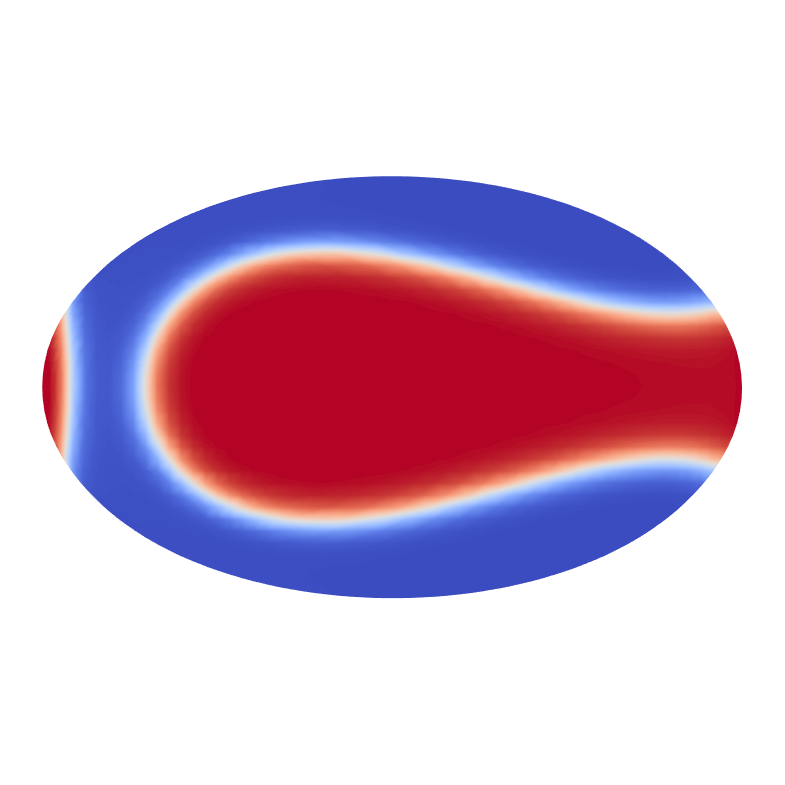} &
					\includegraphics[height=1.7cm]{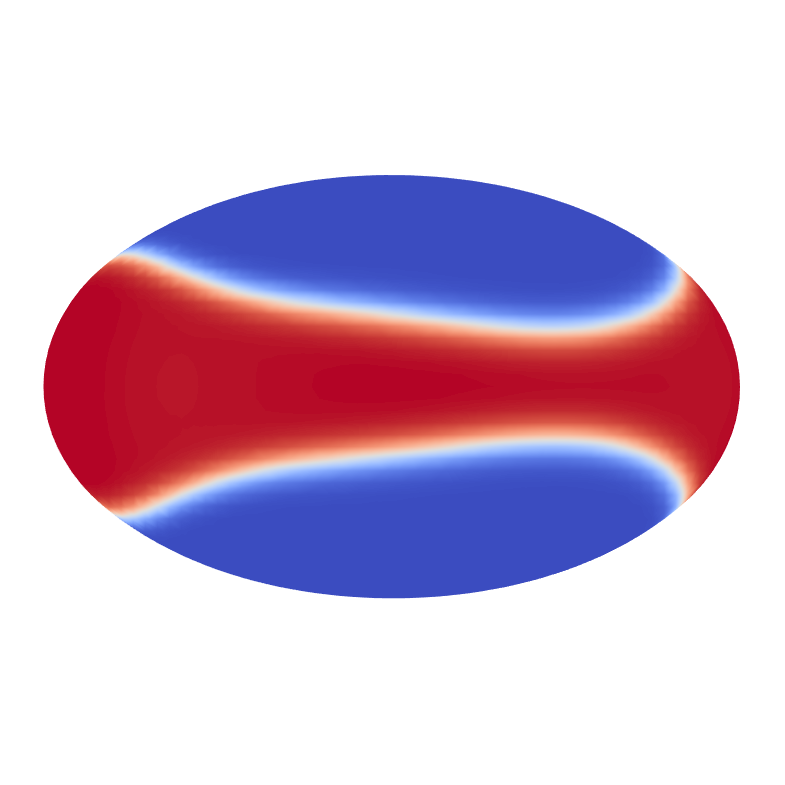} &
					\includegraphics[height=1.7cm]{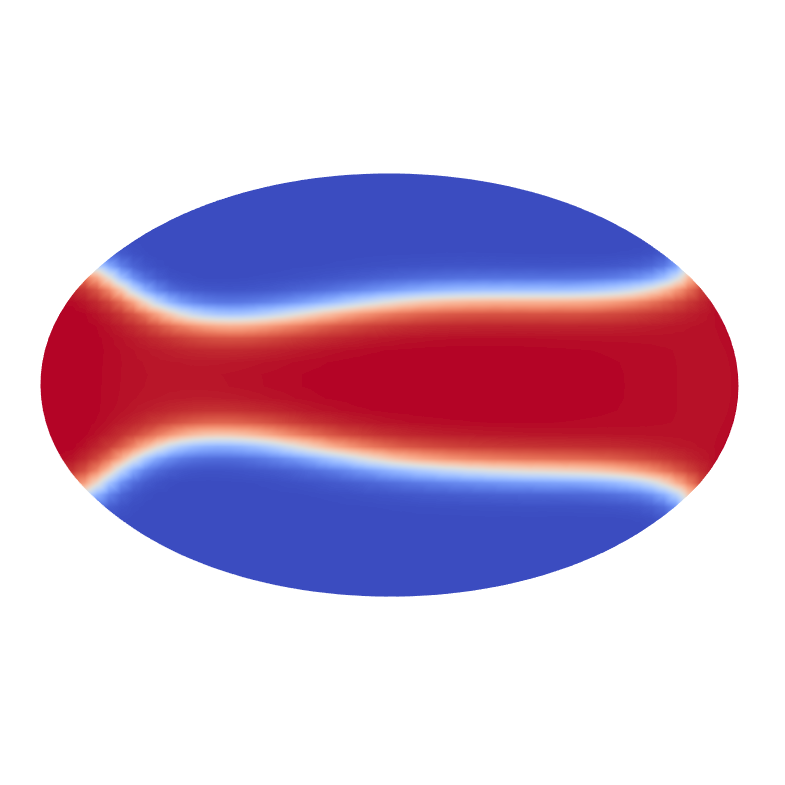} \tabularnewline
			\end{tabular}
		} 
		\subfigure[Bending Energy]{
			\includegraphics[height=5.5cm]{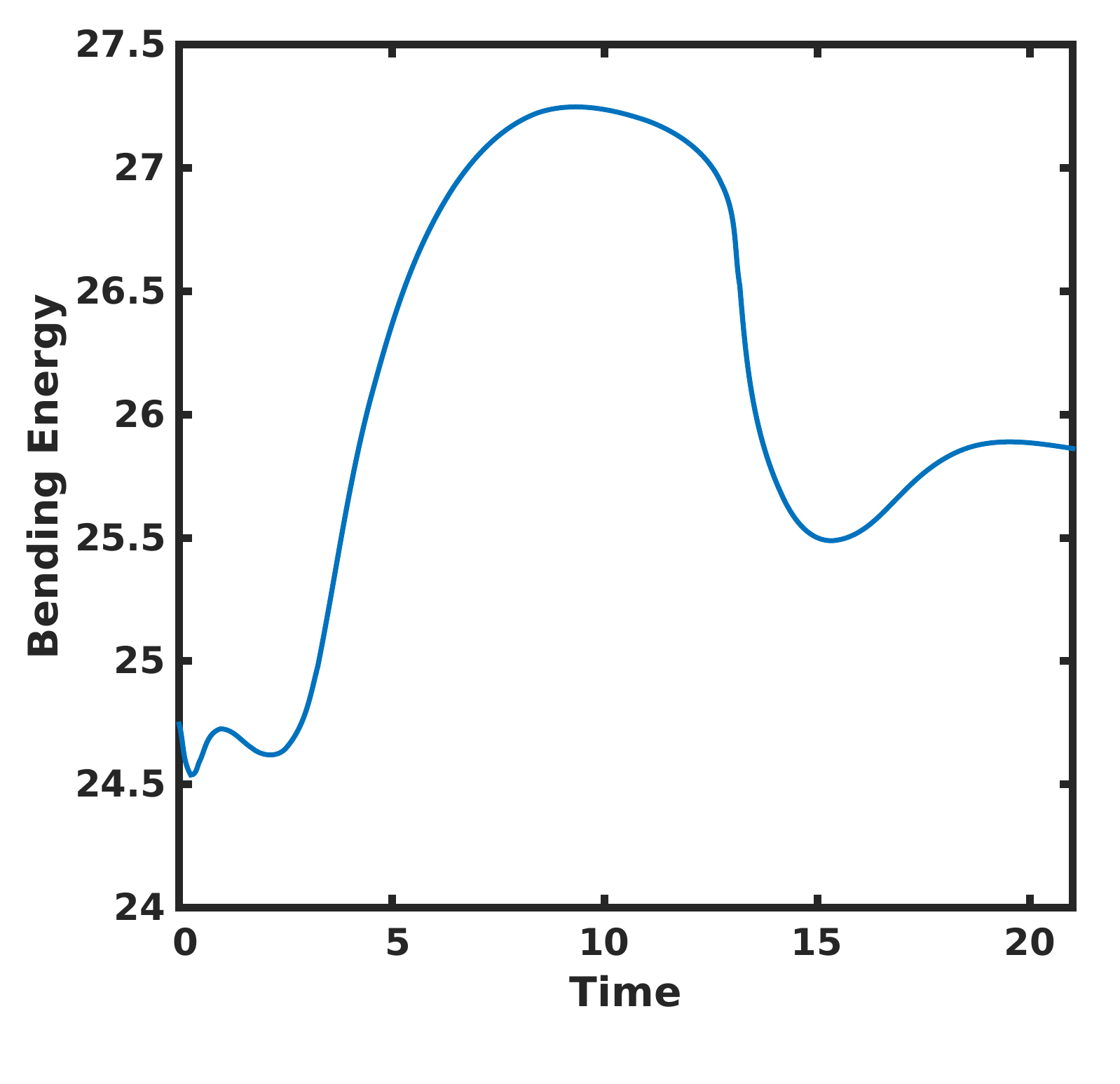}
		} 
		\qquad
		\subfigure[Domain Boundary Energy]{			
			\includegraphics[height=5.5cm]{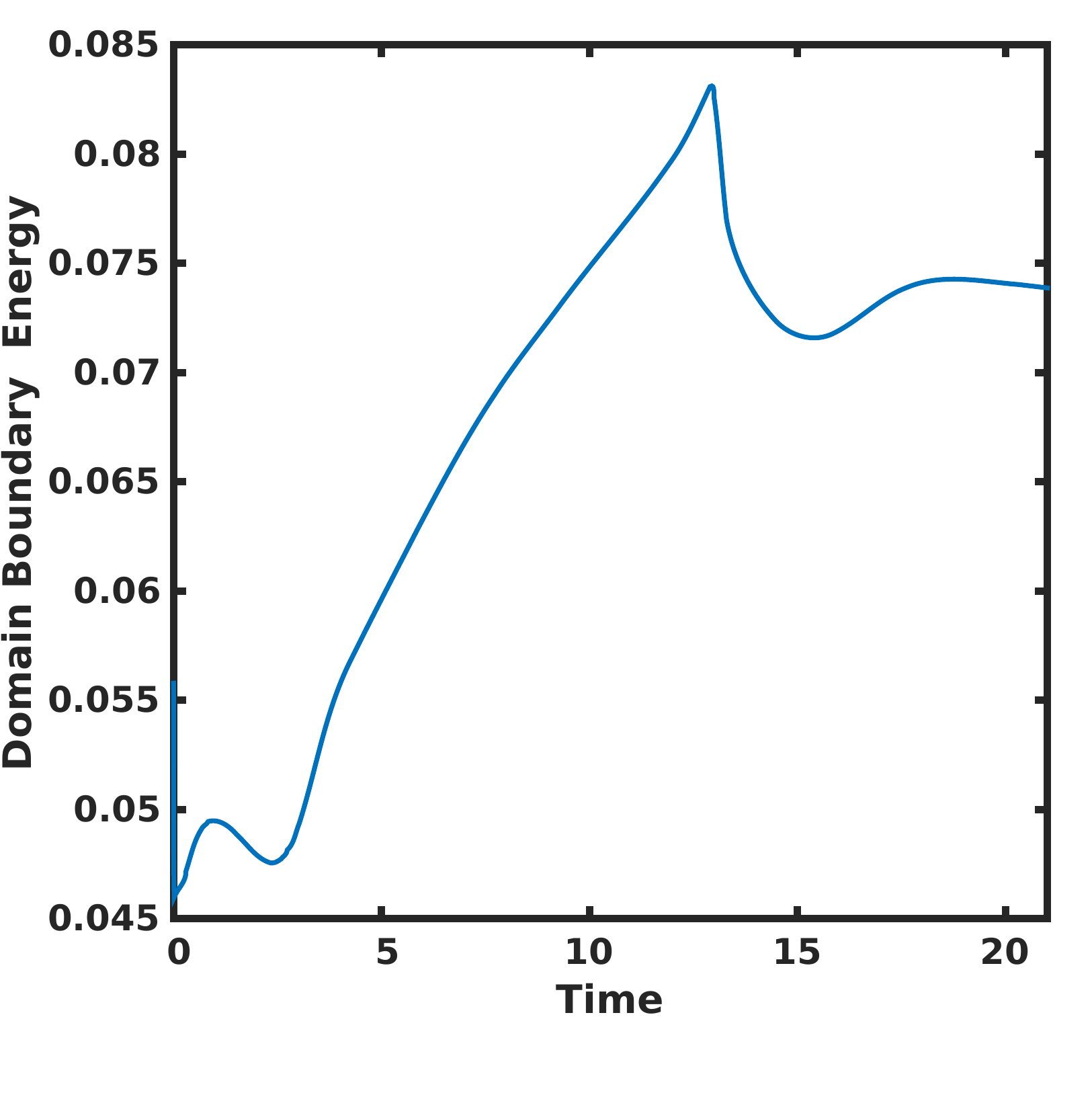}	
		}
	\end{center}
	\caption{Vertical banding dynamics of a vesicle with
		$\bar{c}=0.4$ and $\alpha=20$. The soft
		phase has a bending rigidity of $\kappa_c^B=0.7$ and the Peclet number
		is $\Pe=0.5$. The higher Peclet number allows for the domains to extend and
		eventually merge.}
	\label{fig:VSB}
\end{figure*}

\subsubsection{Stationary/Diffusion Dominated Dynamics} 
\label{sec:diffusionDominated}

Recall that the surface Peclet number indicates how quickly the surface phases 
can adjust to changes in the system energy; smaller values of $\Pe$ indicate
the surface phases can adjust quickly relative to the advection forces.
Alternatively, decreasing the bending rigidity of the soft phase 
decreases the overall energy when the soft phase inhabits the high curvature regions of 
the vesicle. 
For small values of $\Pe$ and $\kappa_c^B$, it has been observed that the domains remain at the 
vesicle tips until one domain grows at the expense of the other (which is 
typical of Cahn-Hilliard models). This type of dynamic is denoted as stationary 
or diffusion dominated, and has been previously seen for two-dimensional multicomponent vesicles~\cite{C6SM02452A}. 

For example, consider a vesicle where the soft-phase bending rigidity is given by $\kappa_c^B=0.6$ 
with a Peclet number of $\Pe=0.2$, Fig~\ref{fig:DD}.
Until a time of $t=2.5$ very little deformation of the domains is observed as the vesicle
reaches the equilibrium inclination angle. From a time of $t=2.5$ to $t=5$,
the domains deform until reaching an equilibrium, elongated shape. 
They remain in this shape until a time of $t=15$, after which 
one domain grows at the expense of the other domain.

This dynamic is confirmed by considering the bending and domain boundary energy, Fig~\ref{fig:DD}.
Initially, there is growth in the bending and
domain boundary energy as the domains on the surface of the vesicle change
from circular to slight elongated. Both energies 
remain relatively constant, until one of the domains grows dramatically
to reduce the domain boundary energy. This merging event 
results in a larger bending energy, as a smaller 
amount of the softer phase is in the high curvature tips. After reaching
the minima, there is a slight increase in domain boundary and bending energy as the
domains slightly elongate, and a major portion of it lies on the low curvature
region of the vesicle. Similar dynamics 
have been observed in the recent two-dimensional work of Liu et al.~\cite{C6SM02452A}.

\begin{figure*}[!ht]
	\begin{center}
		\subfigure[]{
			\begin{tabular}{
				>{\centering}m{1.0cm}>{\centering}m{1.7cm}>{\centering}m{1.7cm}>{\centering}m{1.7cm}>{\centering}m{1.7cm}>{\centering}m{1.7cm}>{\centering}m{1.7cm}>{\centering}m{1.7cm}>{\centering}m{1.7cm}}
				& \multicolumn{8}{c}{Time} \tabularnewline
				\cline{2-9}
				View & $0.0$ & $0.5$ & $9.0$ & $13.0$ & $15.0$ & $17.0$ & $18.0$ & $20.0$ \tabularnewline
				Iso & 
					\includegraphics[height=1.7cm]{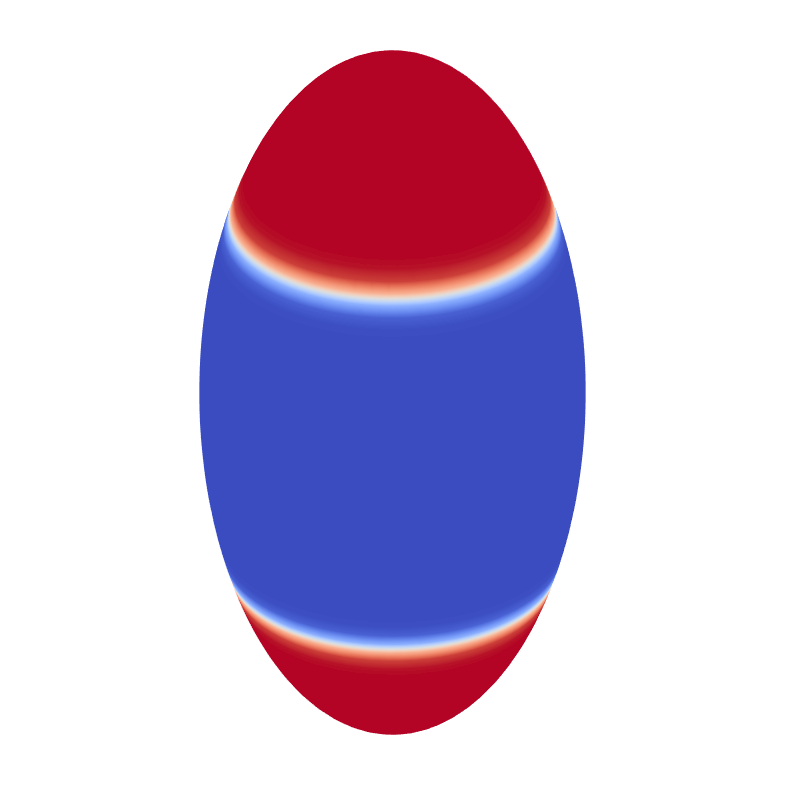} &
					\includegraphics[height=1.7cm]{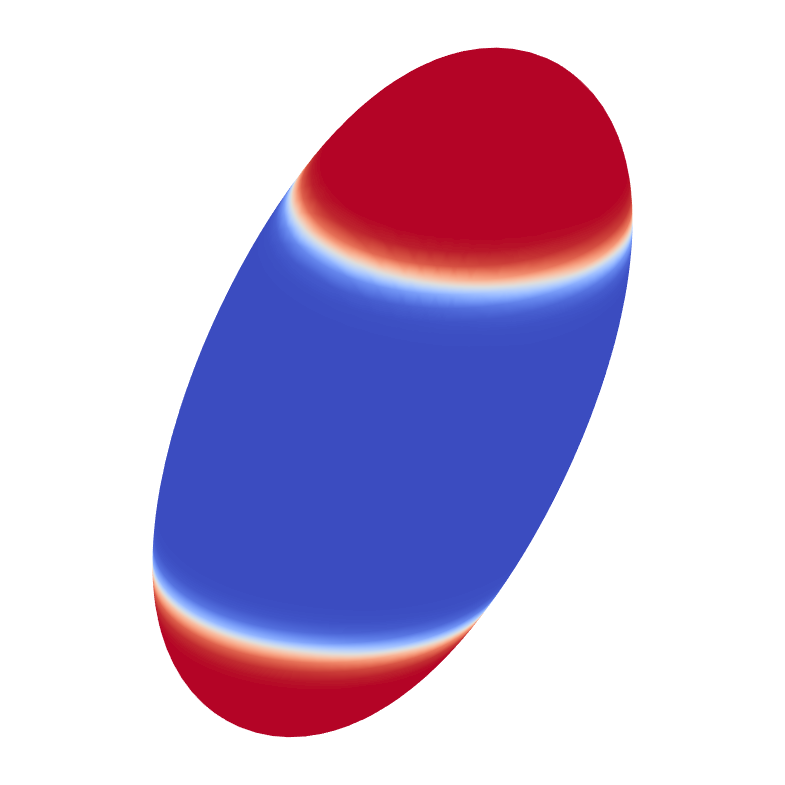} &
					\includegraphics[height=1.7cm]{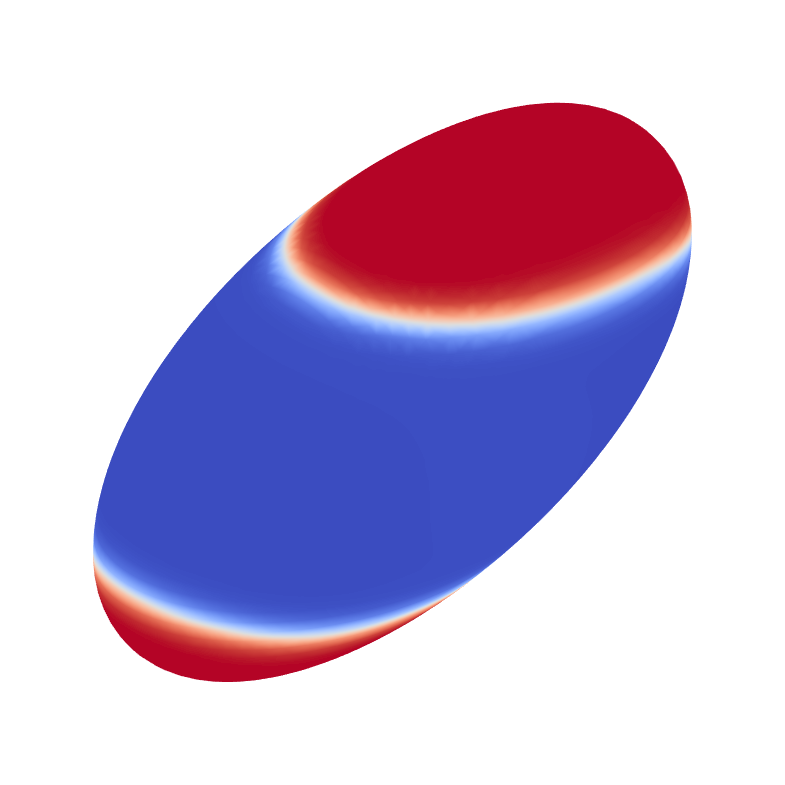} &
					\includegraphics[height=1.7cm]{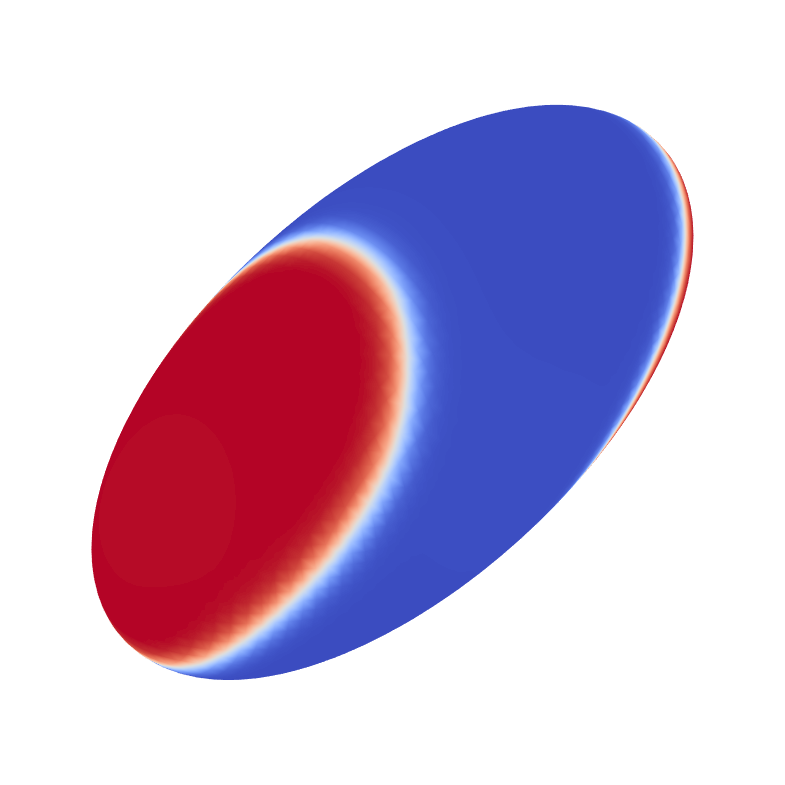} &
					\includegraphics[height=1.7cm]{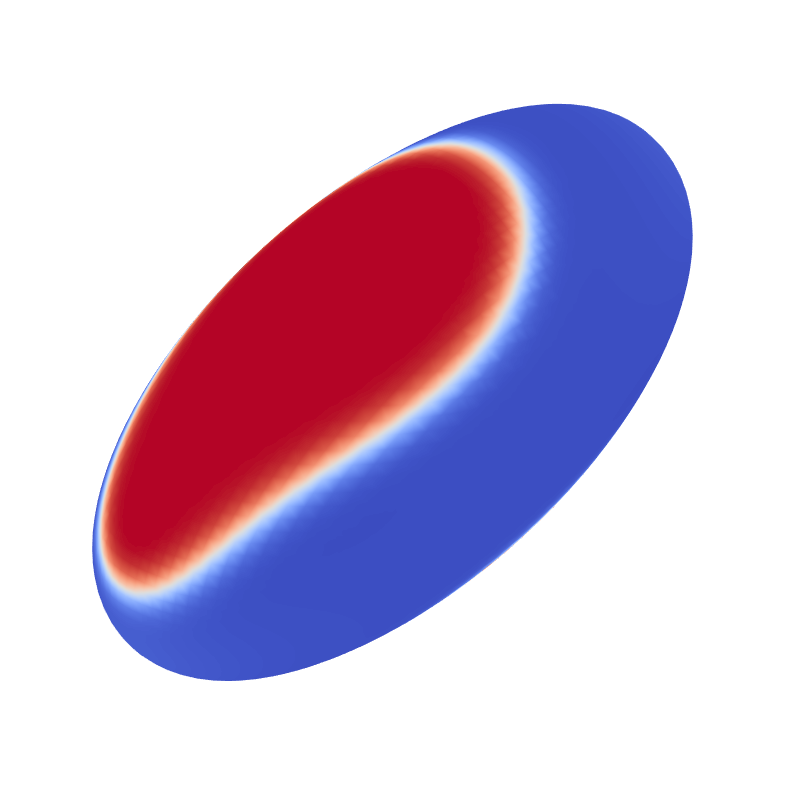} &
					\includegraphics[height=1.7cm]{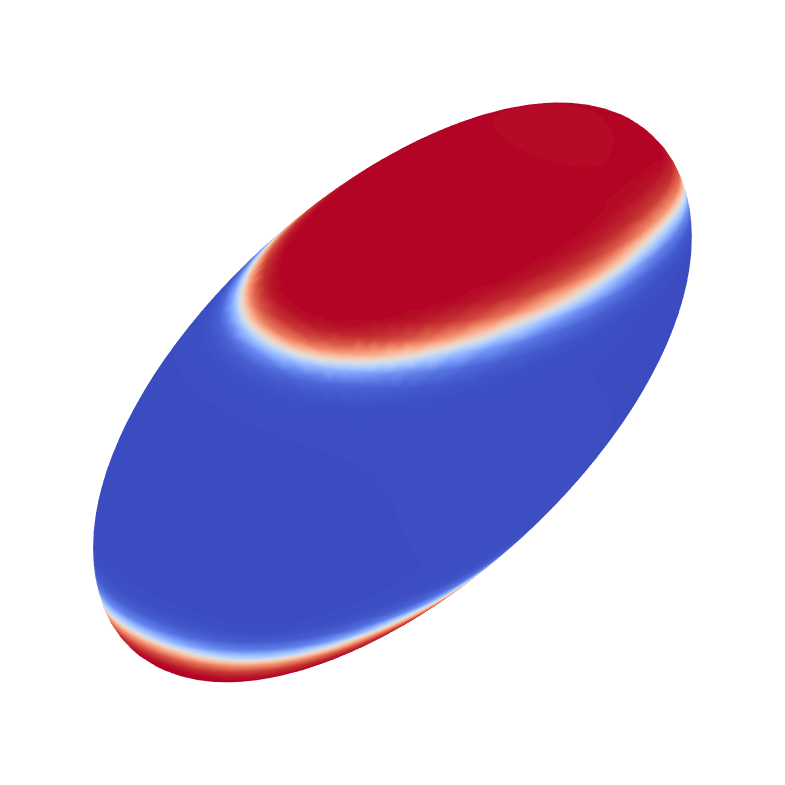} &
					\includegraphics[height=1.7cm]{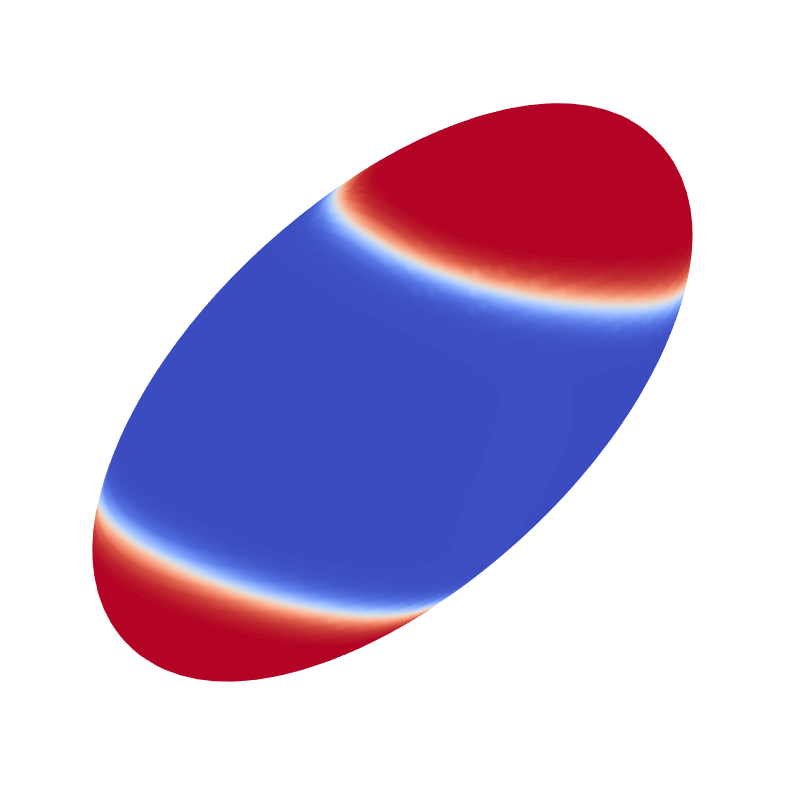} &
					\includegraphics[height=1.7cm]{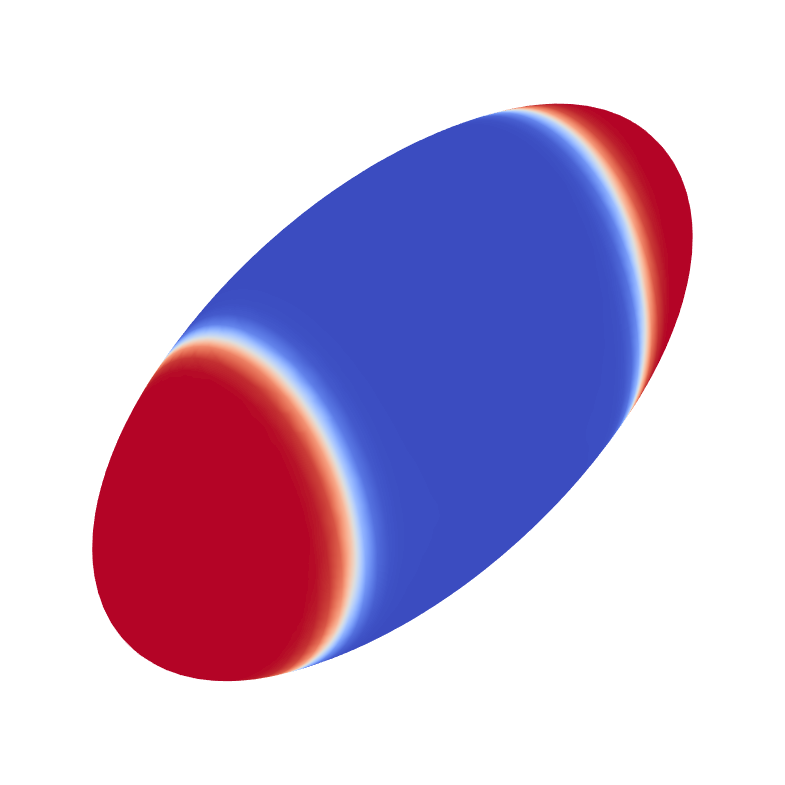} \tabularnewline
				X-Y & 
					\includegraphics[height=1.7cm]{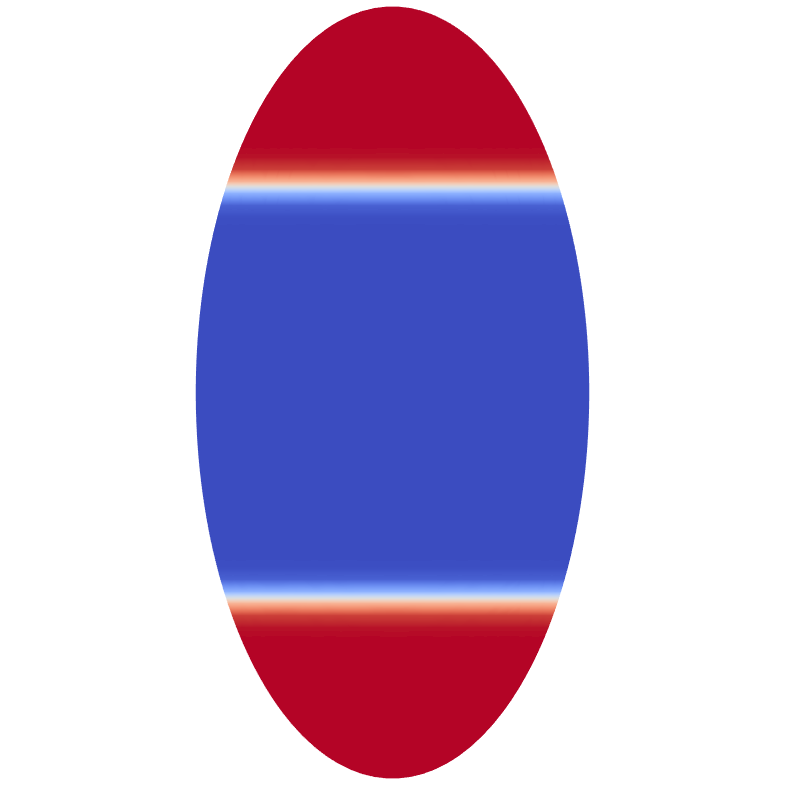} &
					\includegraphics[height=1.7cm]{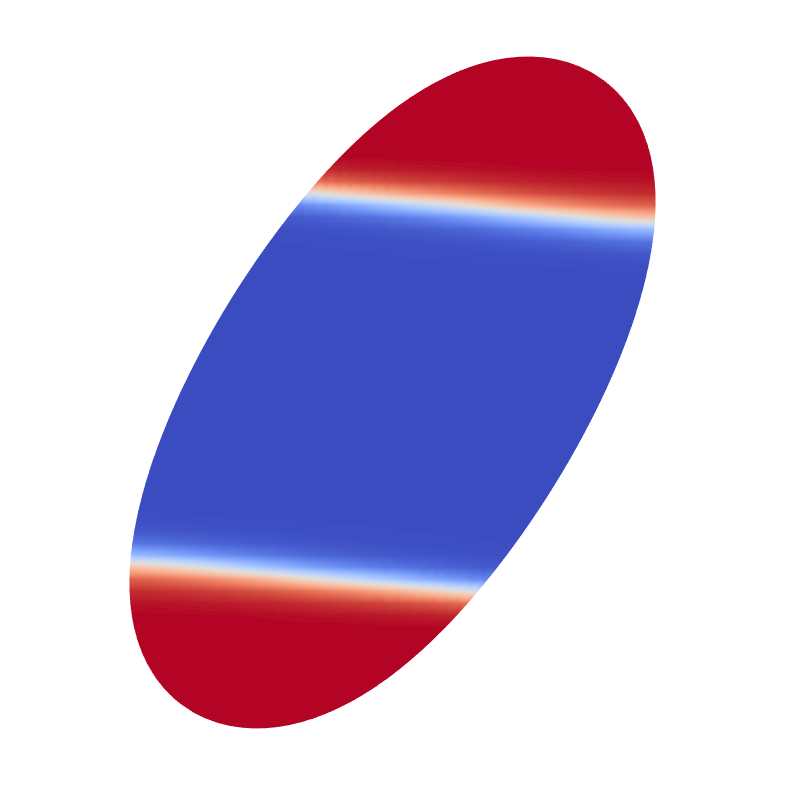} &
					\includegraphics[height=1.7cm]{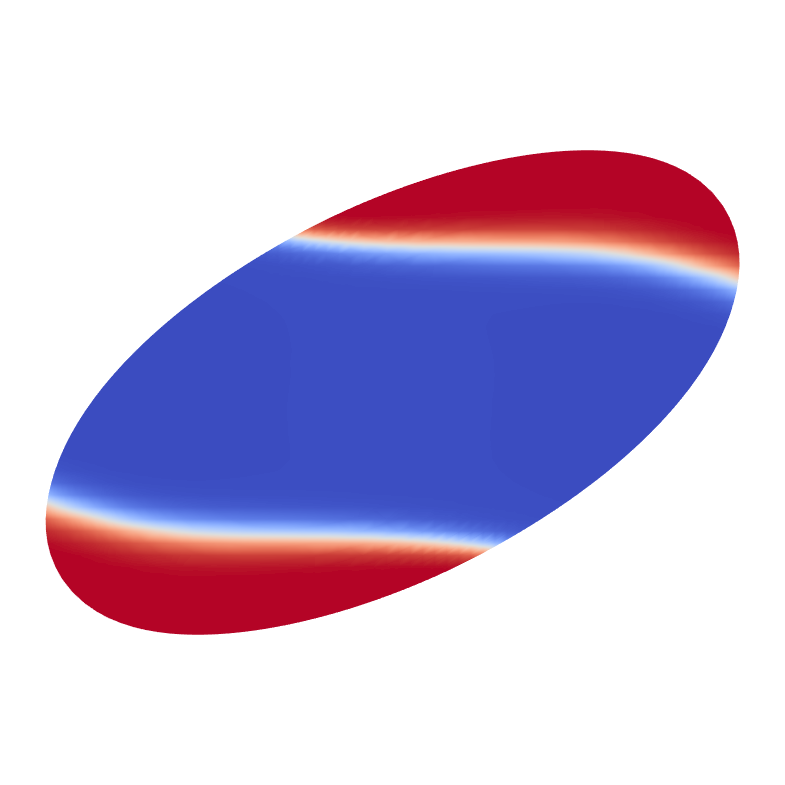} &
					\includegraphics[height=1.7cm]{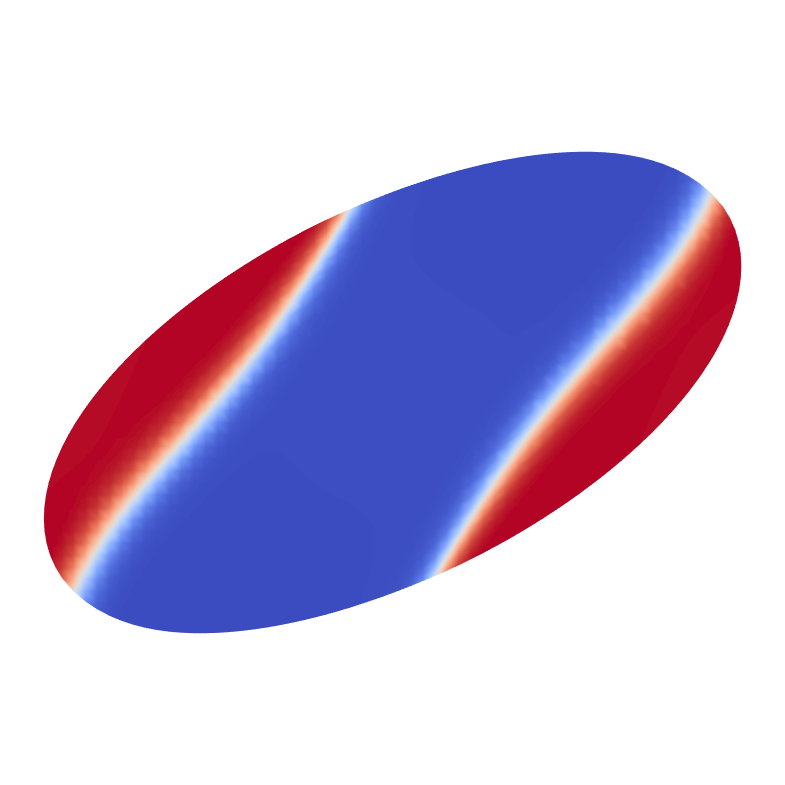} &
					\includegraphics[height=1.7cm]{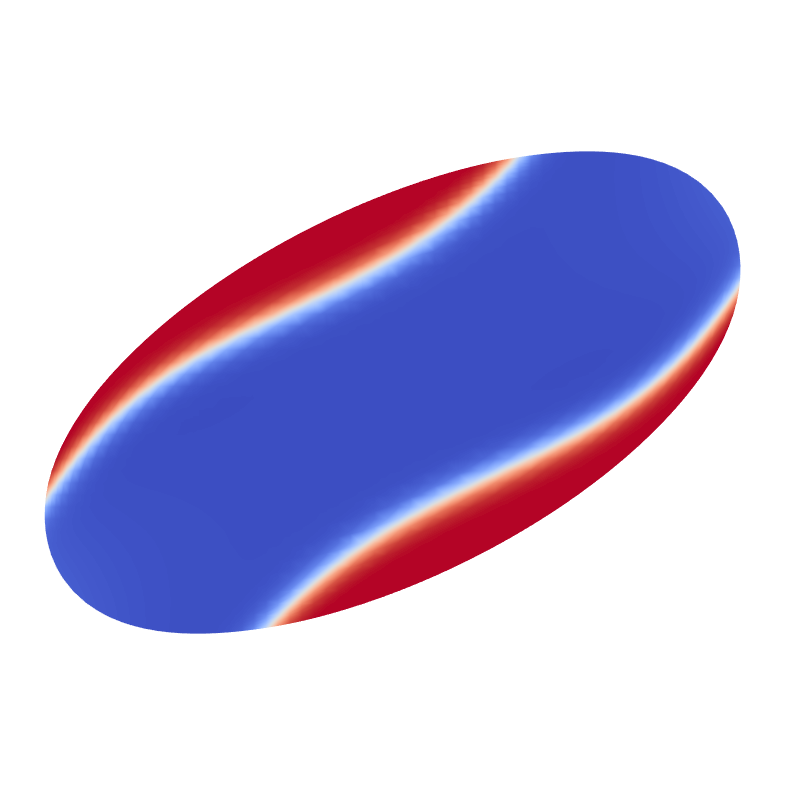} &
					\includegraphics[height=1.7cm]{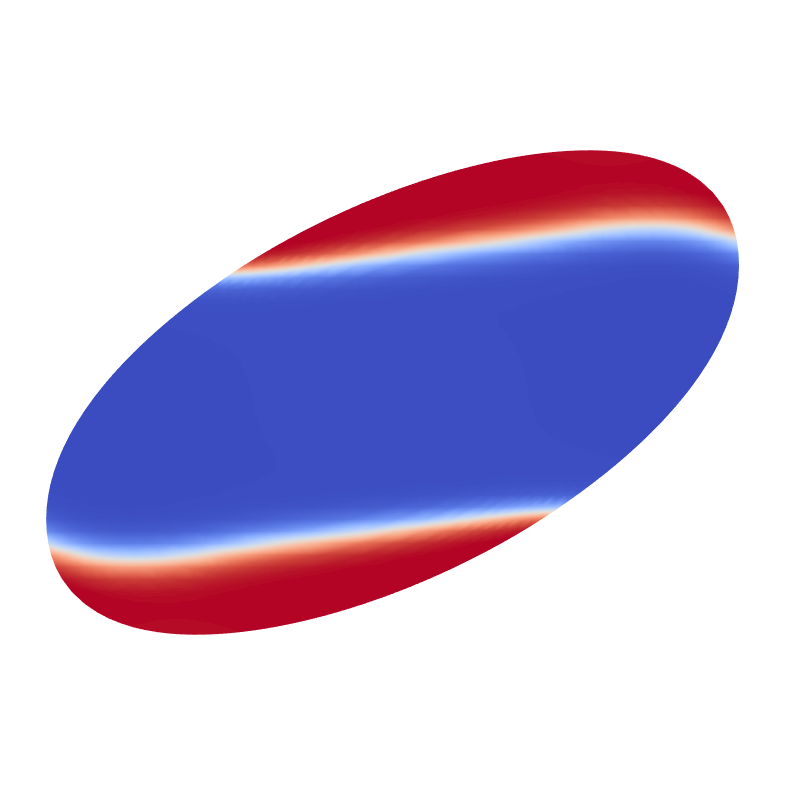} &
					\includegraphics[height=1.7cm]{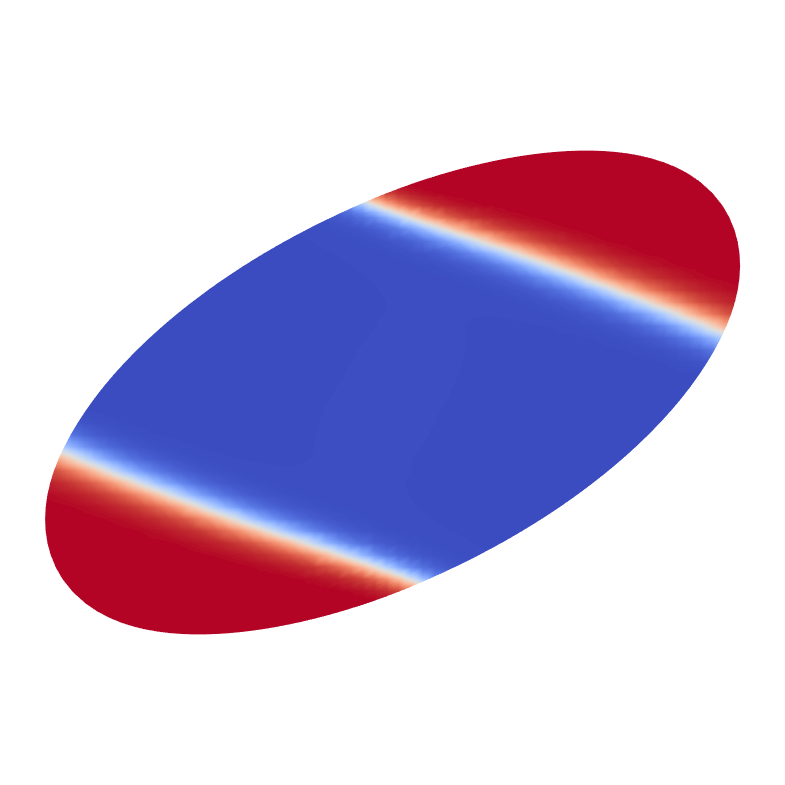} &
					\includegraphics[height=1.7cm]{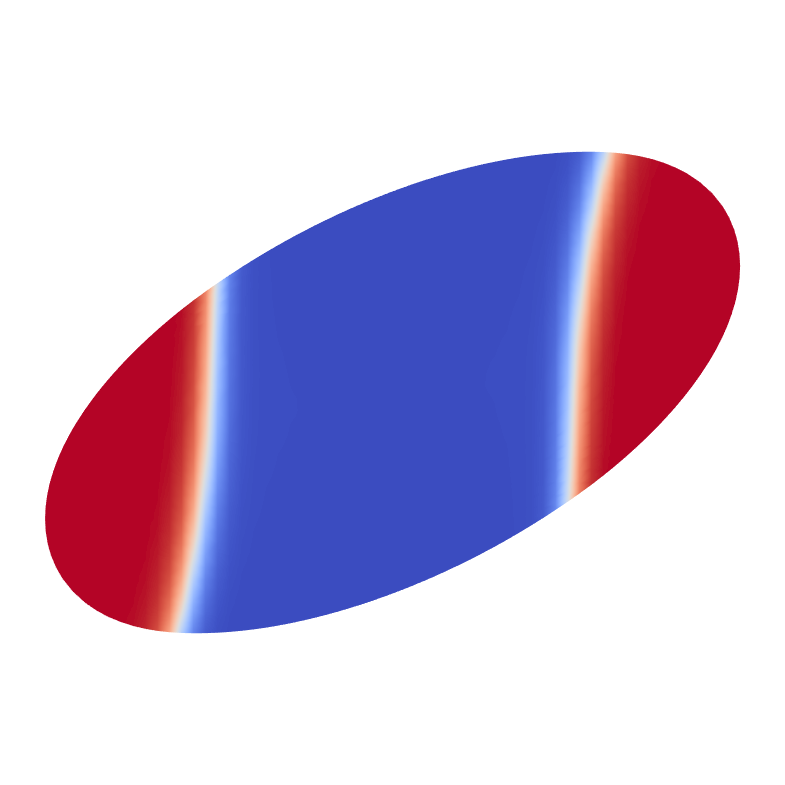} \tabularnewline
			\end{tabular}
		} 
		\subfigure[Bending Energy]{
			\includegraphics[height=5.5cm]{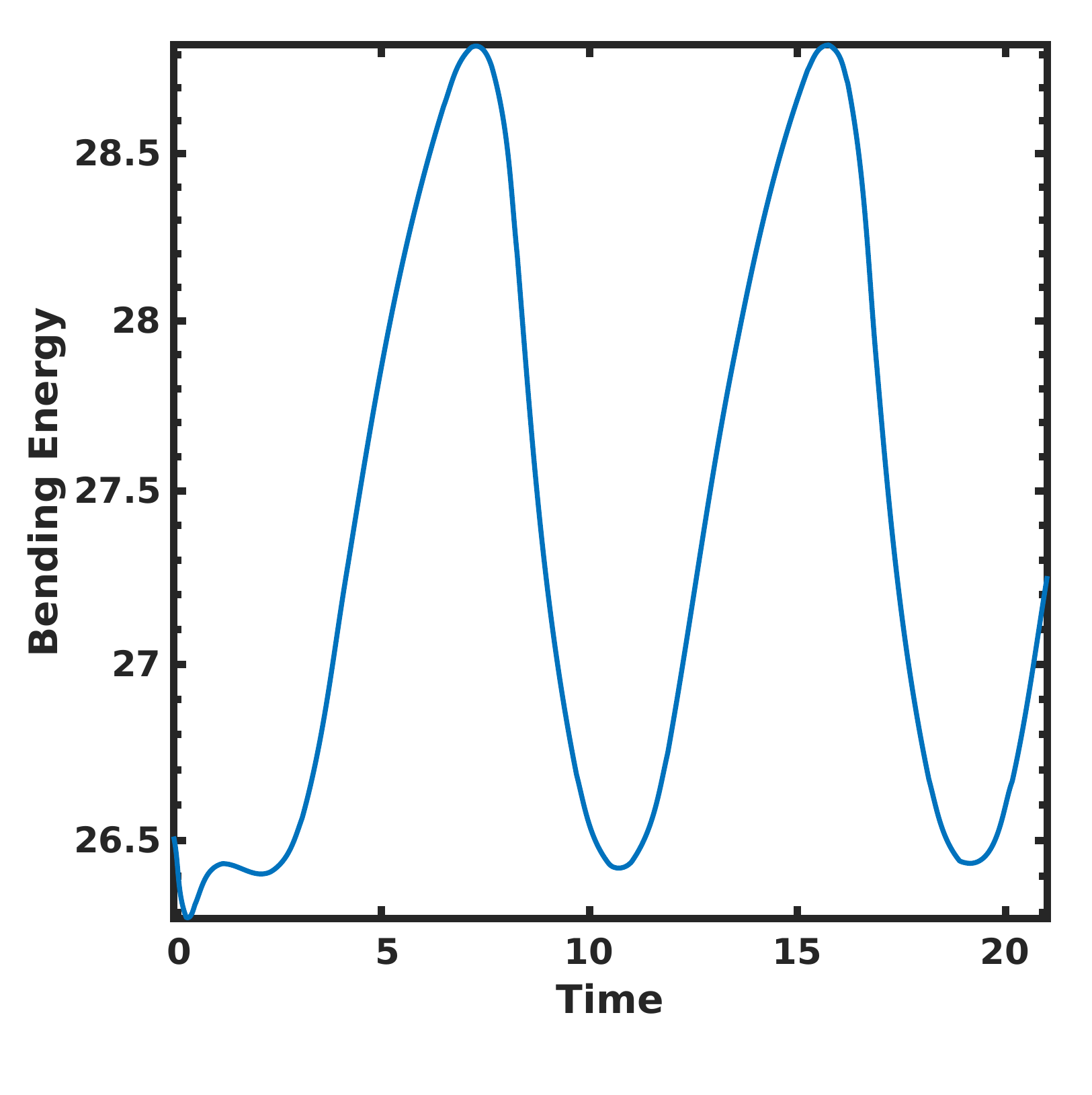}
		} 
		\subfigure[Domain Boundary Energy]{			
			\includegraphics[height=5.5cm]{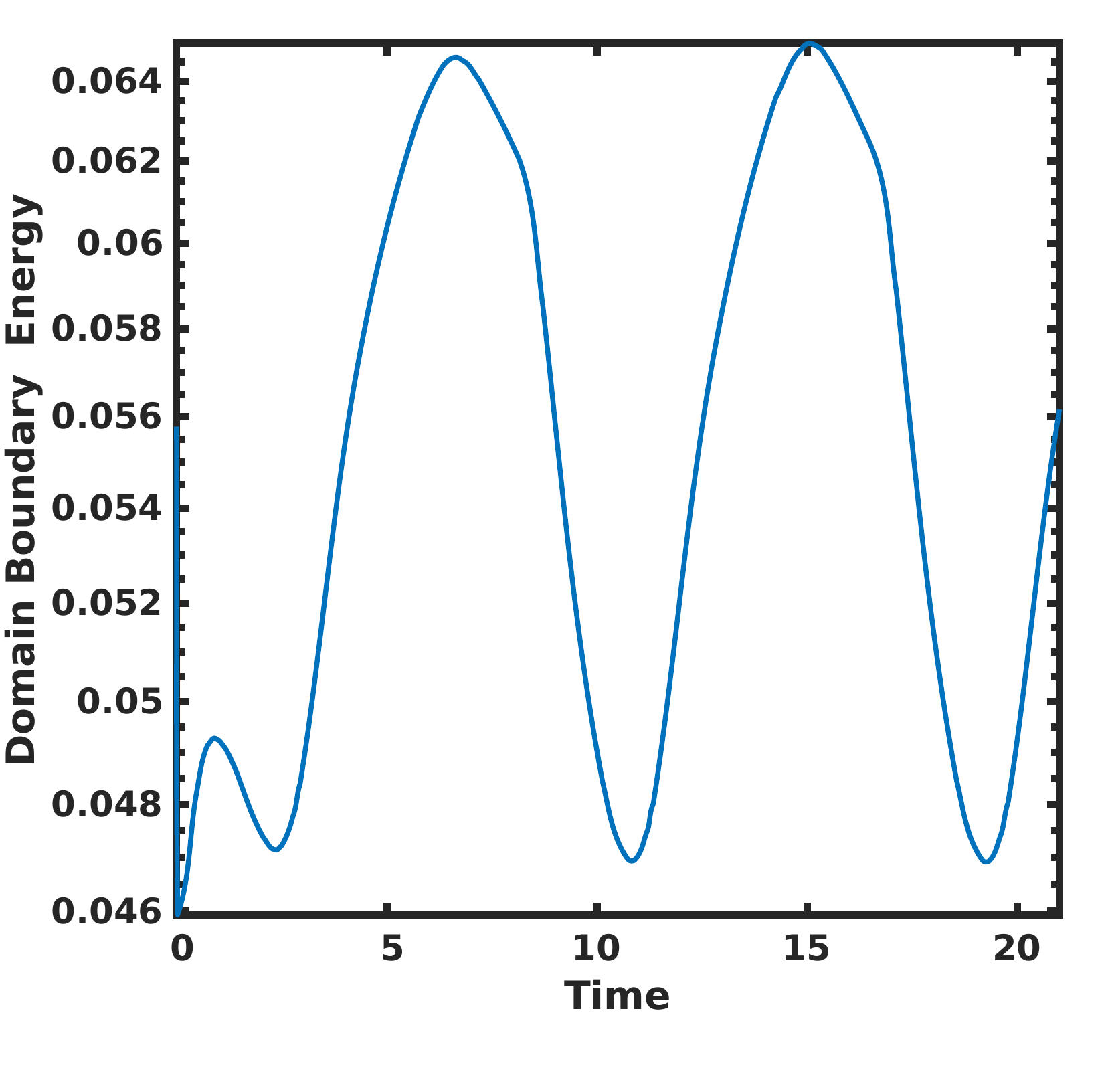}
		}
	\end{center}
		\caption{Phase treading dynamics of a vesicle with
        $\bar{c}=0.4$ and $\alpha=20$. The soft
        phase has a bending rigidity of $\kappa_c^B=0.8$ and the Peclet number
        is $\Pe=1.0$. The Peclet number allows for the domains to tread along the membrane.}        
        
		\label{fig:Tread}
\end{figure*}

\subsubsection{Vertical Stationary Band Dynamics}

When the Peclet number and soft phase rigidity are increased, the
vertical stationary banding dynamic is observed. 
The vertical stationary dynamic
is characterized by the stretching of the domains from the tips
vertically on the surface, with the eventual merging of the domain
resulting in a single and thin domain.

Consider a vesicle with $\Pe=0.5$ and soft phase bending rigidity of $\kappa_c^B=0.7$, 
Fig.~\ref{fig:VSB}. As in the diffusion dominated case shown
previously, the vesicle rotates and achieves a relatively stable inclination
and the domains begin to elongate. Due to the higher Peclet number,
fluid motion forces the domains to elongate further along the vesicle membrane
than in the prior case. This elongation continues until the 
two domains meet and merge into a single domain spanning the vertical plane.

Exploration of the bending and domain boundary energy provides further insights 
to this dynamic. As the domains begin to grow along the vesicle,
both energies increase until $t=12.5$. At this point the two domains merge
and there is a large decrease in the domain boundary energy. The bending energy
decreases more slowly, as a large amount of the softer phase still inhabits 
the low curvature regions away from the tips. Eventually the soft phase
diffuses to the tips, resulting in a further reduction of the bending energy.
After reaching the minima, there is a slight increase in the bending and
domain boundary energy. This is due to some material being advected from
the high curvature tips to the lower curvature center.
Note that this particular dynamic has not been previously reported.

\subsubsection{Phase Treading Dynamics}

Further increasing the Peclet or soft phase bending rigidity 
decreases the restorative surface diffusion forces, which allows
the force exerted by the external fluid to become dominate.
An example of this behavior can be seen in Fig.~\ref{fig:Tread},
where the soft phase has a bending rigidity of $\kappa_c^B=0.8$ 
and the Peclet number is $\Pe=1.0$. As in the prior examples, the vesicle
rotates to become more aligned with the shear flow. During this time 
the domains elongate along the long-axis. Unlike the prior cases,
the domains do not remain attached to the vesicle tips and migrate along the interface until it
reaches the other tip, when the process is then repeated. Similar dynamics 
have been observed in the recent two-dimensional work of Liu et al.~\cite{C6SM02452A}.

The periodic nature of this dynamic can be seen by considering the 
bending and domain boundary energy, Fig.~\ref{fig:Tread}.
After an initially transient period, both the bending and domain boundary
energy quickly increase as the domains leave the tips. 
When the domains reach
the opposite tip, both energies quickly decrease, with the decrease in the domain boundary
energy slightly lagging the drop in the bending energy. 

\begin{figure*}[!ht]
	\begin{center}
		\subfigure[]{
			\begin{tabular}{
				>{\centering}m{1.0cm}>{\centering}m{1.7cm}>{\centering}m{1.7cm}>{\centering}m{1.7cm}>{\centering}m{1.7cm}>{\centering}m{1.7cm}>{\centering}m{1.7cm}>{\centering}m{1.7cm}>{\centering}m{1.7cm}}
				& \multicolumn{8}{c}{Time} \tabularnewline
				\cline{2-9}
				View & $0.5$ & $2.0$ & $4.0$ & $6.0$ & $10.0$ & $14.0$ & $16.0$ & $20.0$ \tabularnewline
				Iso & 
					\includegraphics[height=1.7cm]{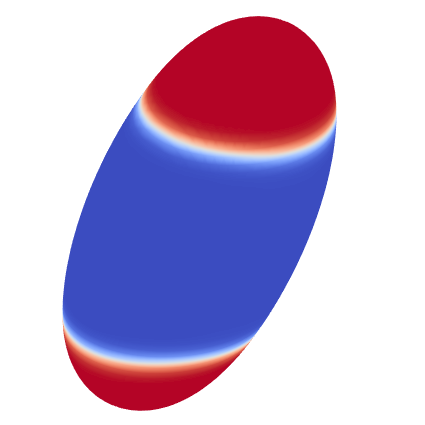} &
					\includegraphics[height=1.7cm]{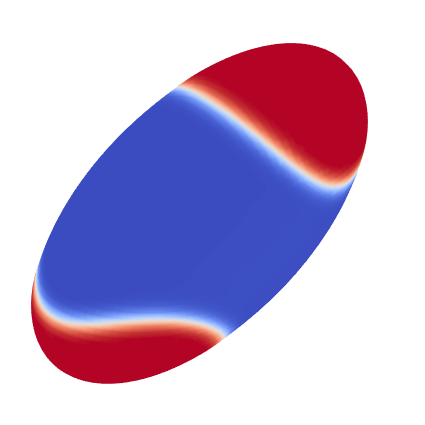} &
					\includegraphics[height=1.7cm]{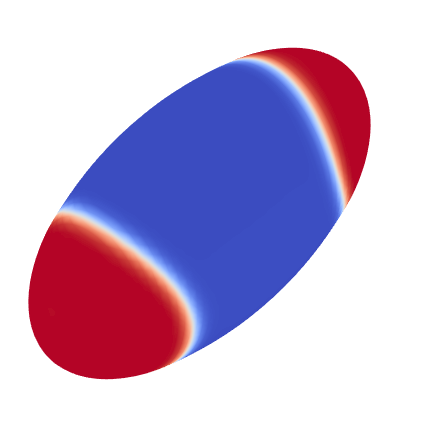} &
					\includegraphics[height=1.7cm]{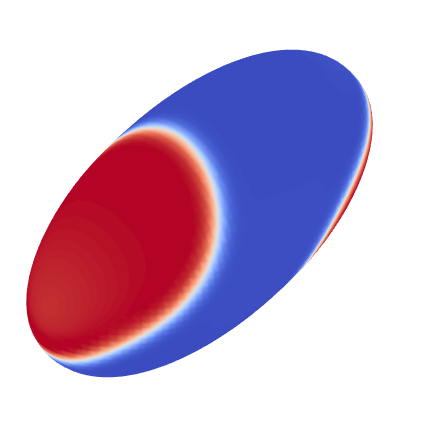} &
					\includegraphics[height=1.7cm]{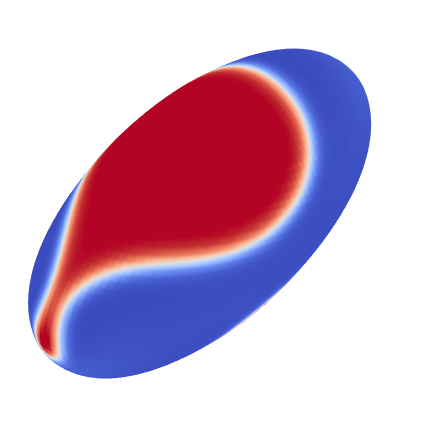} &
					\includegraphics[height=1.7cm]{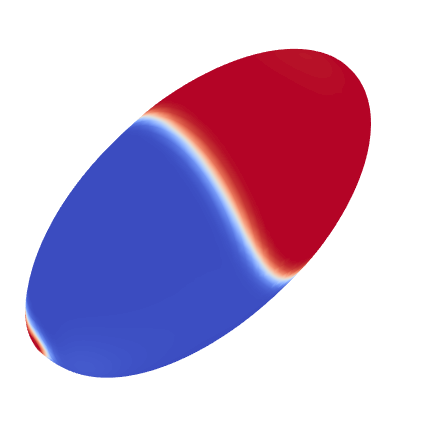} &
					\includegraphics[height=1.7cm]{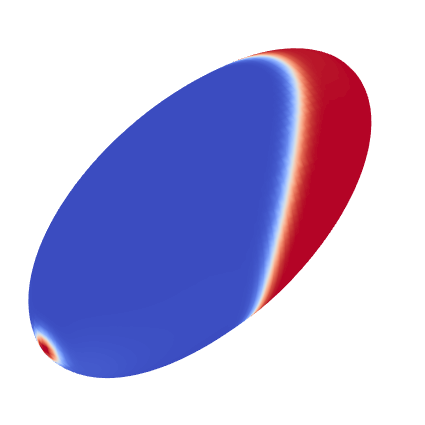} &
					\includegraphics[height=1.7cm]{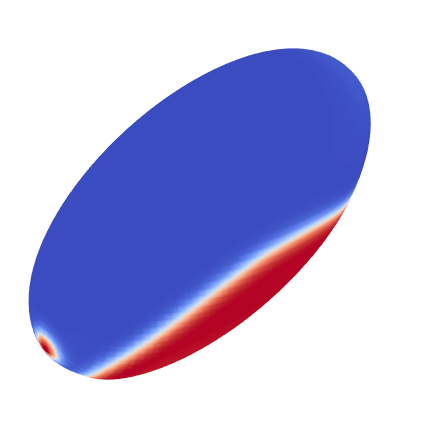} \tabularnewline
				X-Y & 
					\includegraphics[height=1.7cm]{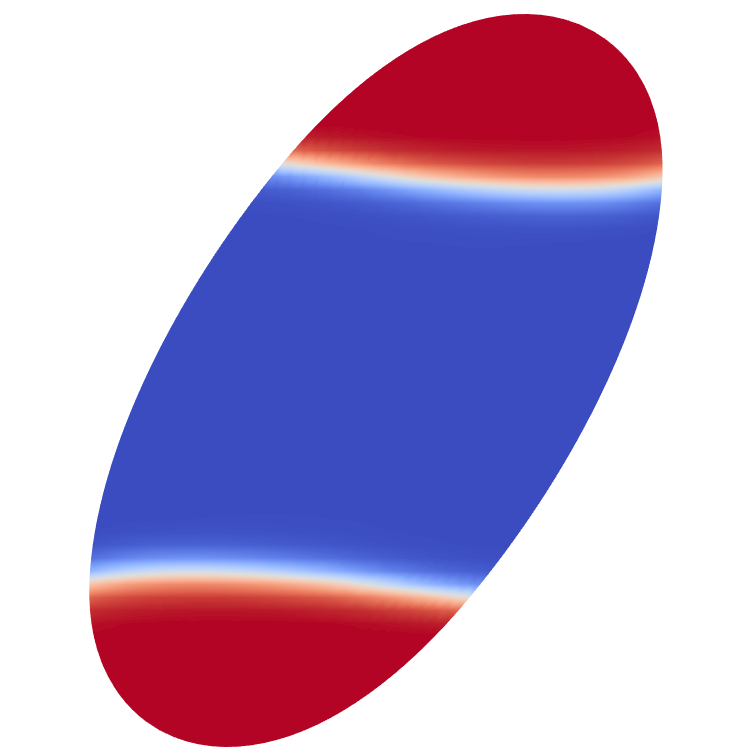} &
					\includegraphics[height=1.7cm]{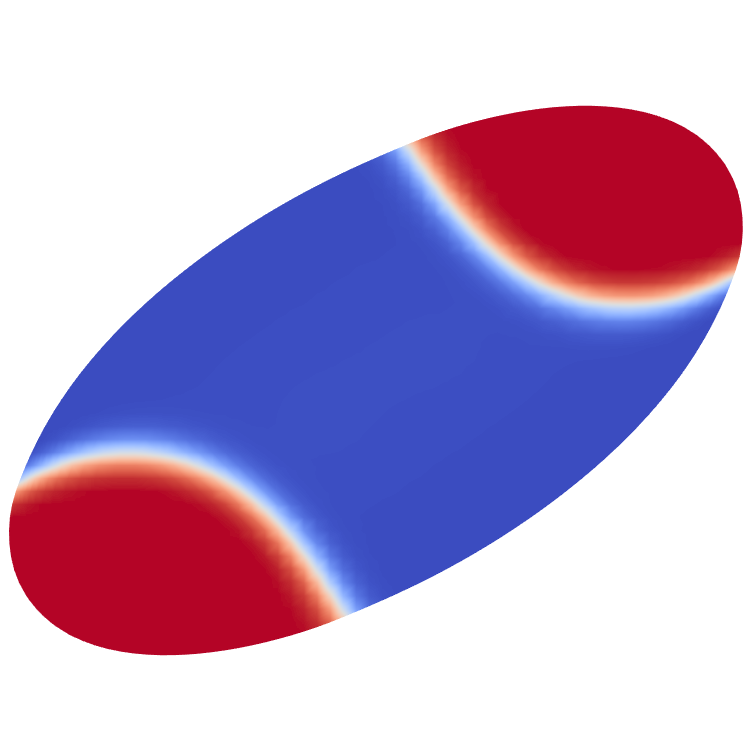} &
					\includegraphics[height=1.7cm]{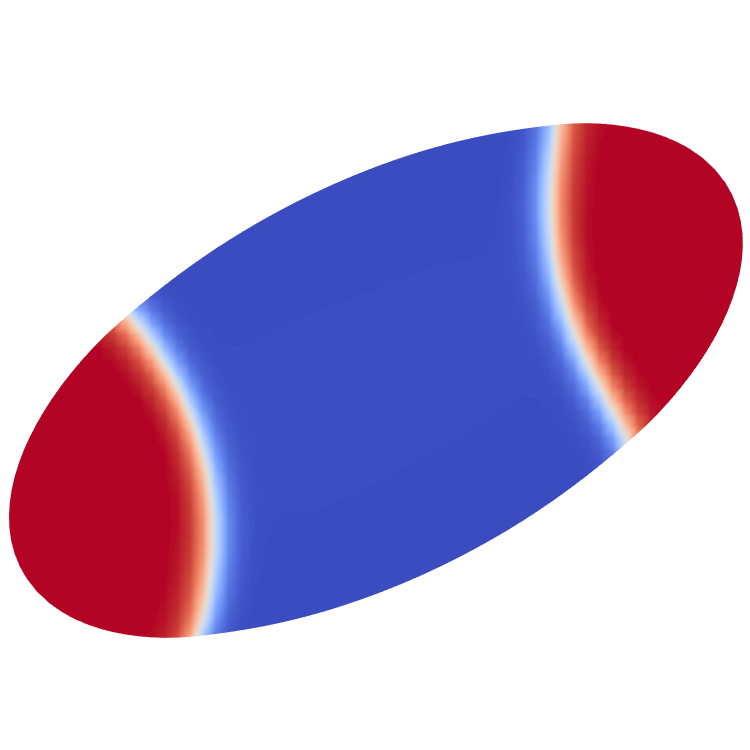} &
					\includegraphics[height=1.7cm]{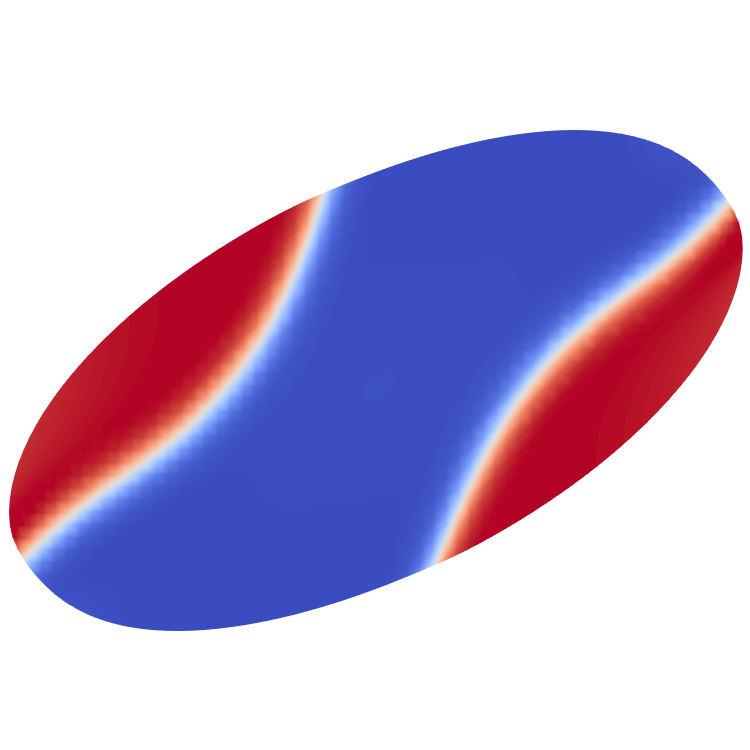} &
					\includegraphics[height=1.7cm]{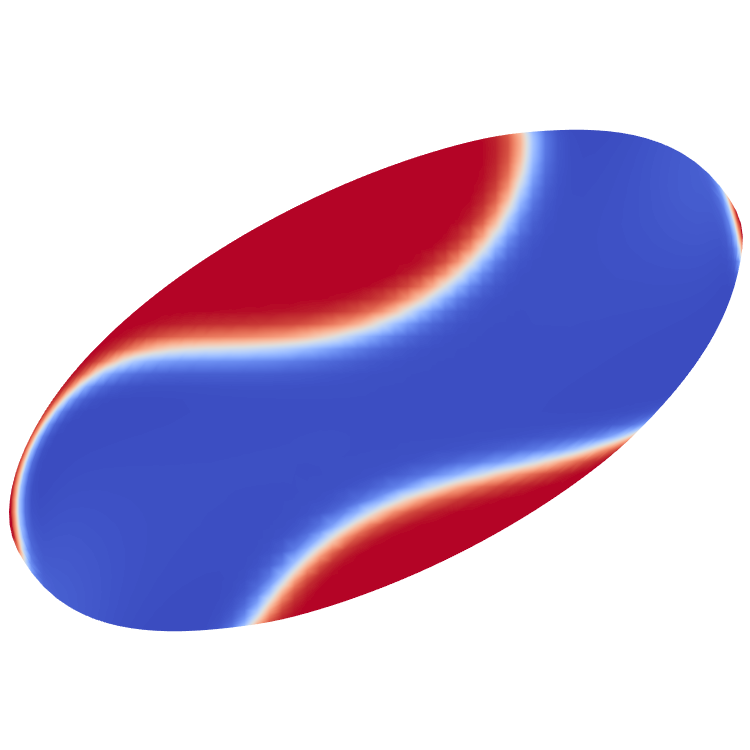} &
					\includegraphics[height=1.7cm]{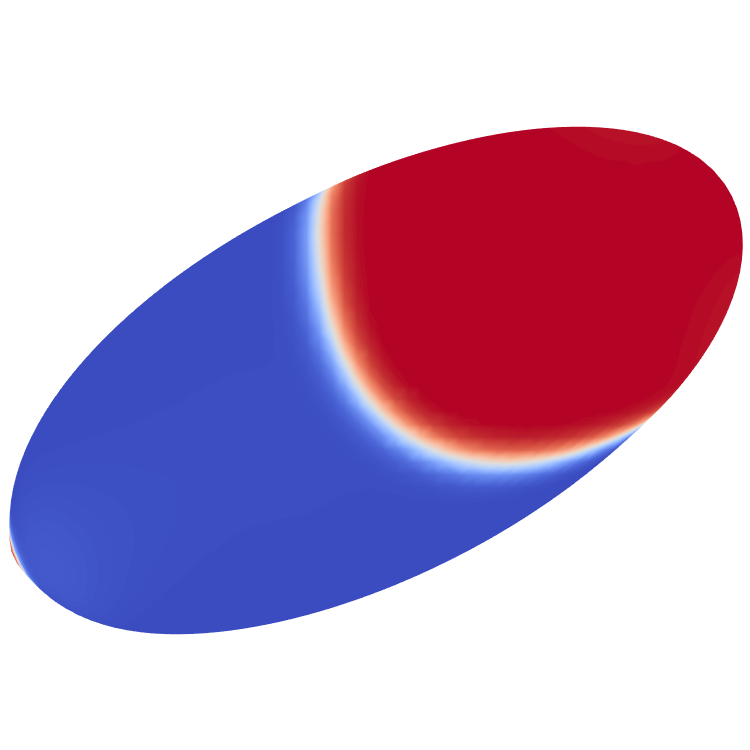} &
					\includegraphics[height=1.7cm]{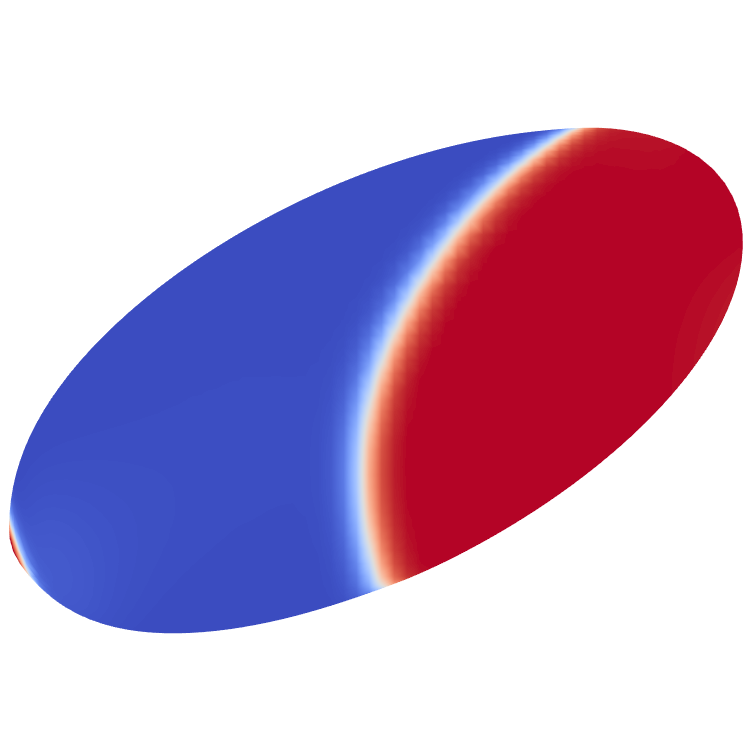} &
					\includegraphics[height=1.7cm]{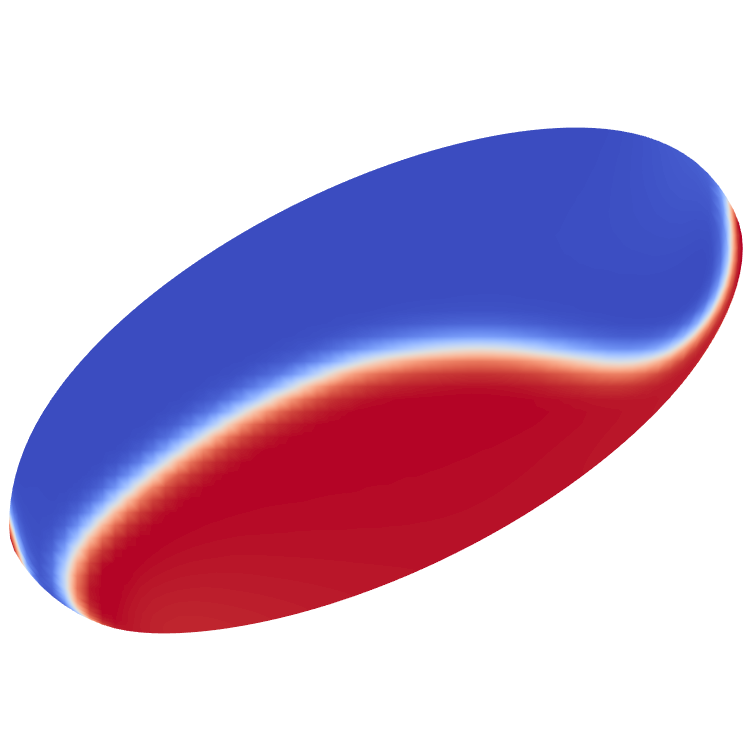} \tabularnewline
			\end{tabular}
		} 
		\subfigure[Bending Energy]{
			\includegraphics[height=5.5cm]{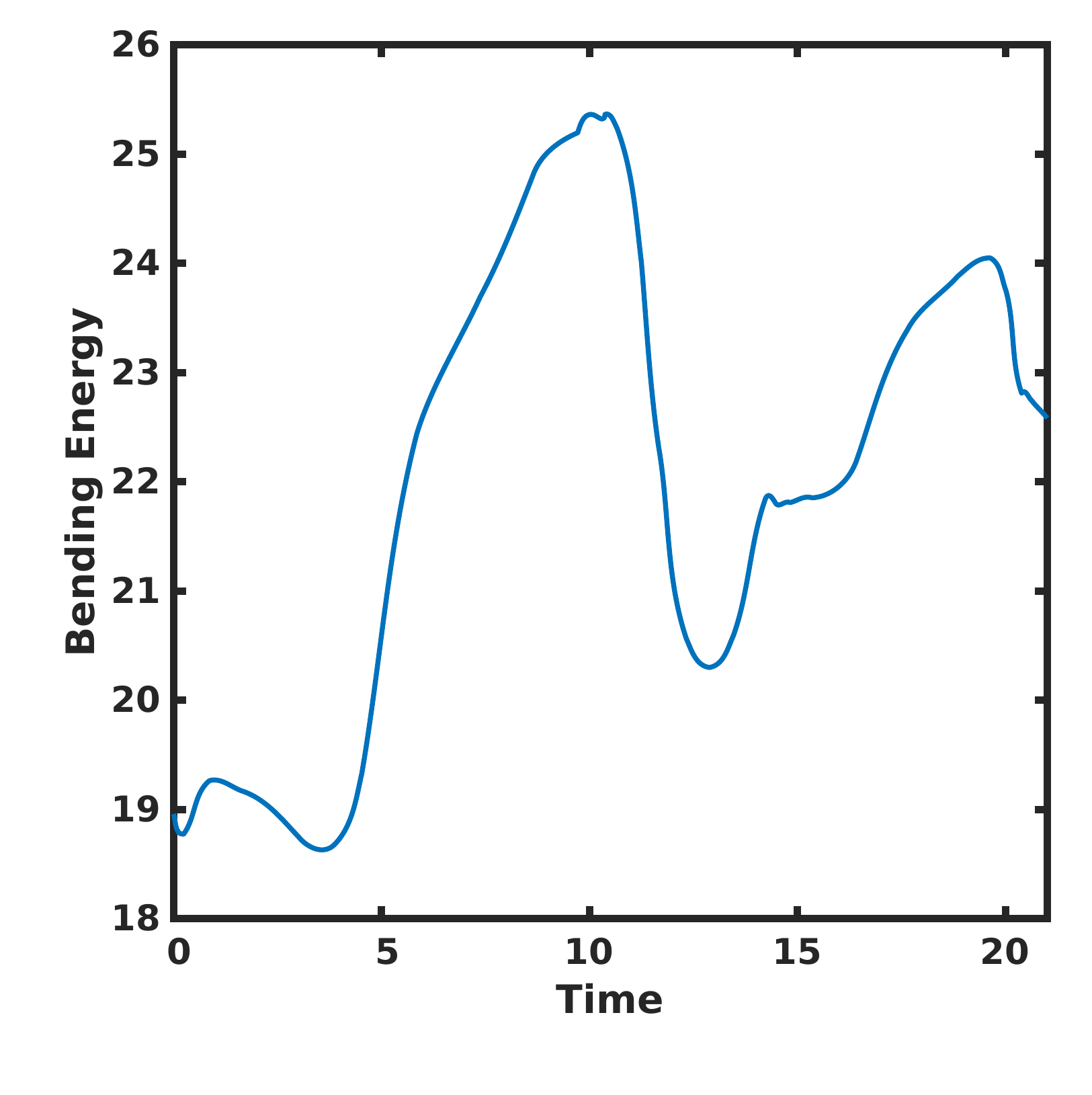}
		} 
		\qquad
		\subfigure[Domain Boundary Energy]{			
			\includegraphics[height=5.5cm]{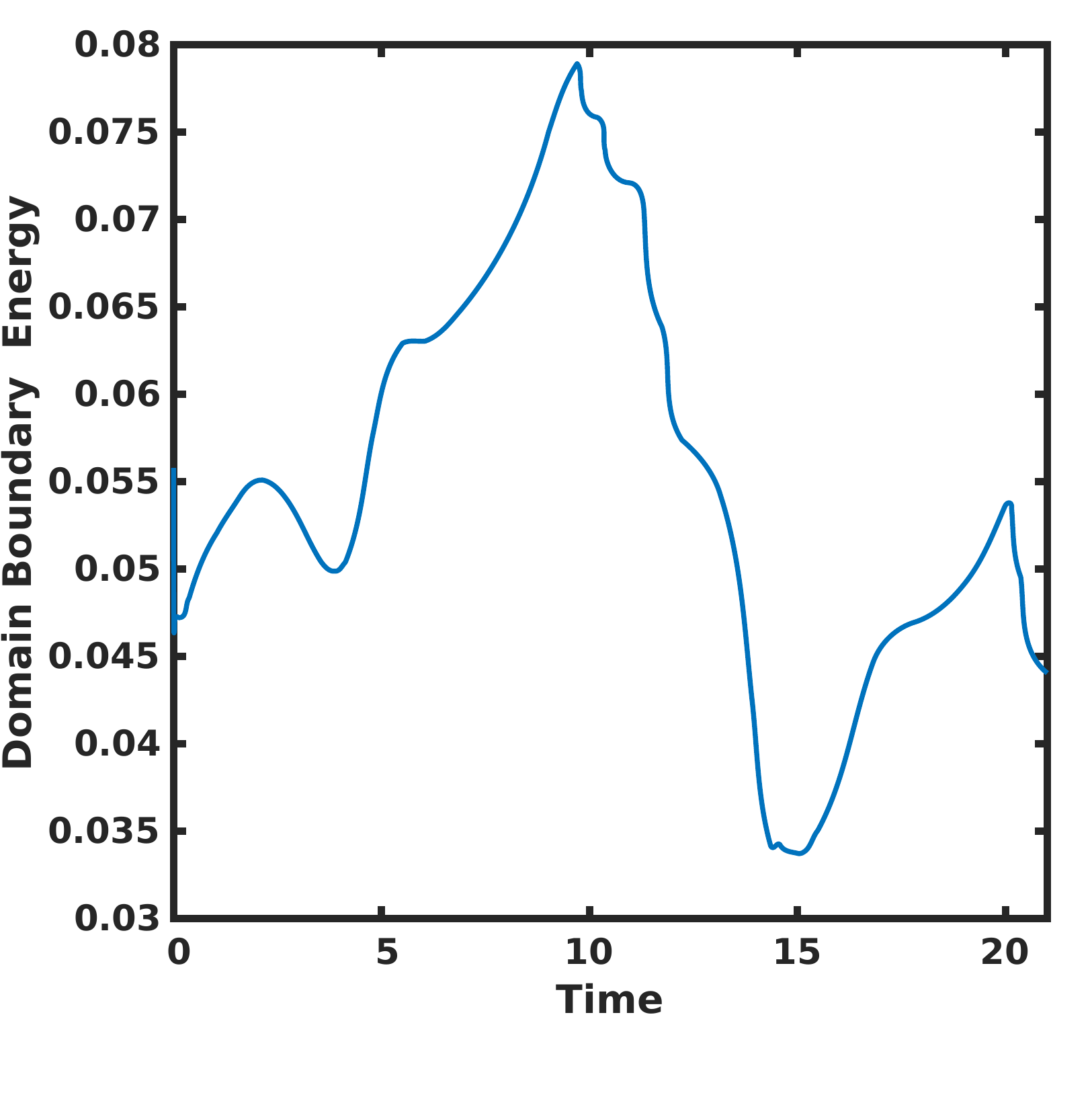}	
		}
	\end{center}
		\caption{Sample of the Tread-1 dynamic for a vesicle with $\bar{c}=0.4$ and $\alpha=20$.
		A softer phase with a bending rigidity of $\kappa_c^B=0.4$ and with a Peclet number
        of $\Pe=0.3$ results in the treading of domains for a certain time before one of the
        domains grows at the expense of the other domain.}
		\label{fig:HalfTread}
\end{figure*}

\textbf{Tread-n Dynamics:} 

An interesting sub-dynamic of phase treading exists when the surface Peclet number
or soft phase bending rigidity are too large to allow for stationary phases or 
vertical banding, but not large enough to allow for long-term phase treading behavior.
In this work, this sub-dynamic is further classified as \textit{Tread-n}, where $n$ 
indicates the number of times that each phase travels from one tip to the other tip
before diffusion dominates and results in a single, large domain.

For example, consider the dynamics with a Peclet number of $\Pe=0.3$ and a soft phase
bending rigidity of $\kappa_c^B=0.4$, Fig.~\ref{fig:HalfTread}.
As shown before, the vesicle begins to align itself with the shear flow and the phases begin to migrate 
along the vesicle membrane. During this migration,
the upper domain grows at the expense of the lower phase.
Once the upper domain reaches the opposite tip, the lower domain has completely disappeared.
This is further confirmed via the energy curves, Fig.~\ref{fig:HalfTread}. It is clear
that a large decrease in the domain boundary energy occurs between a time of $t=10$ and $t=15$,
which corresponds to the domain merging event. As the domains only switched tips once,
this dynamic would be classified as Tread-1. Note that the small soft phase seen in 
Fig.~\ref{fig:HalfTread} is typical of this dynamic, as the advective forces
are not strong enough to completely overcome all of the surface diffusion restorative forces.
This is further discussed in later sections.

\subsection{Dynamics as a function of $\kappa_c^B$ and $\Pe$}

From the prior results, it is clear that the dynamics of the vesicle strongly depend on
both the soft phase bending rigidity and the surface Peclet number. 
In this section, systematic parameter studies investigating
the dynamics for a pre-segregated and multicomponent vesicle with an
average concentration of $\bar{c}=0.4$ as a function of 
$\kappa_c^B$ and $\Pe$ for two characteristic domain line energies 
are performed. In both cases, the dynamics will be reported
via phase diagrams.

\subsubsection{Domain Line Tension of $\alpha=20$}

\begin{figure}
\centering
  \includegraphics[height=8cm]{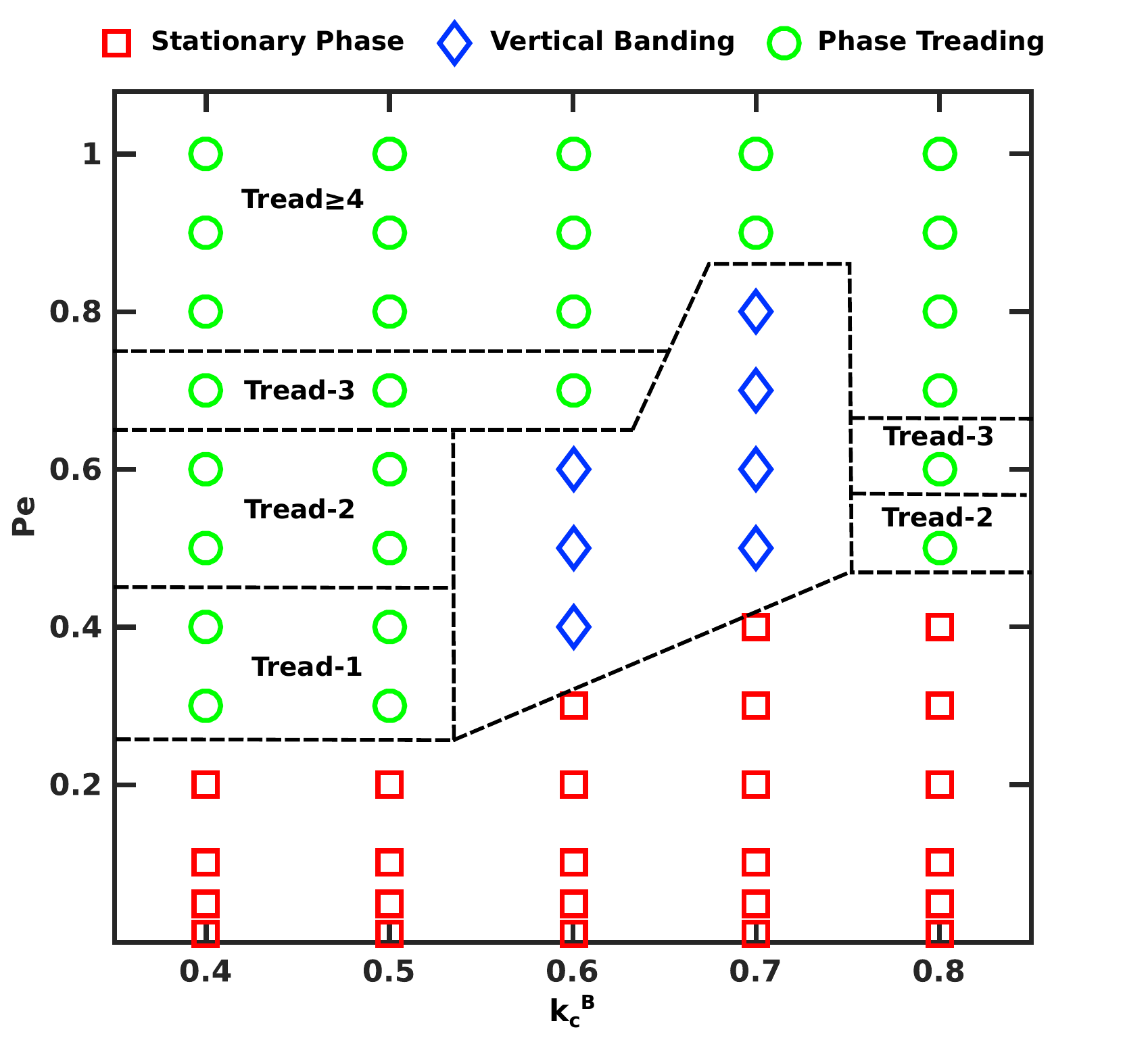}
  \caption{Variation of vesicle behavior with Peclet number and bending rigidity of the soft phase for 
		$\bar{c}=0.4$ and $\alpha=20$.}
  \label{fig:PhaseDiagram_Alpha20}
\end{figure}
\begin{figure*}
	\begin{center}
		\begin{tabular}{
			>{\centering}m{1.0cm}>{\centering}m{1.7cm}>{\centering}m{1.7cm}>{\centering}m{1.7cm}>{\centering}m{1.7cm}>{\centering}m{1.7cm}>{\centering}m{1.7cm}>{\centering}m{1.7cm}>{\centering}m{1.7cm}}
			& \multicolumn{8}{c}{Time} \tabularnewline
			\cline{2-9}
			$\kappa_c^B$ & $5.0$ & $9.0$ & $10.0$ & $12.5$ & $15.0$ & $17.5$ & $20.0$ & $20.5$ \tabularnewline
			0.5 & 
				\includegraphics[height=1.7cm]{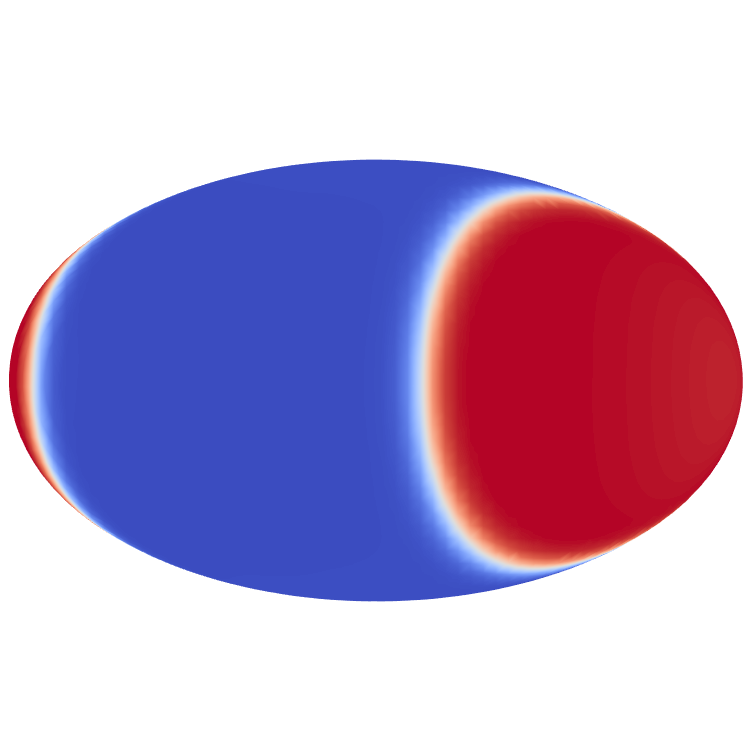} &
				\includegraphics[height=1.7cm]{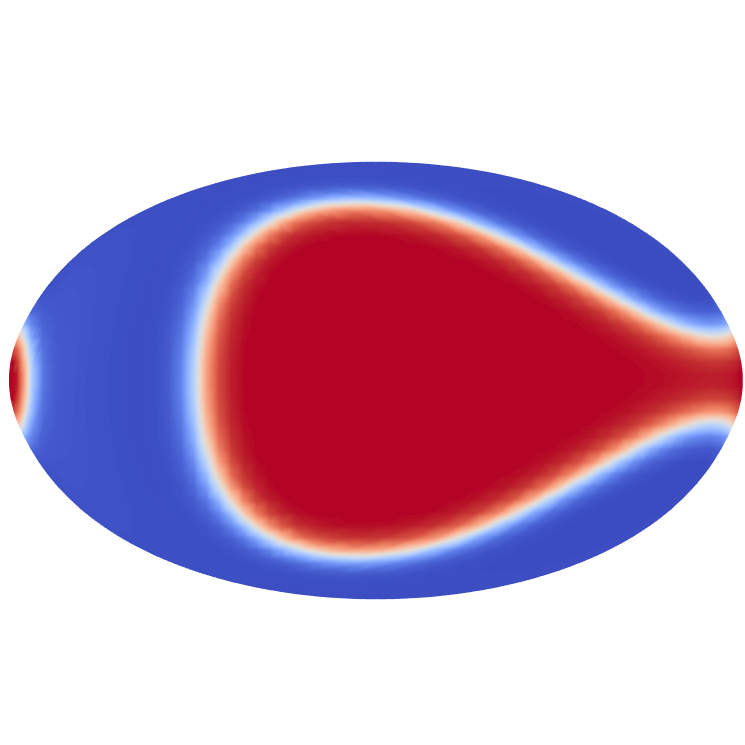} &
				\includegraphics[height=1.7cm]{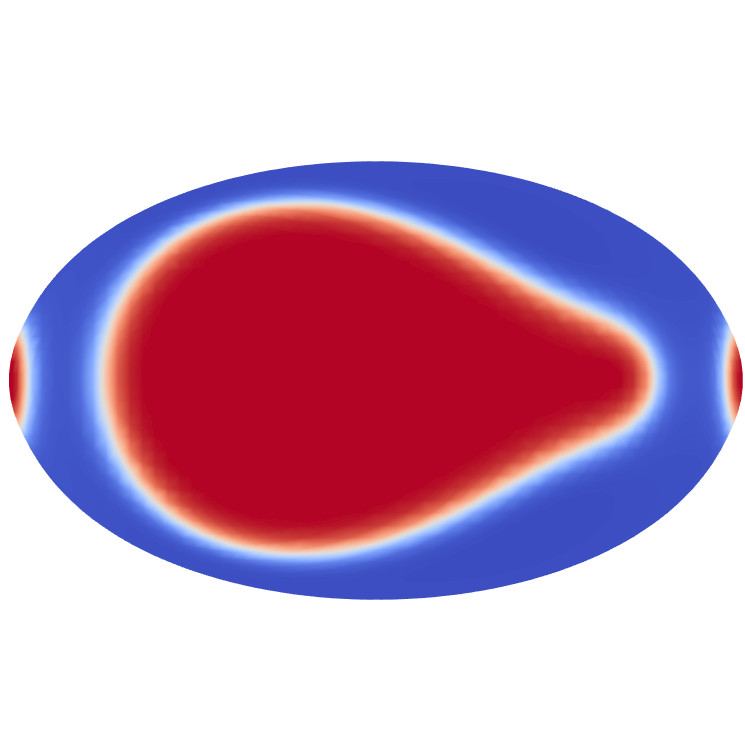} &
				\includegraphics[height=1.7cm]{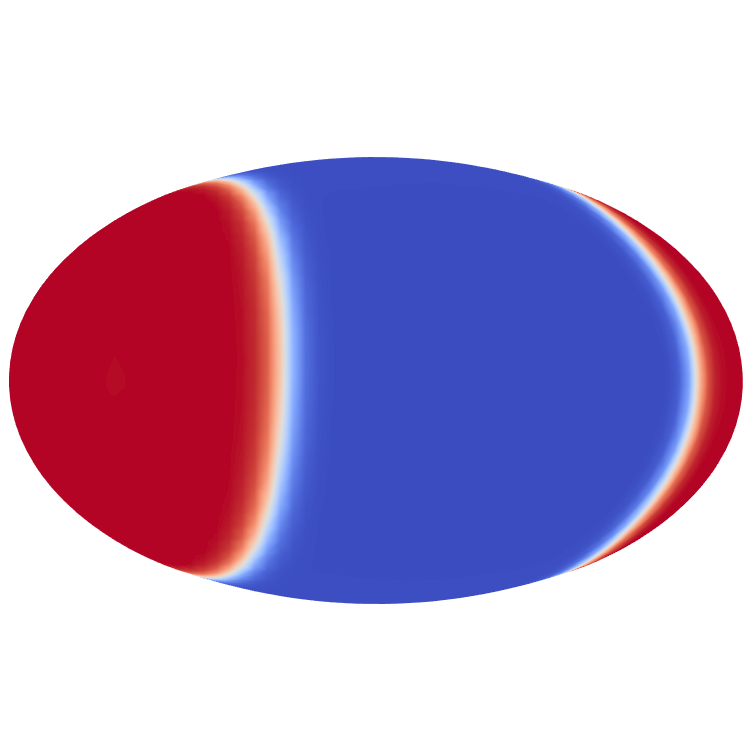} &
				\includegraphics[height=1.7cm]{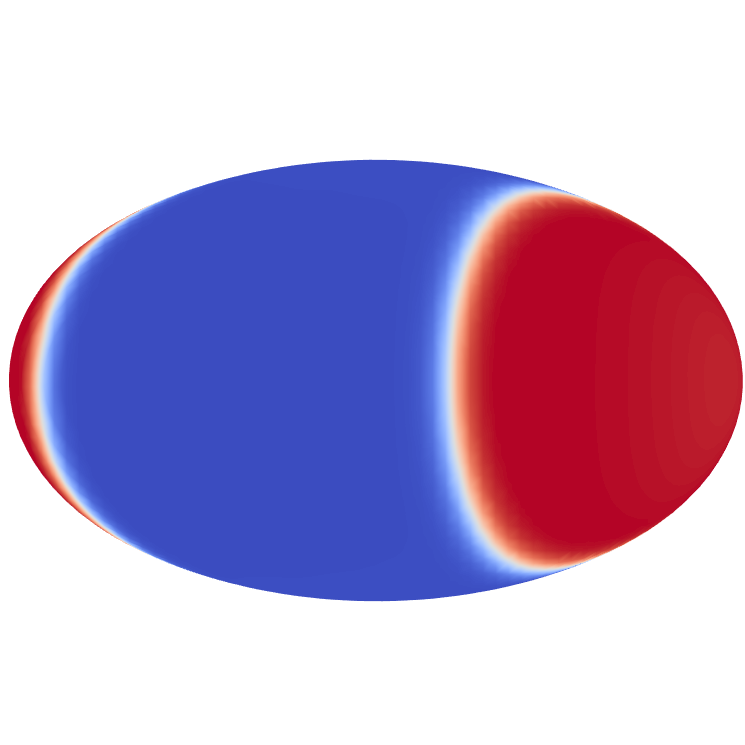} &
				\includegraphics[height=1.7cm]{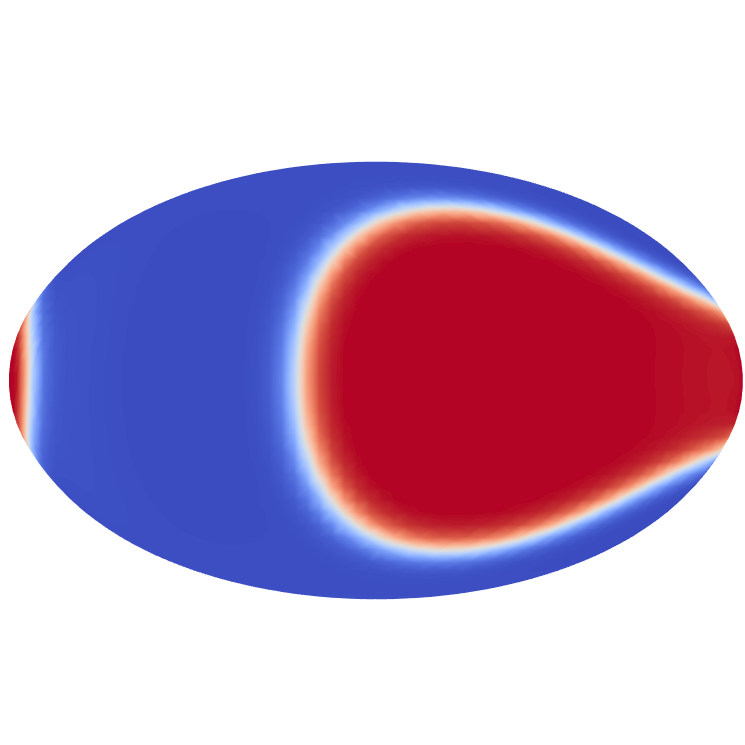} &
				\includegraphics[height=1.7cm]{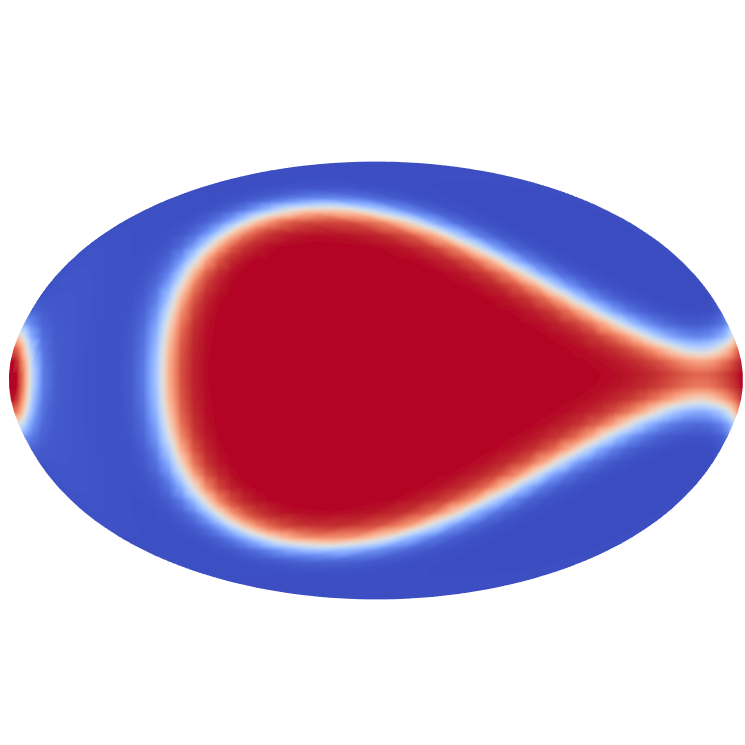} &
				\includegraphics[height=1.7cm]{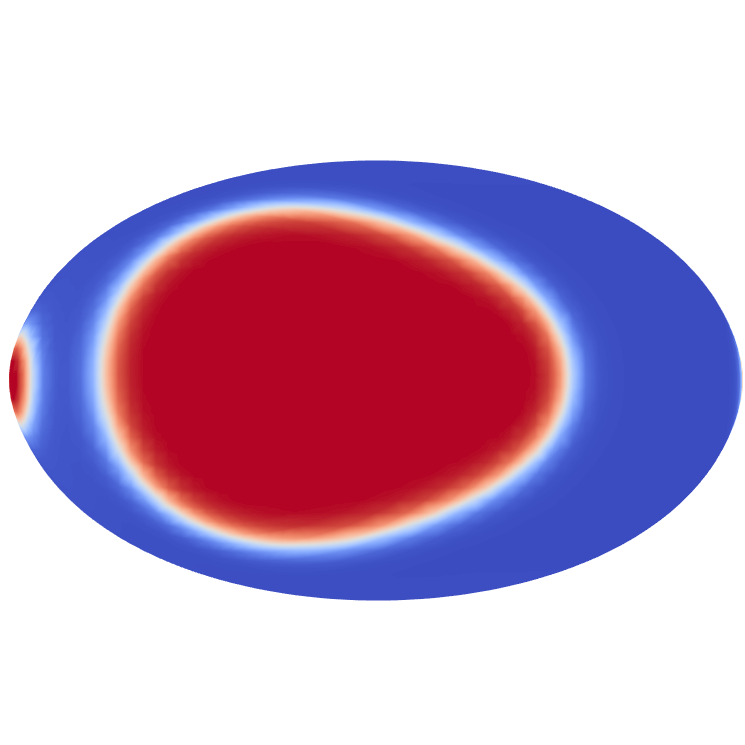} \tabularnewline
			0.7 & 
				\includegraphics[height=1.7cm]{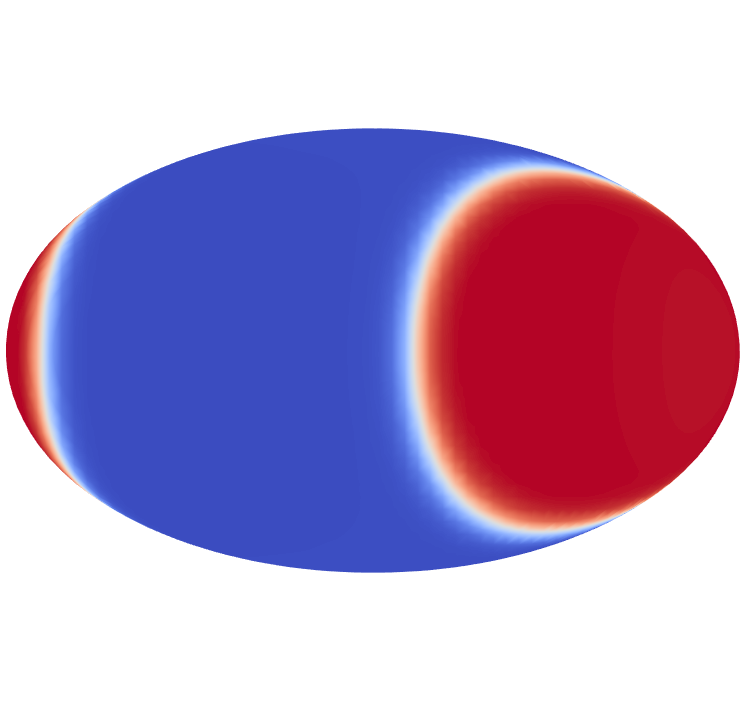} &
				\includegraphics[height=1.7cm]{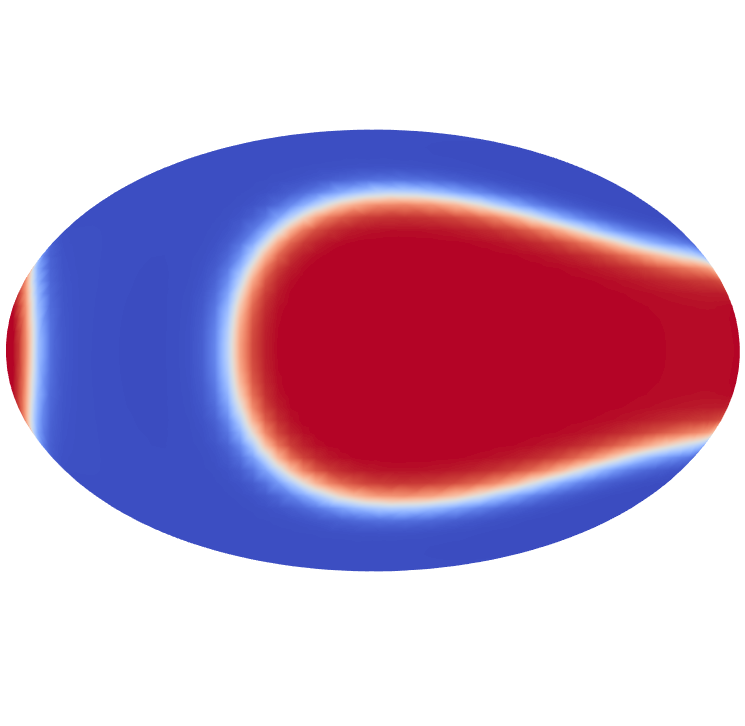} &
				\includegraphics[height=1.7cm]{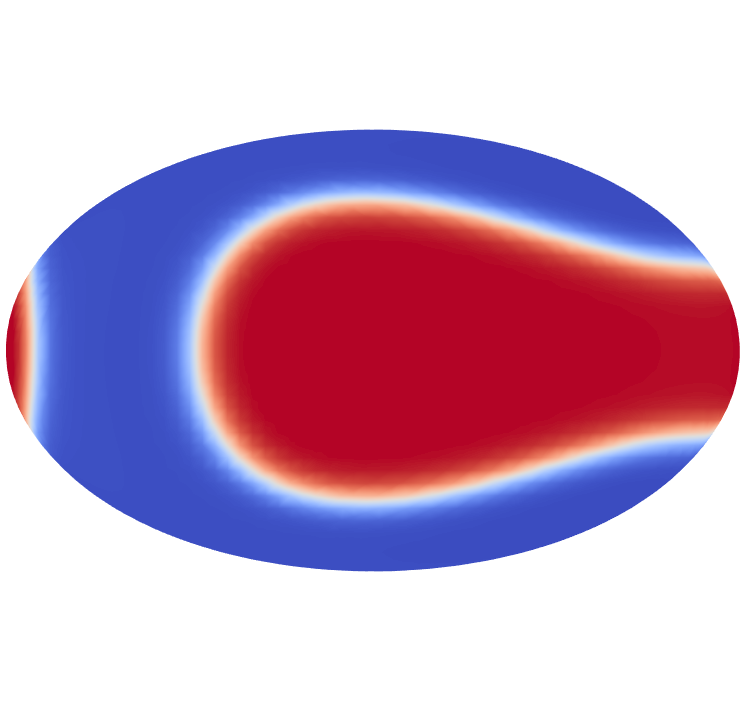} &
				\includegraphics[height=1.7cm]{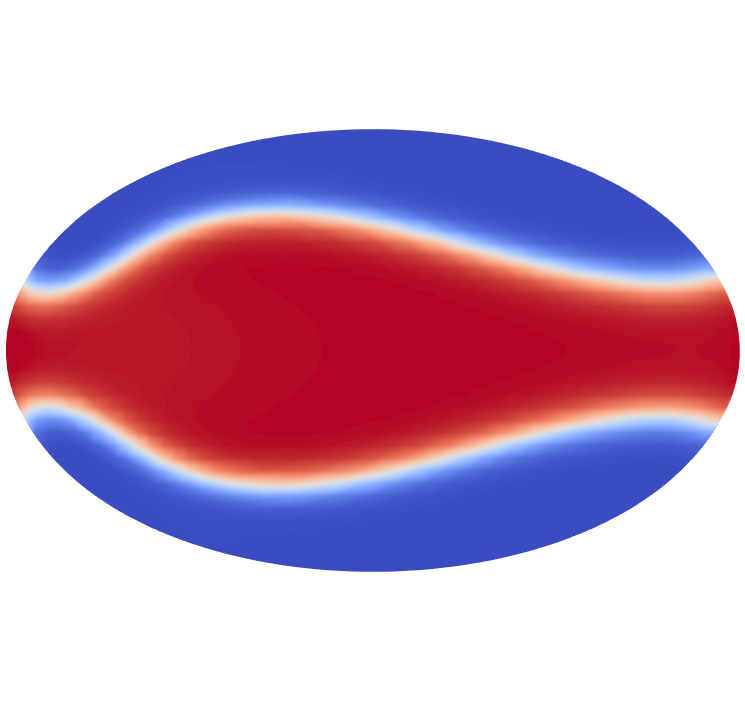} &
				\includegraphics[height=1.7cm]{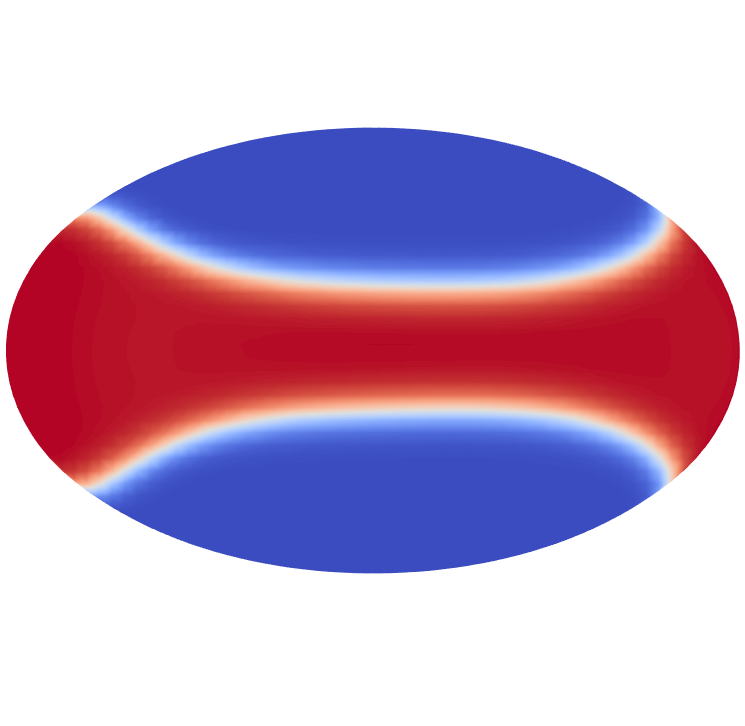} &
				\includegraphics[height=1.7cm]{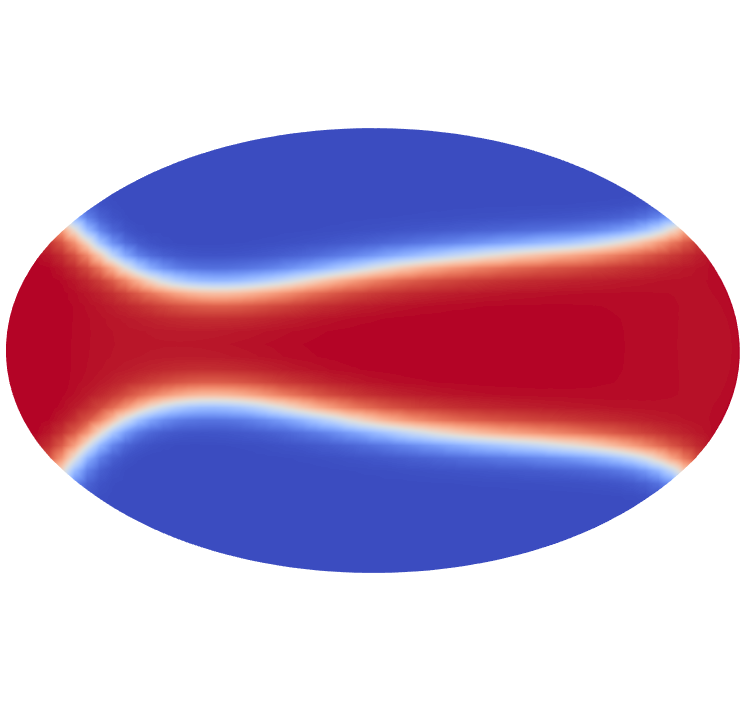} &
				\includegraphics[height=1.7cm]{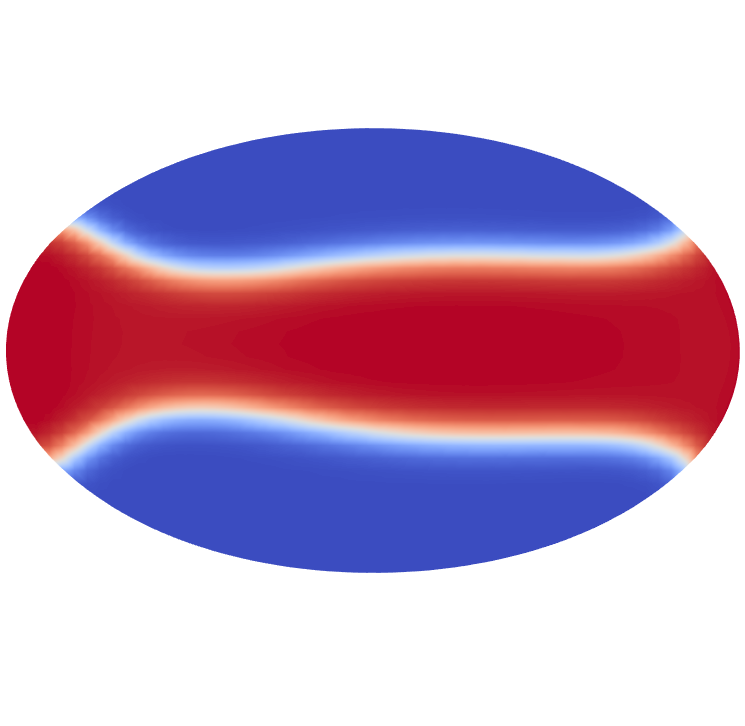} &
				\includegraphics[height=1.7cm]{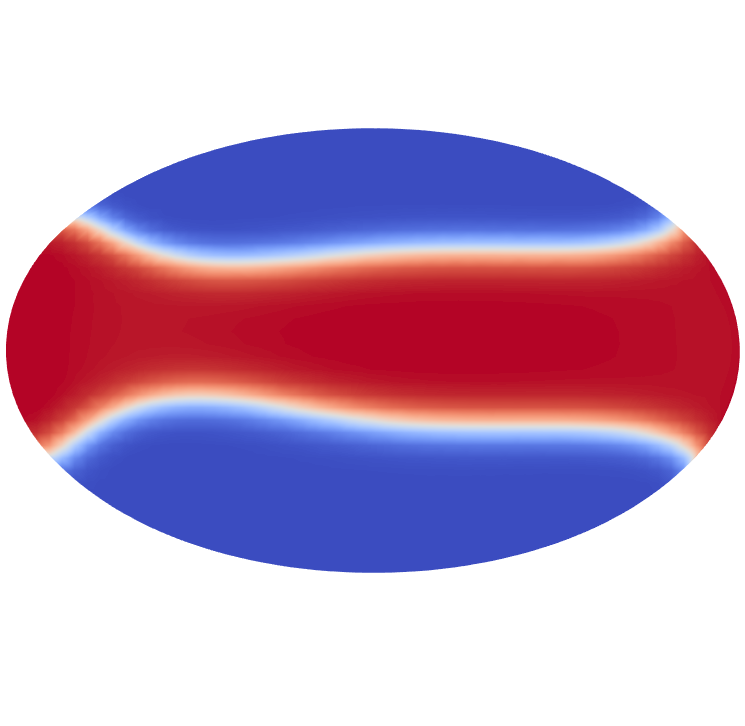} \tabularnewline
		\end{tabular}			
	\end{center}
	\caption{Treading dynamics of lipids on the vesicle in X-Z plane for
		$\bar{c}=0.4$, $\alpha=20$, and $\Pe=0.5$. Two soft phase bending rigidities are shown.}
	\label{fig:Tread_XZ}
\end{figure*}

First consider the phase diagram for $\alpha=20$, Fig.~\ref{fig:PhaseDiagram_Alpha20}.
This diagram clearly shows the three regimes demonstrated previously: stationary phase, vertical banding, and phase treading.
As has been previously demonstrated for two-dimensional vesicles, there is a linear relationship between the critical shear rate
needed for phase treading and the bending rigidity of the soft phase~\cite{C6SM02452A}. The surface Peclet
number scales with the shear rate, \textit{i.e.} as the shear rate increases so does $\Pe$. Therefore,
it should be expected that as the bending rigidity for the soft phase decreases, the critical
Peclet number needed for the stationary phase/treading phase should increase. As can
be seen in the phase diagram, this is not the case. For example,
a multicomponent vesicle
with $\kappa_c^B=0.7$ and $\Pe=0.5$ is in the 
vertical banding regime while one with $\kappa_c^B=0.5$ and $\Pe=0.5$ is in the phase treading regime. 
To explain this counter-intuitive behavior, compare the dynamics between $\kappa_c^B=0.5$ and $\kappa_c^B=0.7$
for $\Pe=0.5$, Fig.~\ref{fig:Tread_XZ}. In general,
the domains with $\kappa_c^B=0.7$ are thinner at the leading edge and thicker at the tail
compared to the domains with $\kappa_c^B=0.5$, see the domains at a time of $t=9$ for an example. 
As the tails for $\kappa_c^B=0.7$ are thicker, 
the domains can extend further along the vesicle and still maintain contact with the
vesicle tips. This allows for the domains to extend and eventually merge,
forming a single, thin domain. In contrast, the thin tail for $\kappa_c^B=0.5$ results in the 
eventual pinch-off of the domains, which allows for continued phase-treading.

\begin{figure}
	\begin{center}
		\subfigure[X-Z plane]{
			\includegraphics[width=3.0cm]{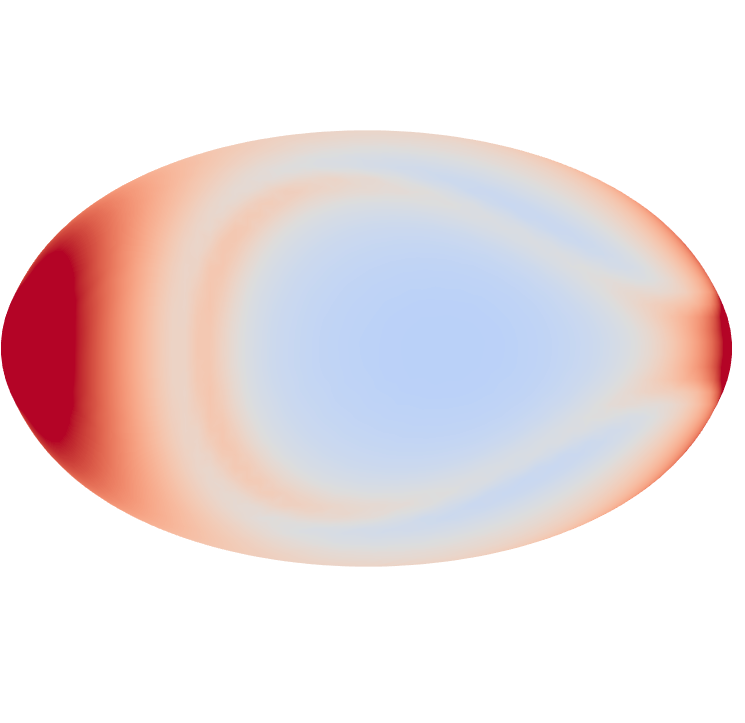}
		} 
		\qquad
		\subfigure[X-Y plane]{			
			\includegraphics[width=3.0cm]{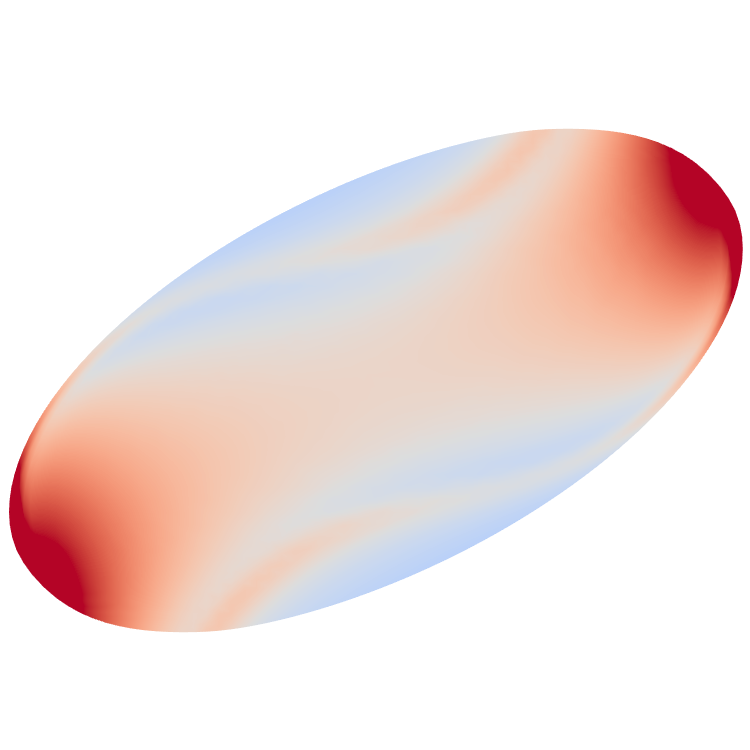}	
		}
		\caption{Curvature on the vesicle shown in Fig.~\ref{fig:Tread_XZ}.
		Red (color online) indicates regions of high curvature. In this figure $\kappa_c^B=0.5$.} 
		\label{fig:Curvature_XZ_XY}
	\end{center}
\end{figure}

The rationale for this behavior can be seen by considering the curvature
on the vesicle membrane at a time of $t=9$, Fig.~\ref{fig:Curvature_XZ_XY}.
As can be seen, due to the deformation of the vesicle in shear flow,
the membrane has a higher curvature along the horizontal edges 
of the vesicle than along the vertical edges. A higher curvature 
along the horizontal edges of the vesicle is true in general,
and only the particular curvature values depend on the bending rigidity of the soft phase. 
When the soft phase bending rigidity is $\kappa_c^B=0.5$, material is drawn 
towards these high curvature regions on the edge. 
As the domain grows towards the edge, the material required
is taken from the tail, which results in a thin tail.
On the other hand, when $\kappa_c^B=0.7$, the force drawing the soft
phase down towards the edge is small, which allows it to maintain 
a thick tail and thus contact with the vesicle tip.

\begin{figure*}
	\begin{center}
		\subfigure[]{
			\begin{tabular}{
				>{\centering}m{1.0cm}>{\centering}m{1.1cm}>{\centering}m{1.1cm}>{\centering}m{1.1cm}>{\centering}m{1.1cm}>{\centering}m{1.1cm}>{\centering}m{1.1cm}>{\centering}m{1.1cm}>{\centering}m{1.1cm}>{\centering}m{1.1cm}>{\centering}m{1.1cm}>{\centering}m{1.1cm}}
				& \multicolumn{11}{c}{Time} \tabularnewline
				\cline{2-12}
				View & $4.0$ & $6.0$ & $10.0$ & $12.5$ & $14.0$ & $16.0$ & $20.0$ & $22.5$ & $25.0$ & $30.0$ & $35.0$ \tabularnewline
				Iso & 
					\includegraphics[height=1.1cm]{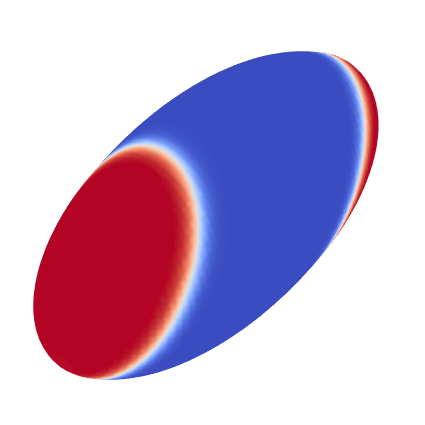} &
					\includegraphics[height=1.1cm]{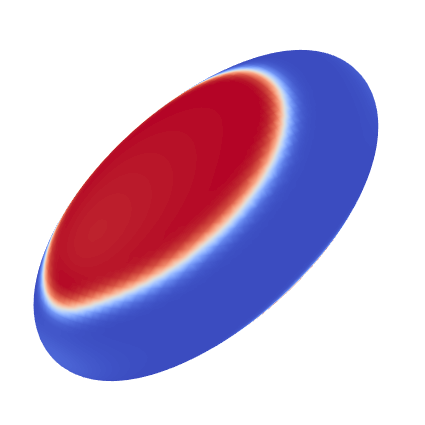} &
					\includegraphics[height=1.1cm]{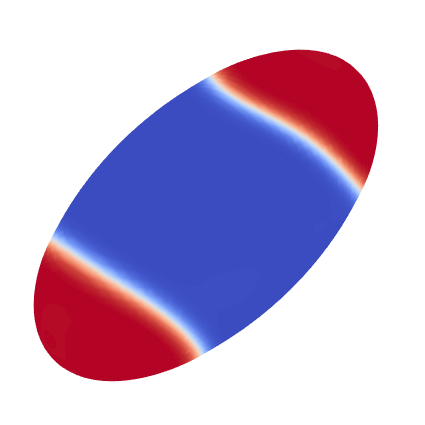} &
					\includegraphics[height=1.1cm]{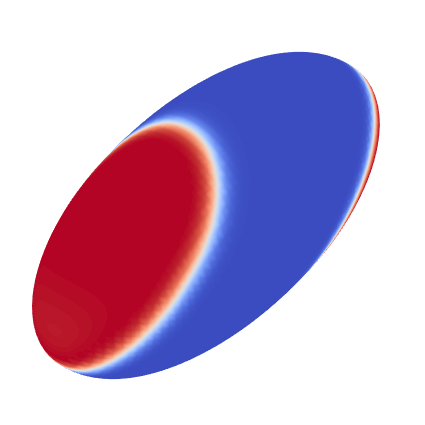} &
					\includegraphics[height=1.1cm]{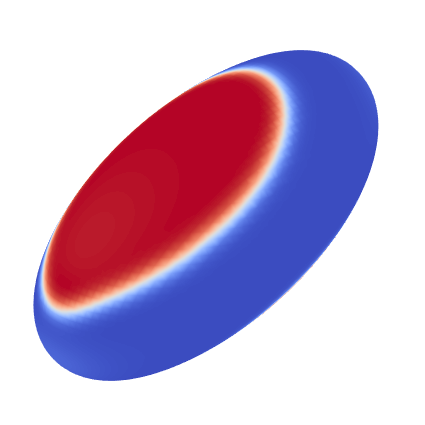} &
					\includegraphics[height=1.1cm]{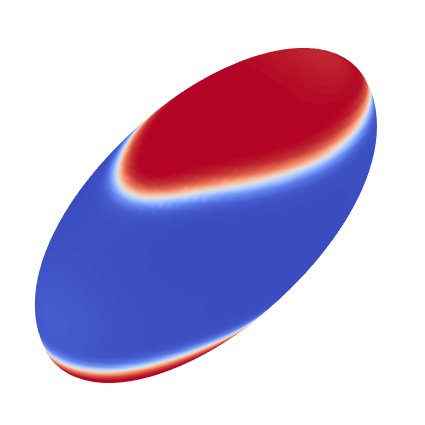} &
					\includegraphics[height=1.1cm]{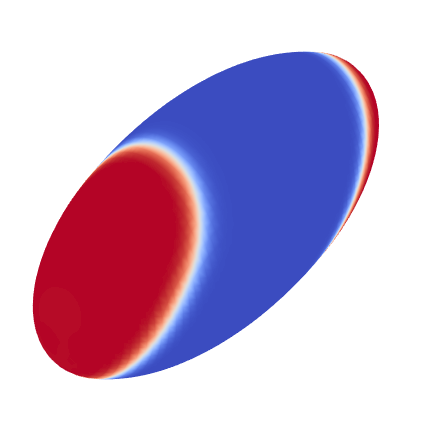} &
					\includegraphics[height=1.1cm]{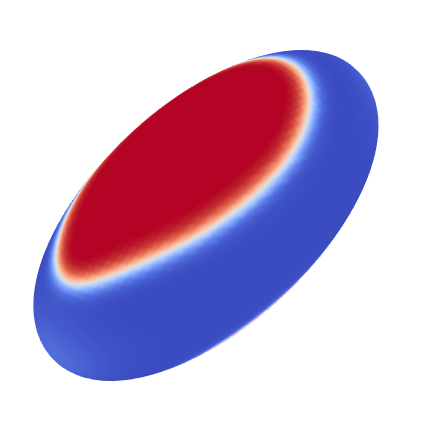} &
					\includegraphics[height=1.1cm]{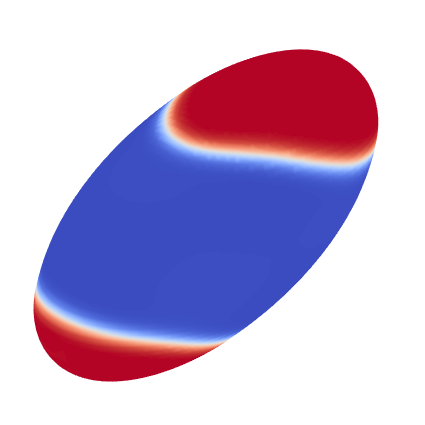} &
					\includegraphics[height=1.1cm]{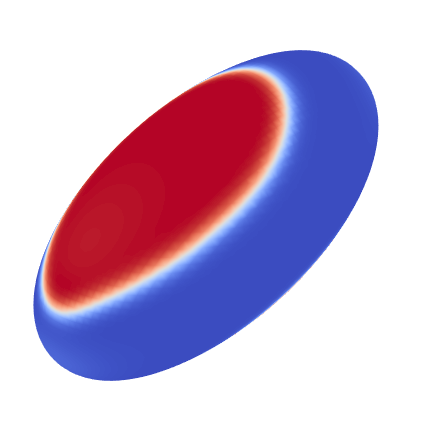} &
					\includegraphics[height=1.1cm]{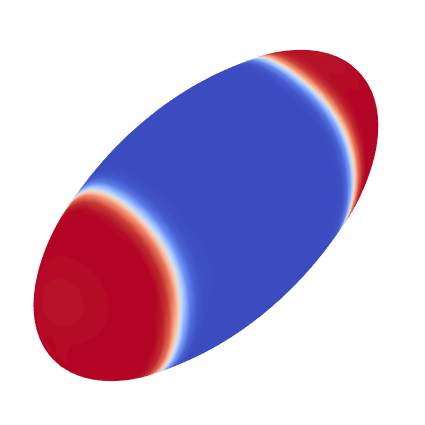} \tabularnewline
				X-Y & 
					\includegraphics[height=1.1cm]{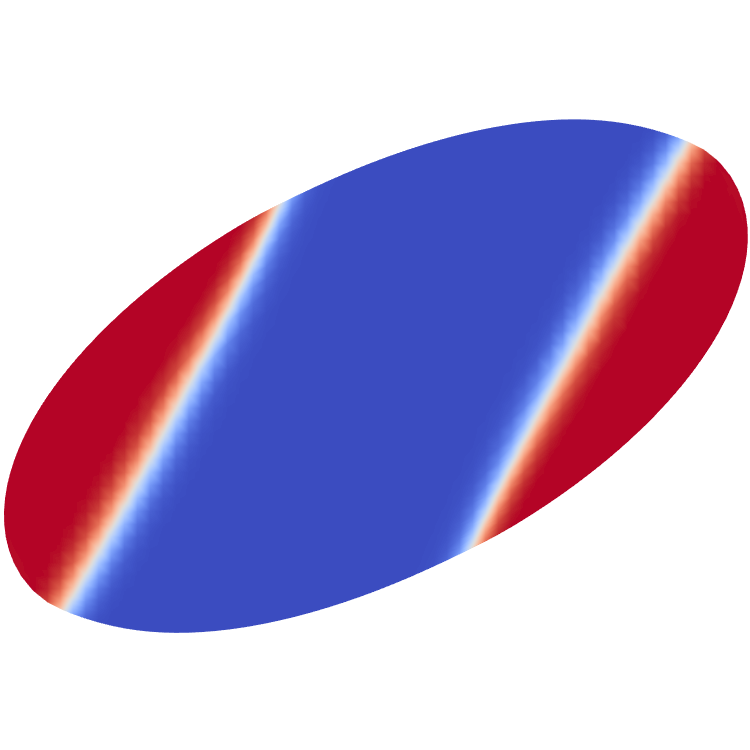} &
					\includegraphics[height=1.1cm]{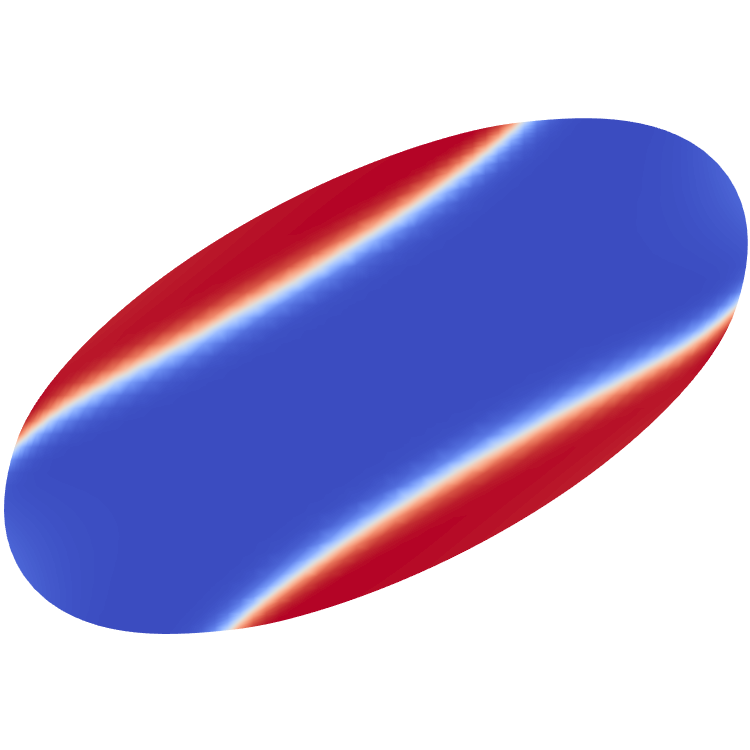} &
					\includegraphics[height=1.1cm]{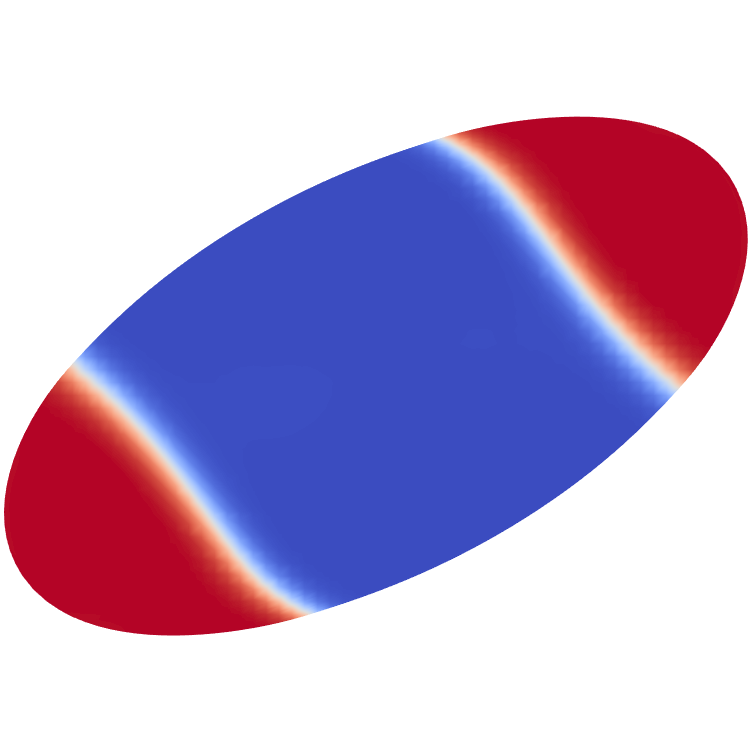} &
					\includegraphics[height=1.1cm]{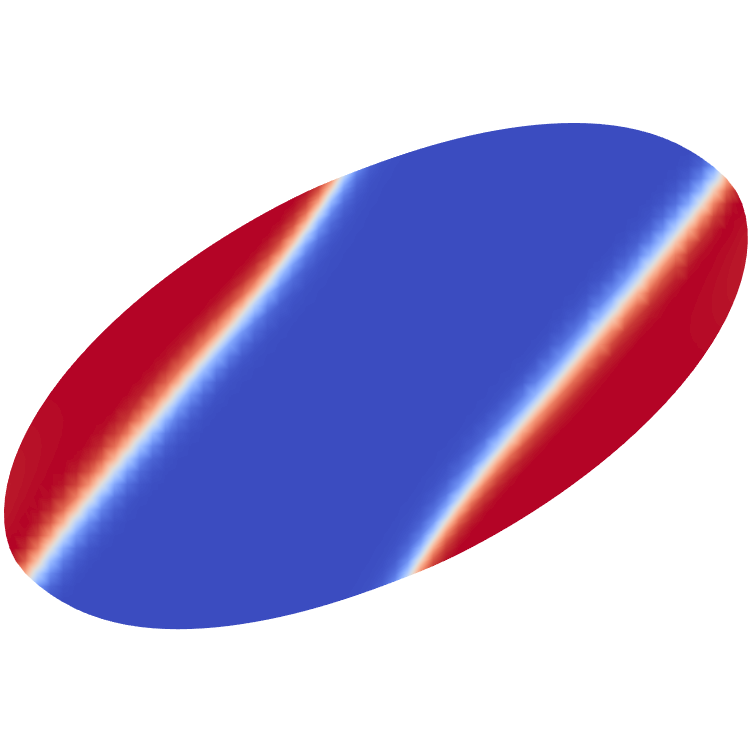} &
					\includegraphics[height=1.1cm]{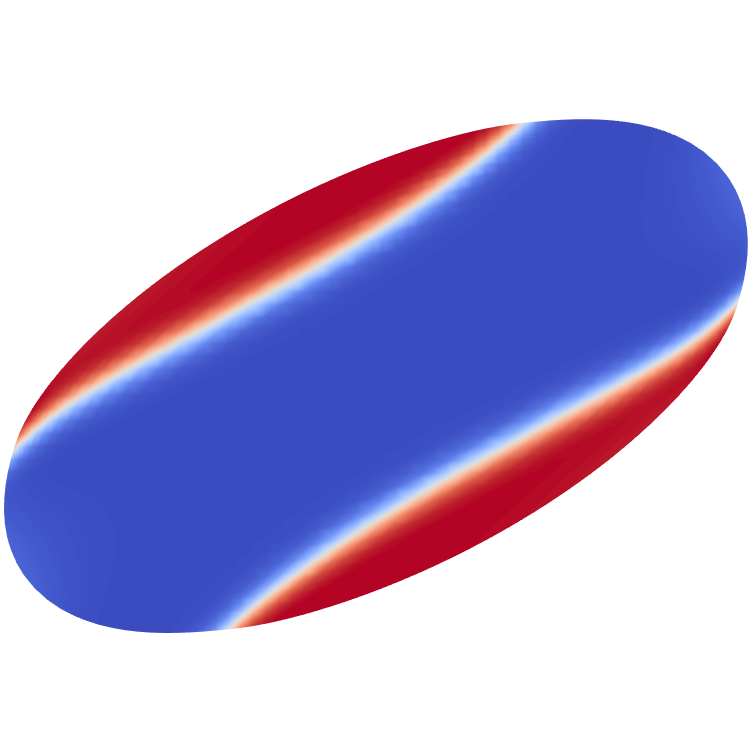} &
					\includegraphics[height=1.1cm]{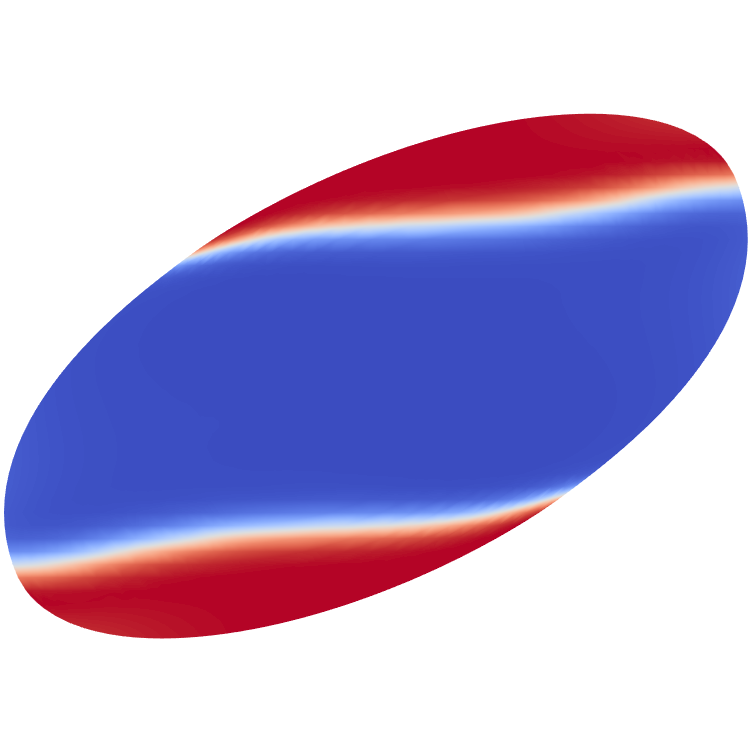} &
					\includegraphics[height=1.1cm]{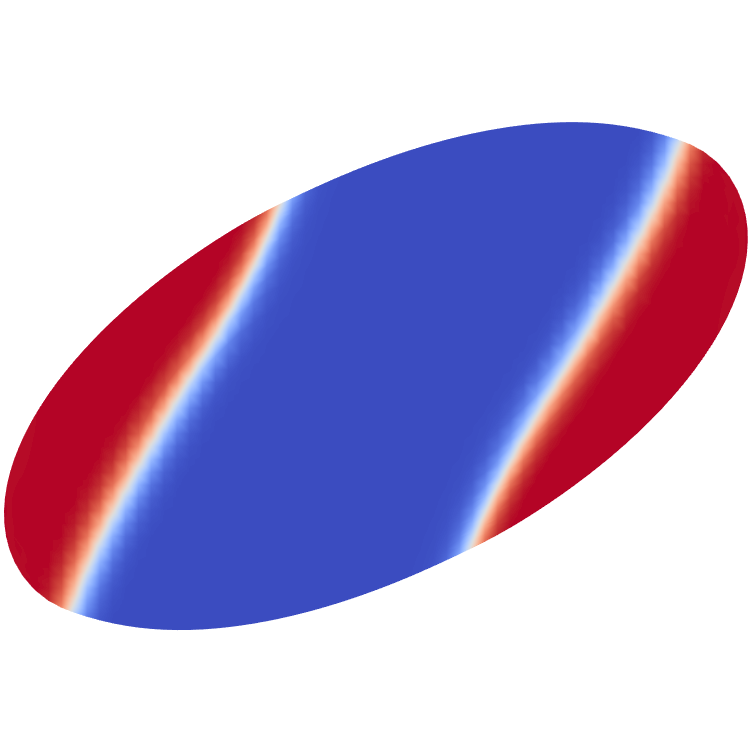} &
					\includegraphics[height=1.1cm]{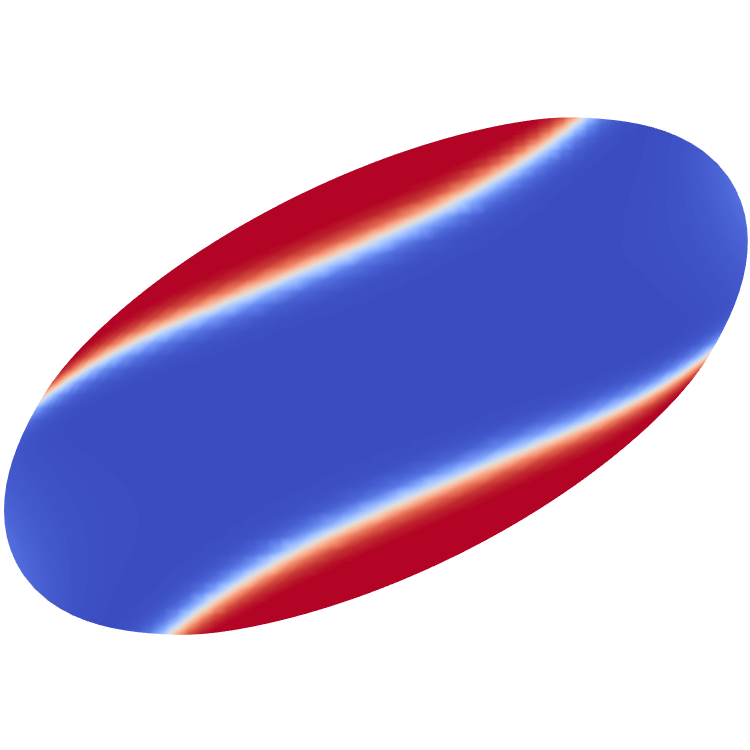} &
					\includegraphics[height=1.1cm]{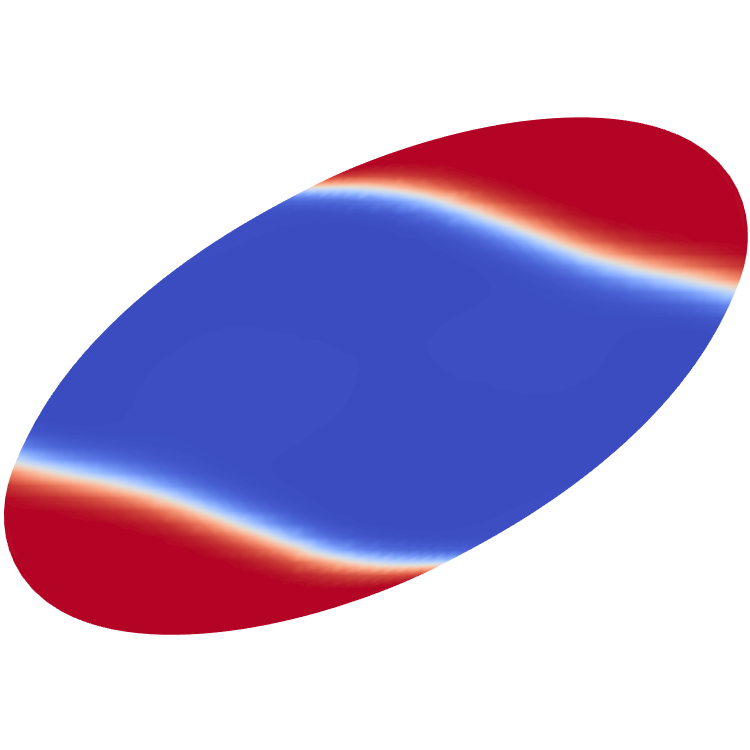} &
					\includegraphics[height=1.1cm]{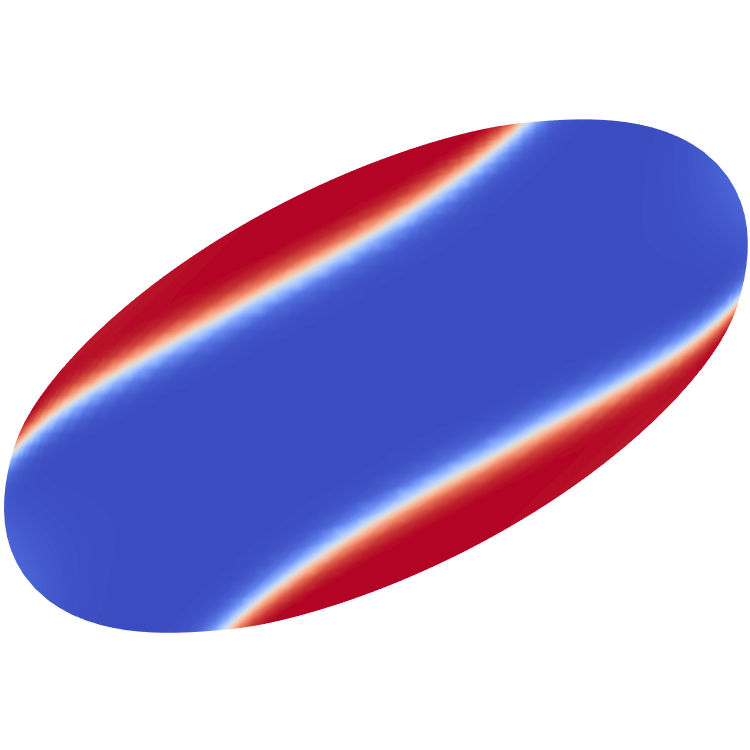} &
					\includegraphics[height=1.1cm]{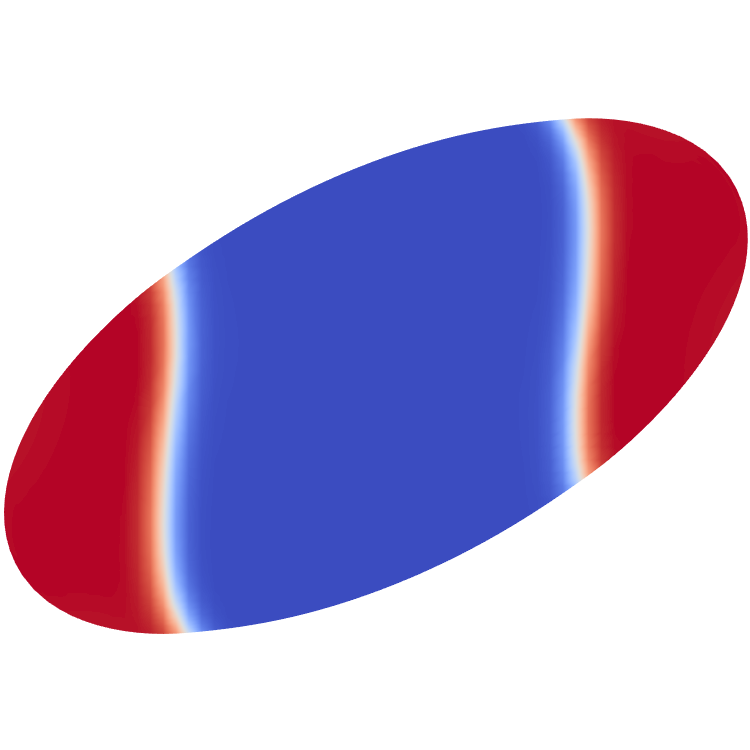}
			\end{tabular}
		}
		\subfigure[Bending Energy]{
			\includegraphics[height=5.5cm]{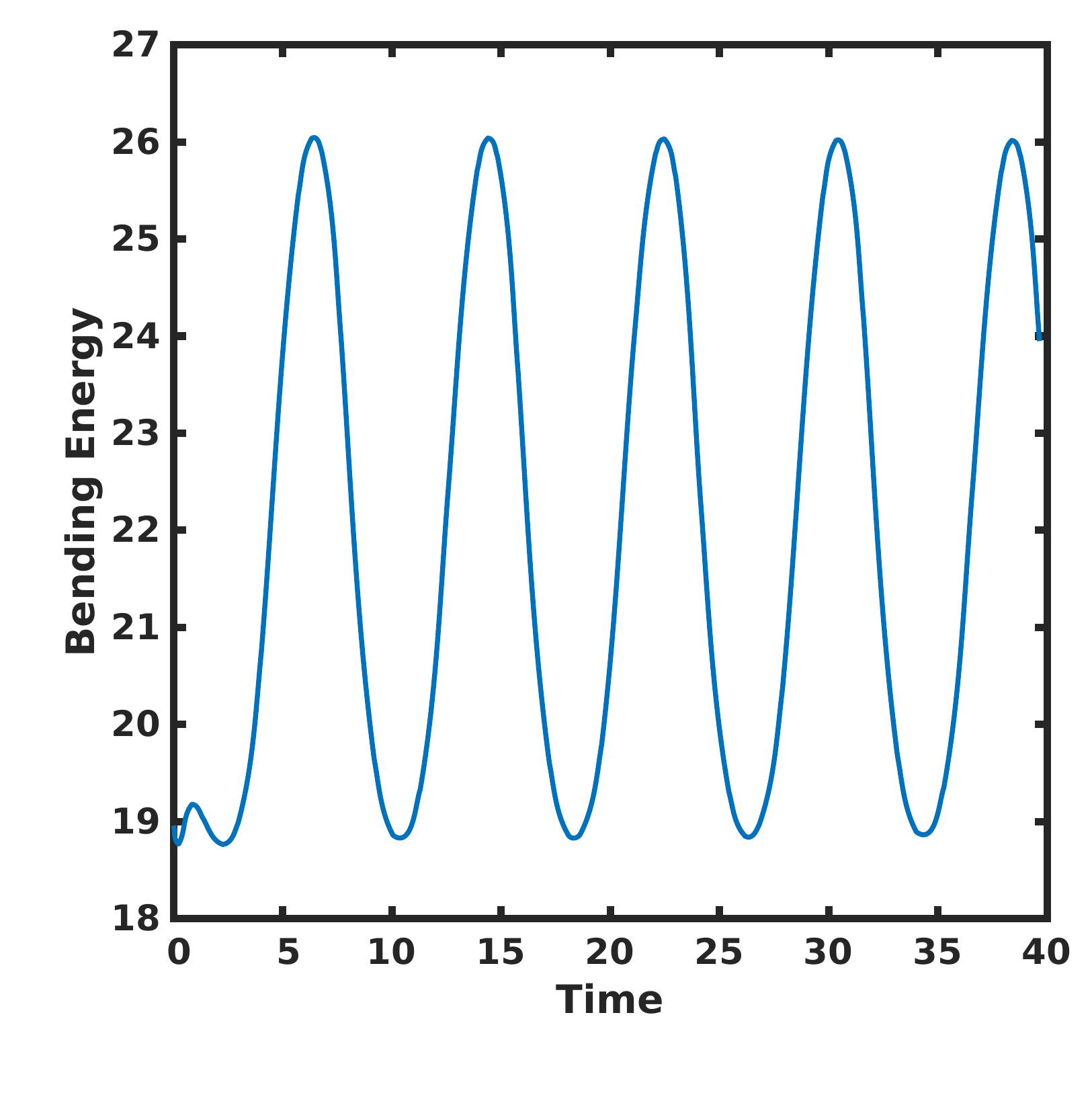}
		} 
		\subfigure[Domain Boundary Energy]{			
			\includegraphics[height=5.5cm]{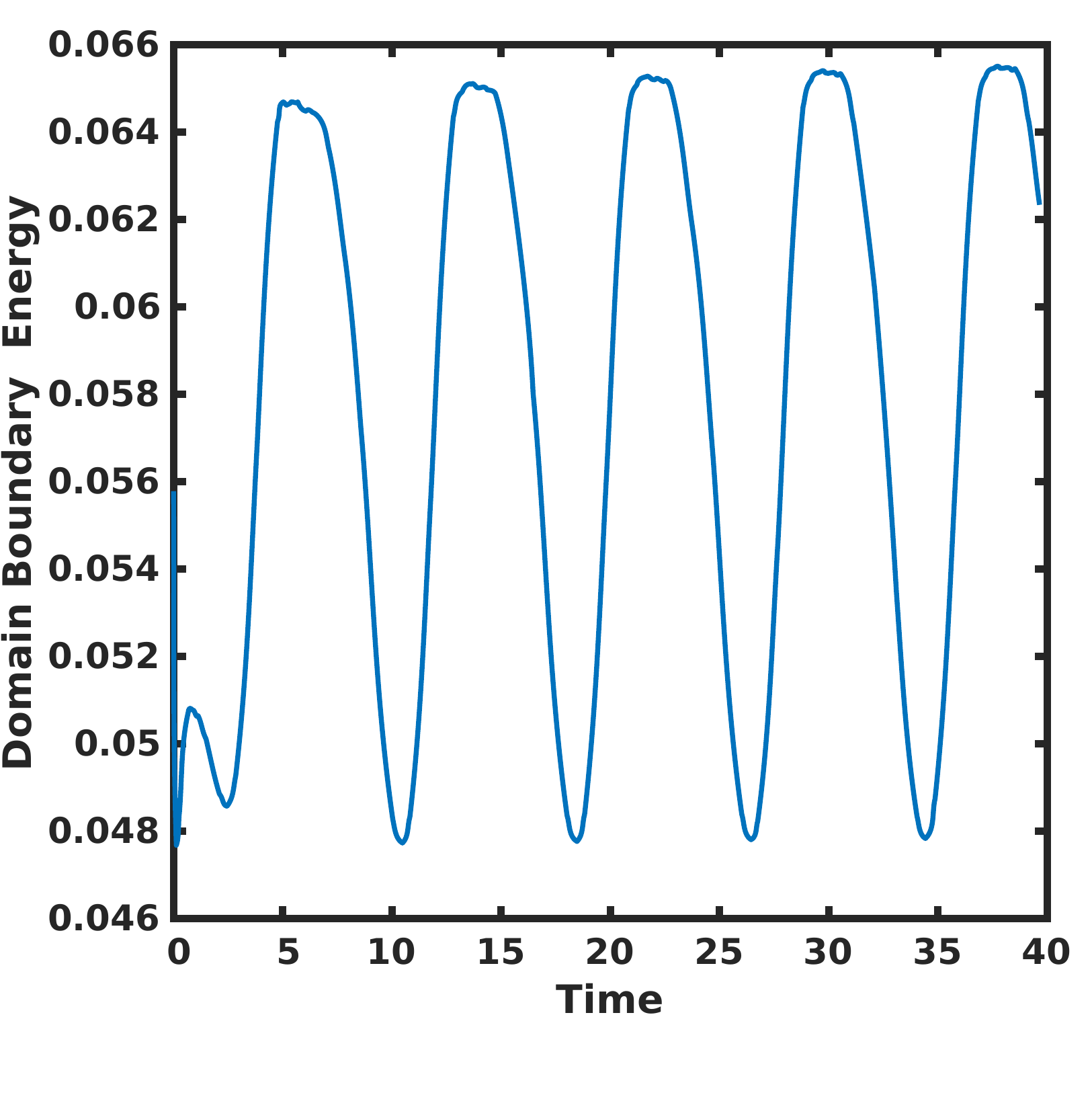}	
		}
	\end{center}
	\caption{{Dynamics of lipids on the vesicle for
        $\bar{c}=0.4$ and $\alpha=20$. The softer
        phase has a bending rigidity of $\kappa_c^B=0.4$ and with a Peclet number
        of $\Pe=20.0$ results in treading of domains.}}
	\label{fig:Tread_Pe20Alpha20}
\end{figure*}

The small domains seen at the vesicle tips, such as those shown in Fig.~\ref{fig:HalfTread},
can also be explained by considering the membrane curvature.
The highest curvature regions occur at the membrane tips,
Fig.~\ref{fig:Curvature_XZ_XY}, and to reduce the overall
bending energy,
the soft phase will preferentially aggregate in these regions if possible.
This aggregation will occur when either bending rigidity
of the soft phase, $\kappa_c^B$, or the surface Peclet number, $\Pe$ are small.
This can be seen in Fig.~(\ref{fig:HalfTread}) where $\kappa_c^B=0.4$, however as it
increases to $\kappa_c^B=0.8$, the domains on the tips are no longer observed, Fig.~(\ref{fig:Tread}). 

The influence of $\Pe$ on this behavior is verified by considering the result with a larger
Peclet number, $\Pe=20$, as seen in Fig.~\ref{fig:Tread_Pe20Alpha20}. 
While the soft phase
bending rigidity is the same as before, $\kappa_c^B=0.4$, these
small domains are no longer present. Additionally,
due to the slower surface diffusion, the merging even 
occurs much later. This is shown by the periodic
nature of the bending and domain boundary energy for the $\Pe=20$
case, Fig.~\ref{fig:Tread_Pe20Alpha20}, which 
still has not merged at a time of $t=40$.

\subsubsection{Domain Line Tension of $\alpha=0.5$}

\begin{figure}[!h]
\centering
  \includegraphics[height=8cm]{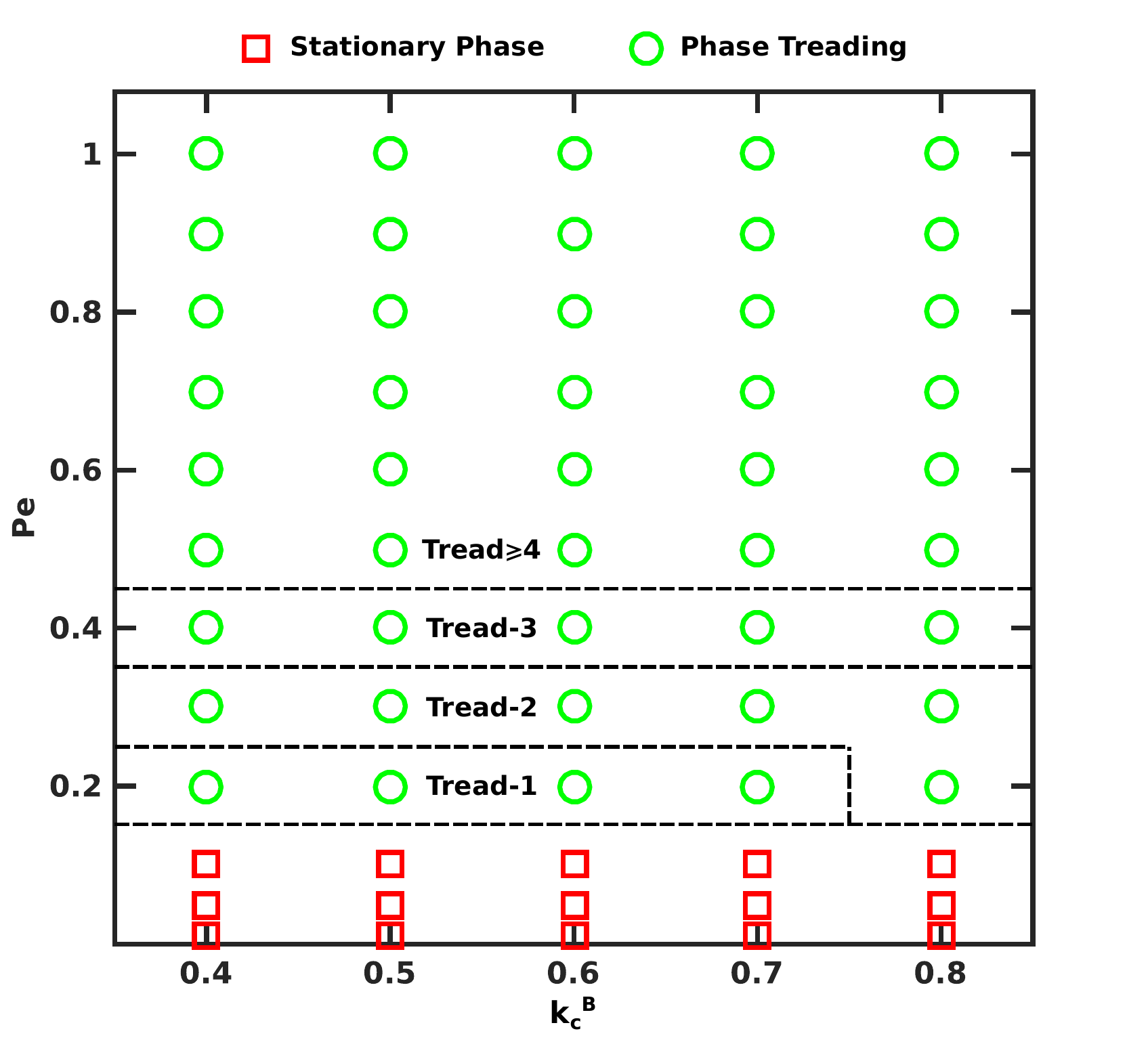}
  \caption{Variation of vesicle behavior with Peclet number and bending rigidity
  of the soft phase for $\bar{c}=0.4$ and
  $\alpha=0.5$.}
  \label{fig:PhaseDiagram_Alphap5}
\end{figure}

Now consider the dynamics for $\alpha=0.5$. In this case, the domain line forces
is stronger than the bending forces. Performing a systematic parameter study,
the resulting phase diagram can be seen in Fig.~\ref{fig:PhaseDiagram_Alphap5}.
It is interesting to note that in this case only two dynamics are observed:
stationary phases and phase treading. Compared to the $\alpha=20$ situation
shown previously, the contribution of the domain line energy to the dynamics
is 40-times larger, and thus the system will attempt to 
minimize the total domain boundary length. It is therefore not possible with $\alpha=0.5$
to obtain the vertical banding dynamic seen previously. Additionally,
it is important to understand the influence of $\alpha$ on the evolution
of the surface domains and the fluid field, Eqs.~\eqref{eq:nonDimNS} and \eqref{eq:nonDimChemPotential}.
When $\alpha=0.5$, the influence of the bending rigidity difference on 
the diffusion of the surface phases becomes weak, while the influence of the
variation of the surface phases on the fluid field becomes strong. Therefore,
it should be expected that the influence of the bending rigidity difference 
as a function of surface Peclet number should be decreased; which is what is observed
in the phase diagram seen in Fig.~\ref{fig:PhaseDiagram_Alphap5}.

\begin{figure*}[!hp]	
	\begin{center}
		\stackunder[15pt]{\includegraphics[height=1.75cm]{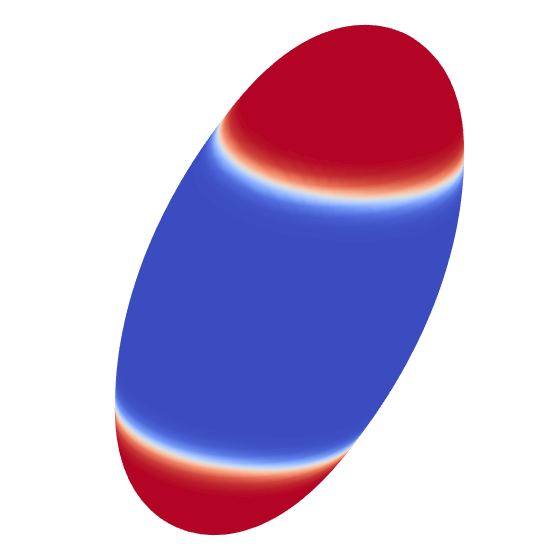}}{$t=2.0$}\hspace{3pt}
		\stackunder[15pt]{\includegraphics[height=1.75cm]{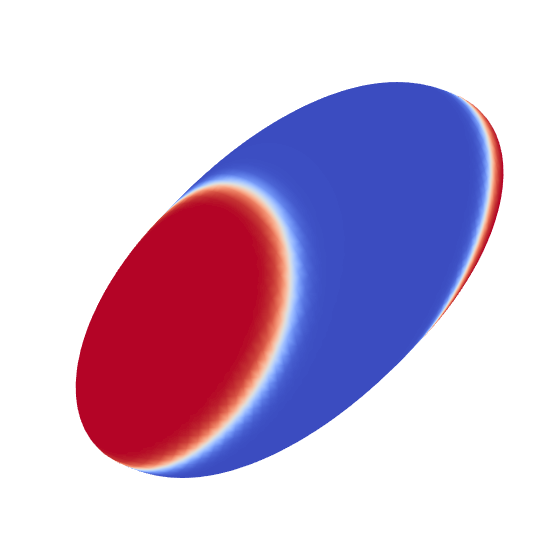}}{$t=4.0$}\hspace{3pt}
		\stackunder[15pt]{\includegraphics[height=1.75cm]{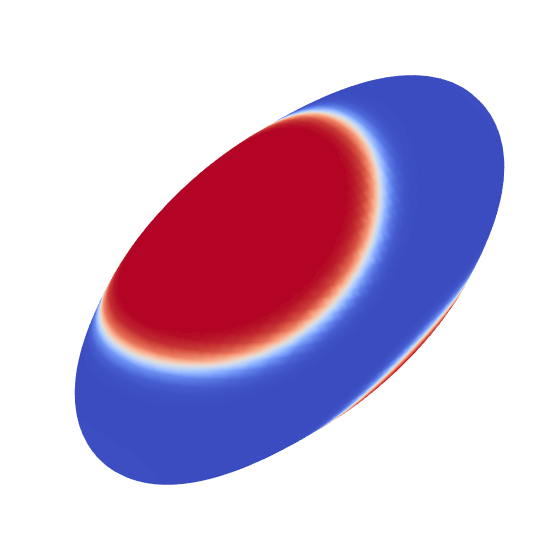}}{$t=6.0$}\hspace{3pt}
		\stackunder[15pt]{\includegraphics[height=1.75cm]{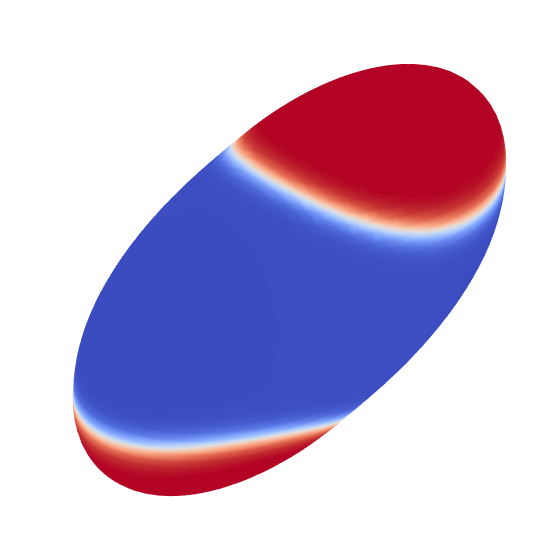}}{$t=10.0$}\hspace{3pt}
		\stackunder[15pt]{\includegraphics[height=1.75cm]{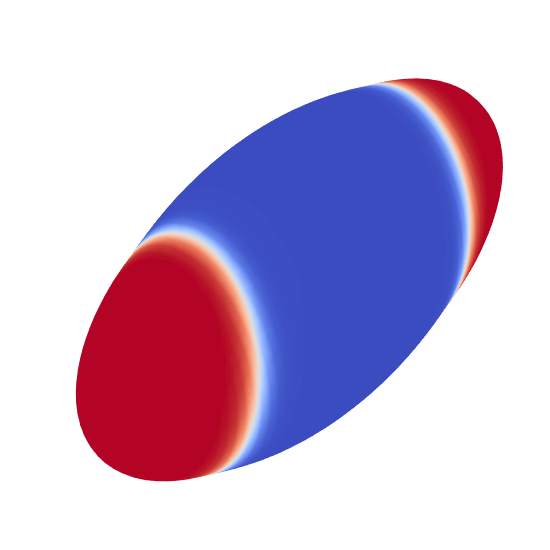}}{$t=12.5$}\hspace{3pt}
		\stackunder[15pt]{\includegraphics[height=1.75cm]{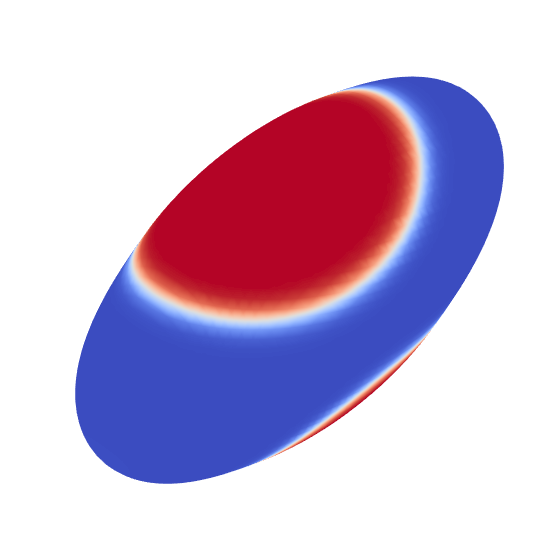}}{$t=17.5$}\hspace{3pt}
		\stackunder[15pt]{\includegraphics[height=1.75cm]{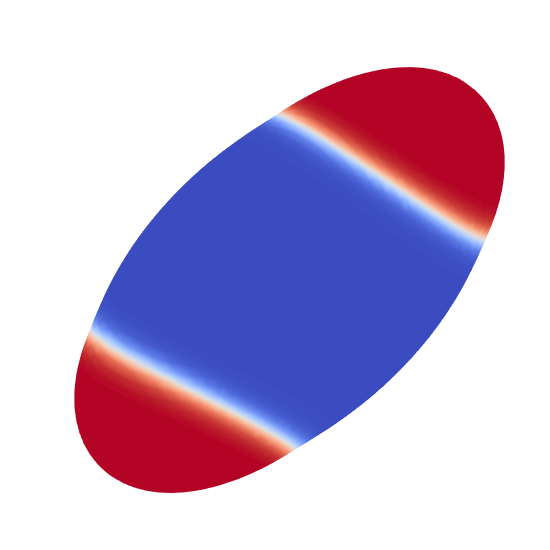}}{$t=20.0$}\hspace{3pt}
		\stackunder[15pt]{\includegraphics[height=1.75cm]{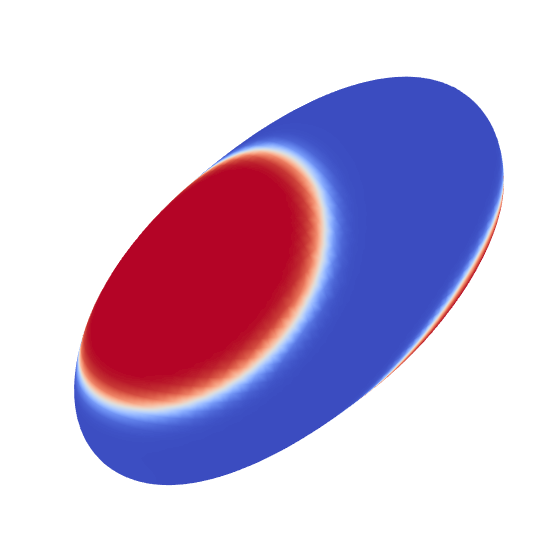}}{$t=25.0$}
		\\\bigskip
		\stackunder[15pt]{\includegraphics[height=1.75cm]{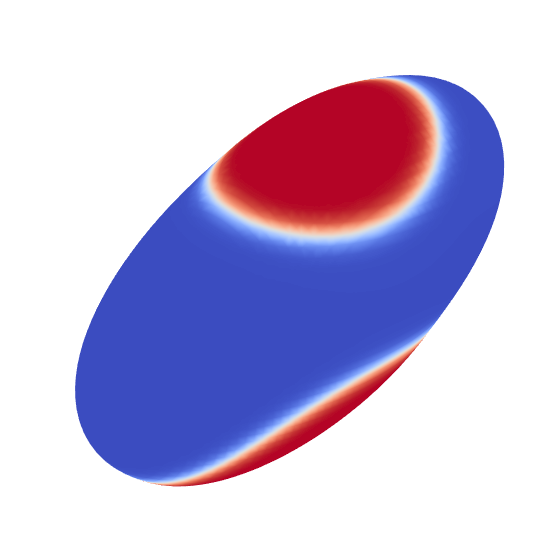}}{$t=27.5$}\hspace{3pt}
		\stackunder[15pt]{\includegraphics[height=1.75cm]{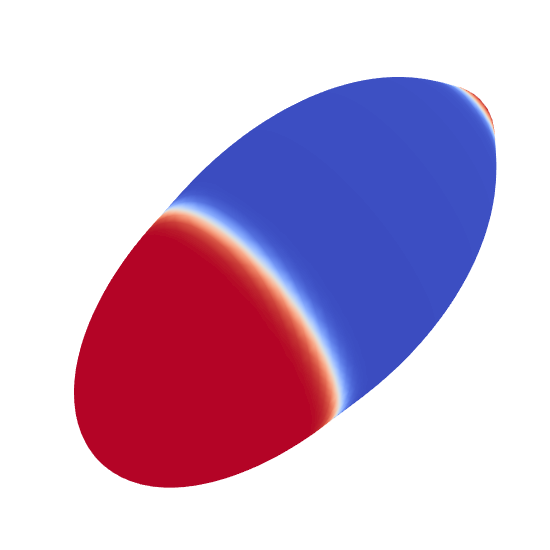}}{$t=30.0$}\hspace{3pt}
		\stackunder[15pt]{\includegraphics[height=1.75cm]{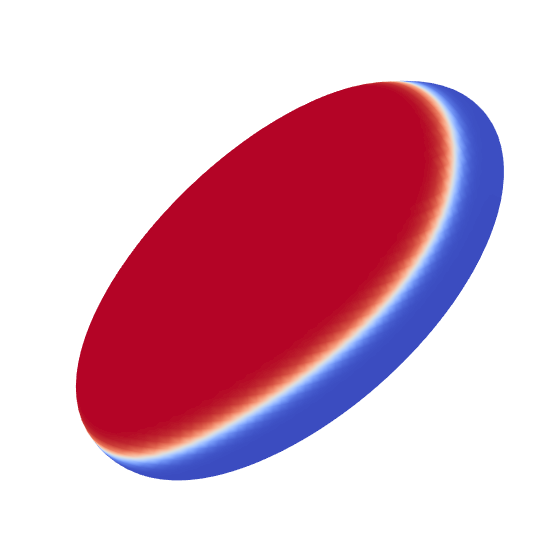}}{$t=35.0$}\hspace{3pt}
		\stackunder[15pt]{\includegraphics[height=1.75cm]{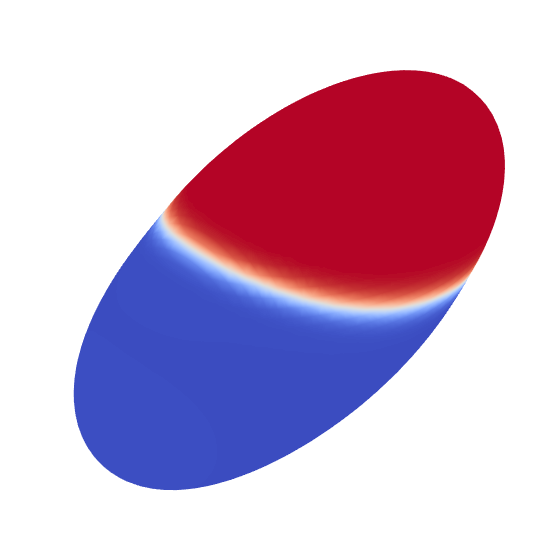}}{$t=37.5$}\hspace{3pt}
		\stackunder[15pt]{\includegraphics[height=1.75cm]{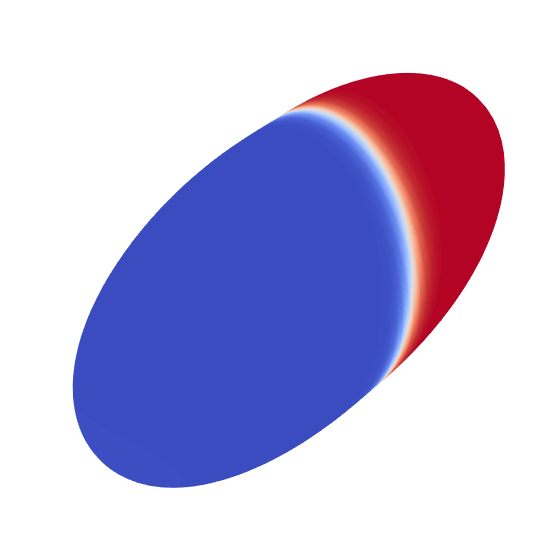}}{$t=40.0$}\hspace{3pt}
		\stackunder[15pt]{\includegraphics[height=1.75cm]{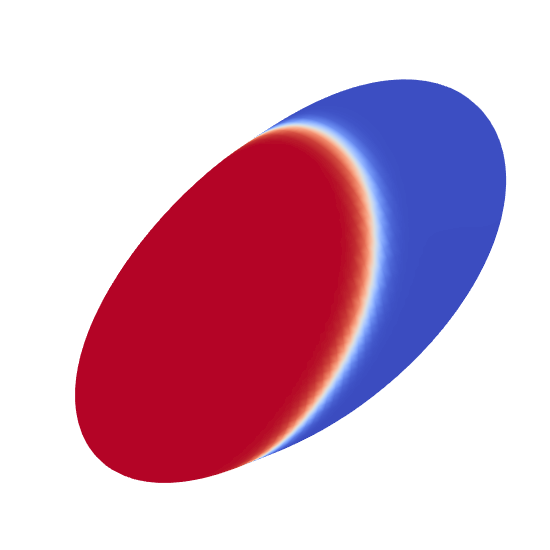}}{$t=50.0$}\hspace{3pt}
		\stackunder[15pt]{\includegraphics[height=1.75cm]{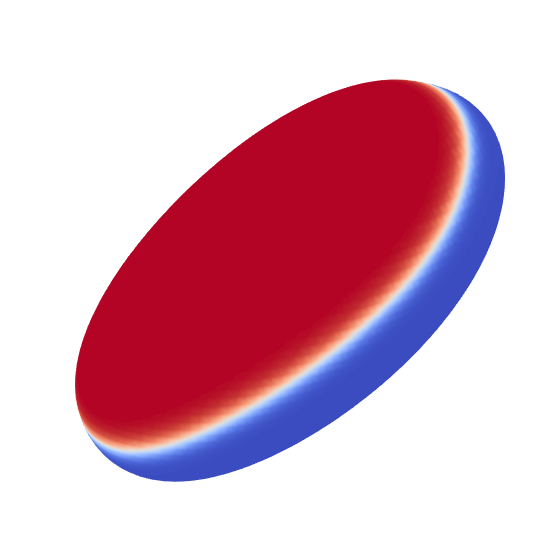}}{$t=52.5$}\hspace{3pt}
		\stackunder[15pt]{\includegraphics[height=1.75cm]{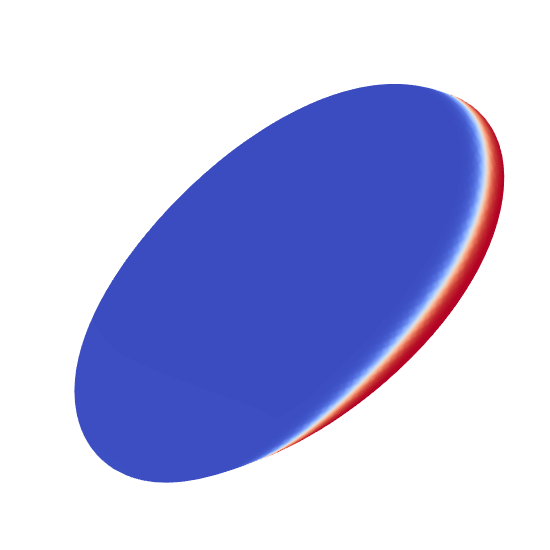}}{$t=60.0$}%
	\end{center}

	\caption{Tread-2 dynamics of vesicle with 
        $\bar{c}=0.4$ and $\alpha=0.5$. The soft
		phase has a bending rigidity of $\kappa_c^B=0.4$ while the Peclet number
        is $\Pe=0.3$. 
        These figures are centered on the vesicle.}
		\label{fig:Tread_Pep3Alphap5}
	
\end{figure*}

\begin{figure*}[!hp]	
	\begin{center}
		\stackunder[2pt]{$t=20.0$}{\includegraphics[height=3.0cm]{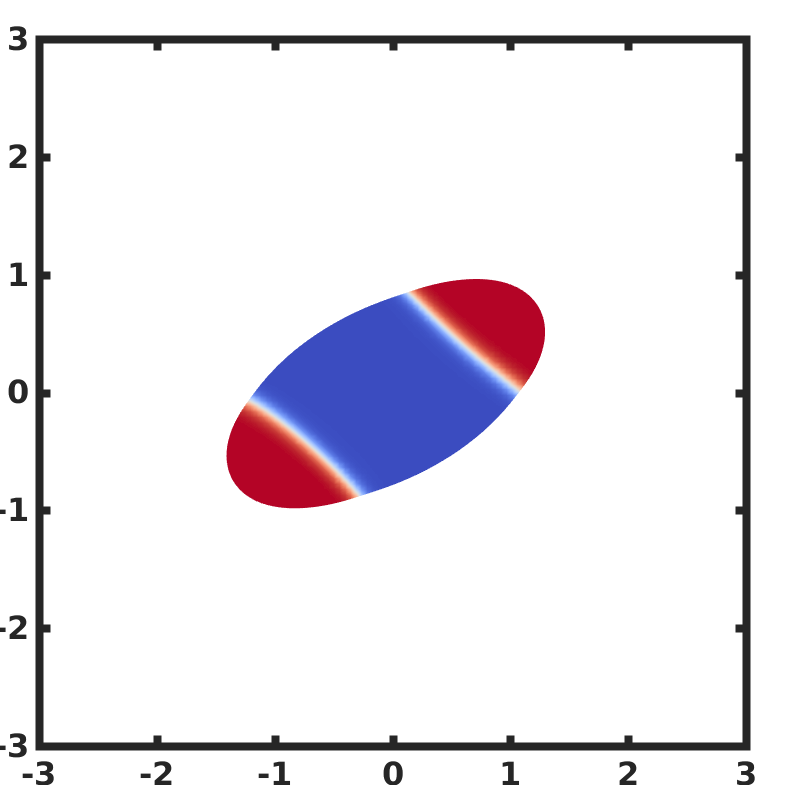}}\hspace{3pt}
		\stackunder[2pt]{$t=25.0$}{\includegraphics[height=3.0cm]{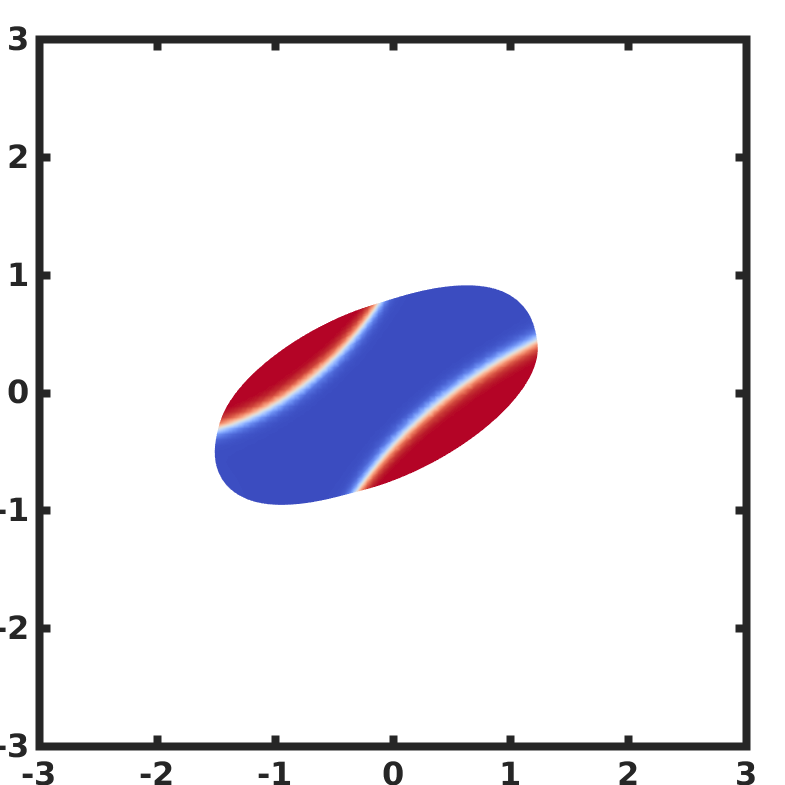}}\hspace{3pt}
		\stackunder[2pt]{$t=30.0$}{\includegraphics[height=3.0cm]{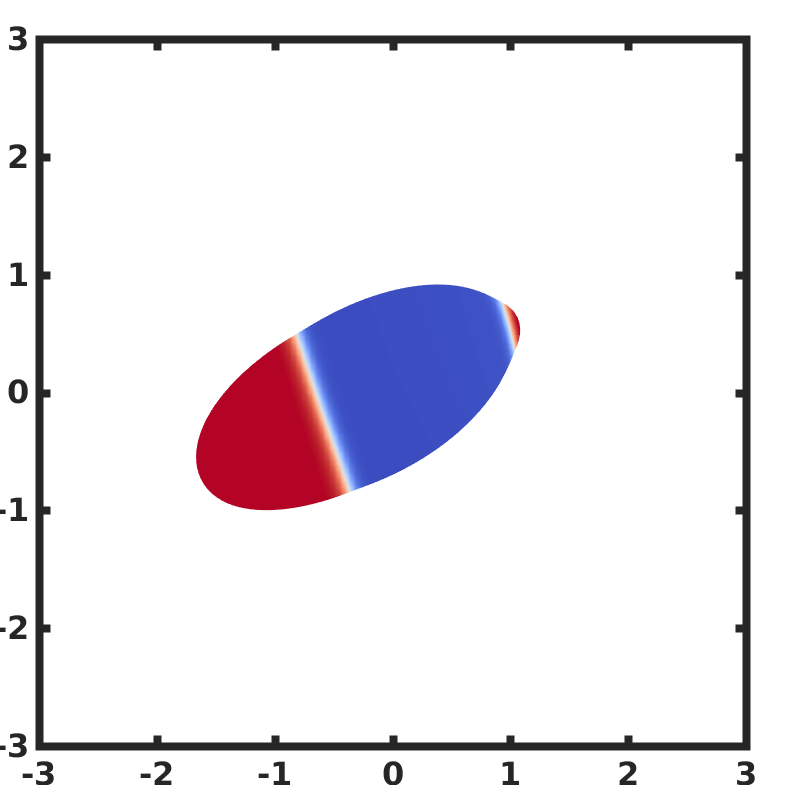}}\hspace{3pt}
		\stackunder[2pt]{$t=35.0$}{\includegraphics[height=3.0cm]{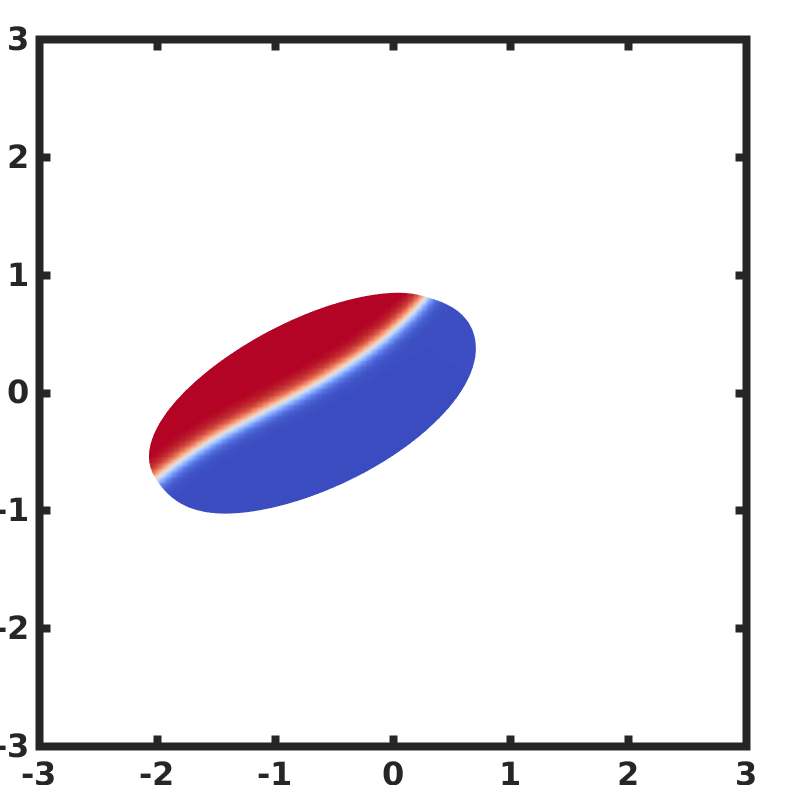}}\hspace{3pt}
		\stackunder[2pt]{$t=40.0$}{\includegraphics[height=3.0cm]{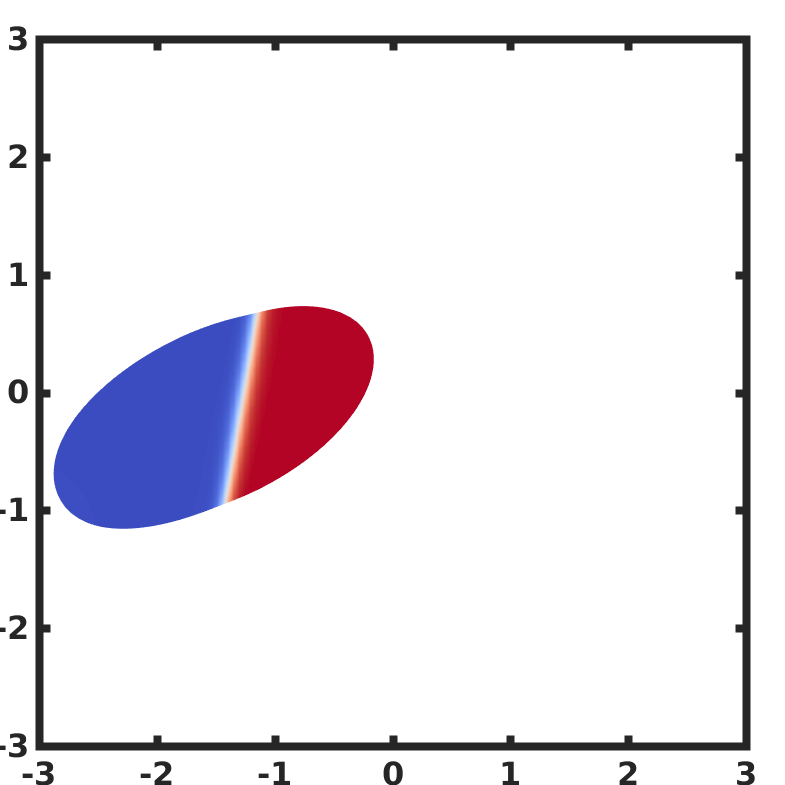}}
		\\\bigskip
		\stackunder[2pt]{$t=47.5$}{\includegraphics[height=3.0cm]{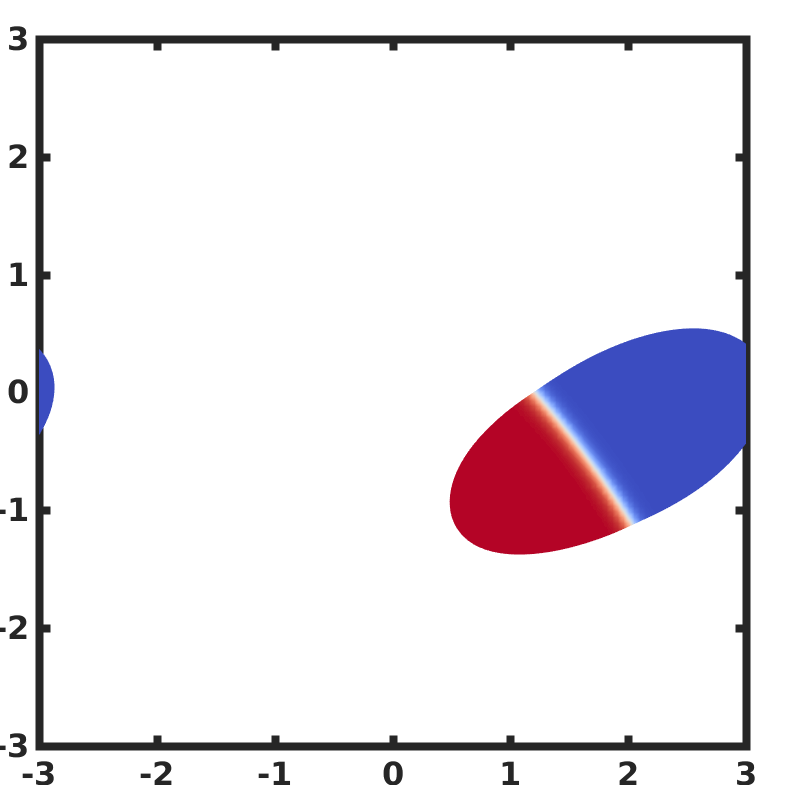}}\hspace{3pt}
		\stackunder[2pt]{$t=50.0$}{\includegraphics[height=3.0cm]{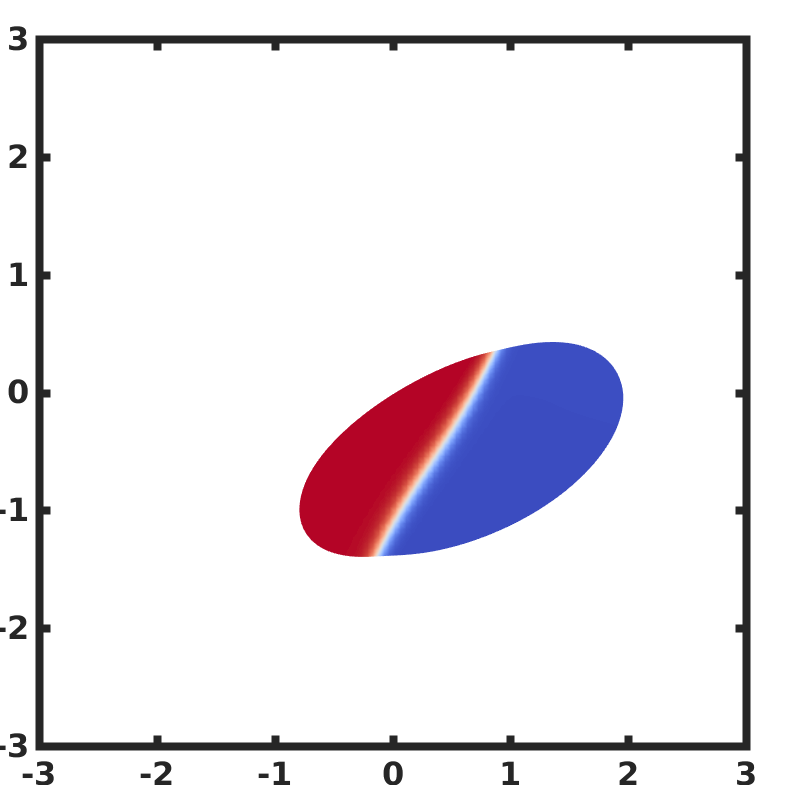}}\hspace{3pt}
		\stackunder[2pt]{$t=53.0$}{\includegraphics[height=3.0cm]{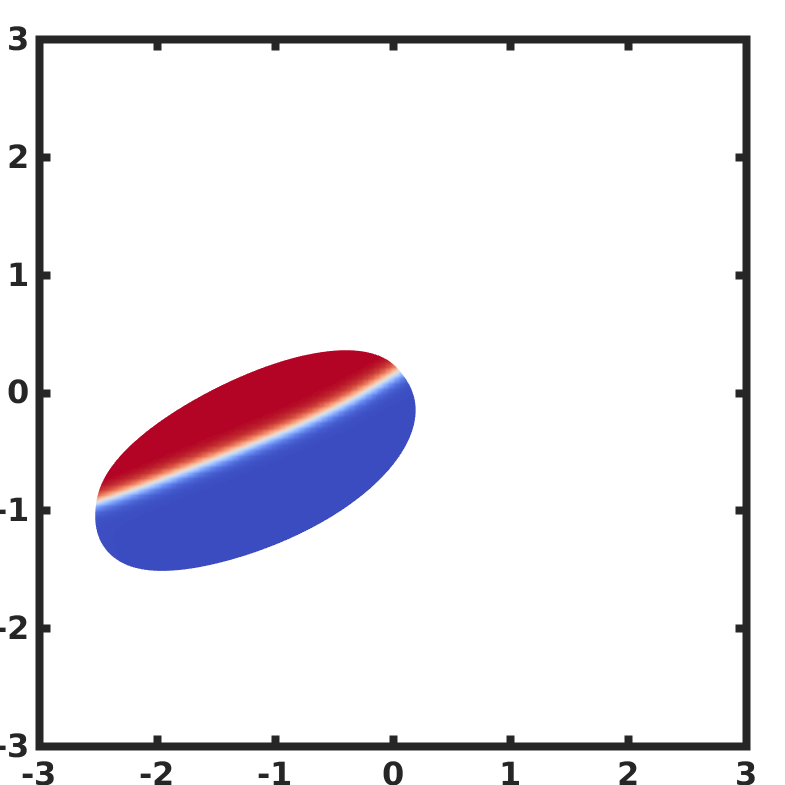}}\hspace{3pt}		
		\stackunder[2pt]{$t=57.5$}{\includegraphics[height=3.0cm]{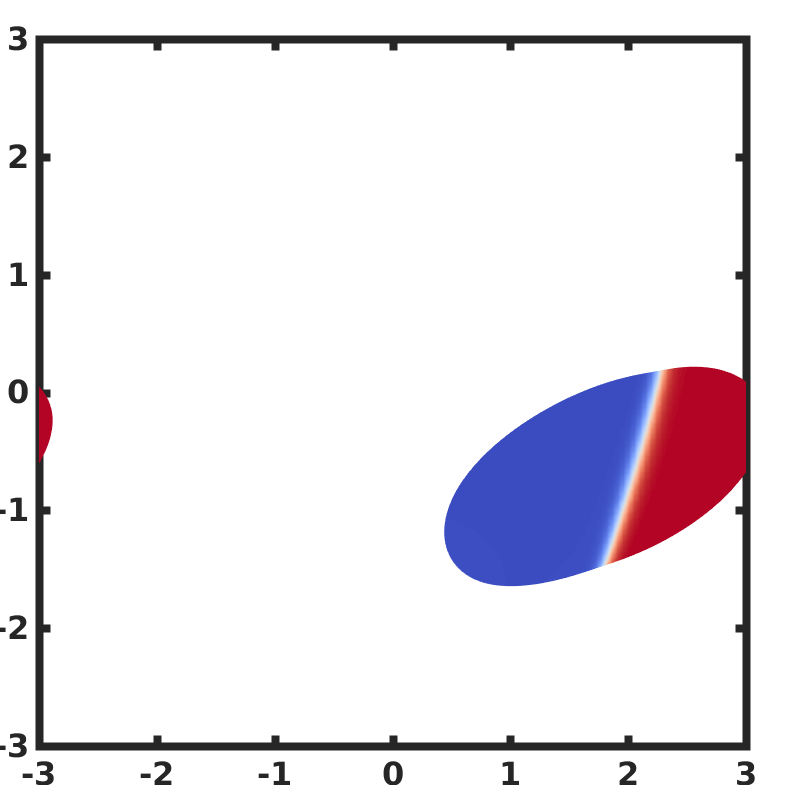}}\hspace{3pt}
		\stackunder[2pt]{$t=60.0$}{\includegraphics[height=3.0cm]{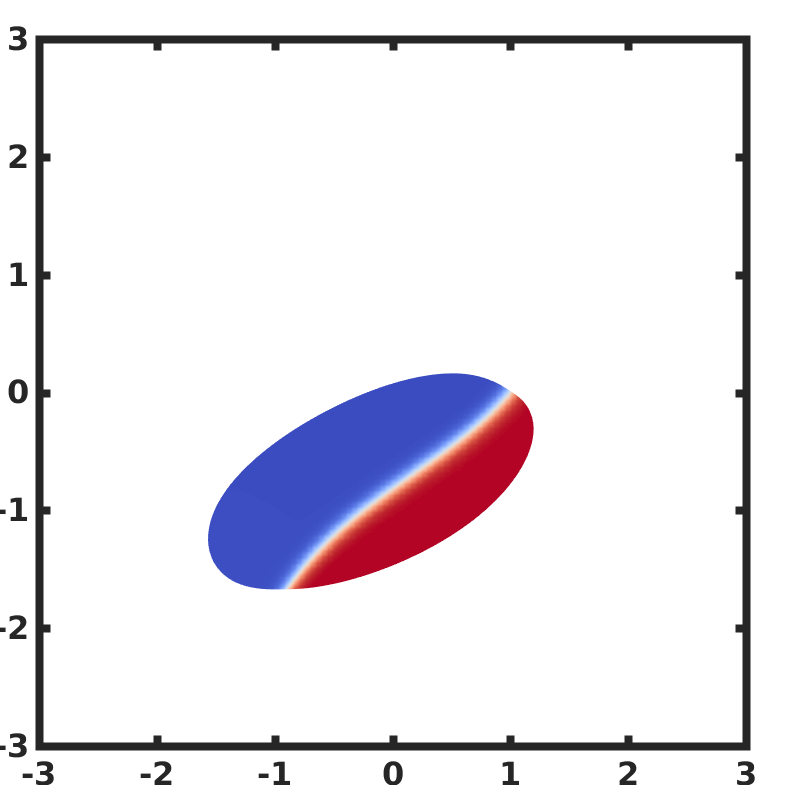}}\hspace{3pt}	
	\end{center}
	\caption{The X-Y plane of the Tread-2 dynamics seen in Fig.~\ref{fig:Tread_Pep3Alphap5}.
		The parameters are $\bar{c}=0.4$, $\alpha=0.5$, $\kappa_c^B =0.4$, and $\Pe=0.3$. 
		A breaking of the symmetry by dissimilar bending rigidities induces motion.}
		\label{fig:Tread_Pep3Alphap5_XY_boxed}
	
\end{figure*}

\begin{figure}[!hp]
	\begin{center}
		\subfigure[Bending Energy]{
			\includegraphics[height=5.5cm]{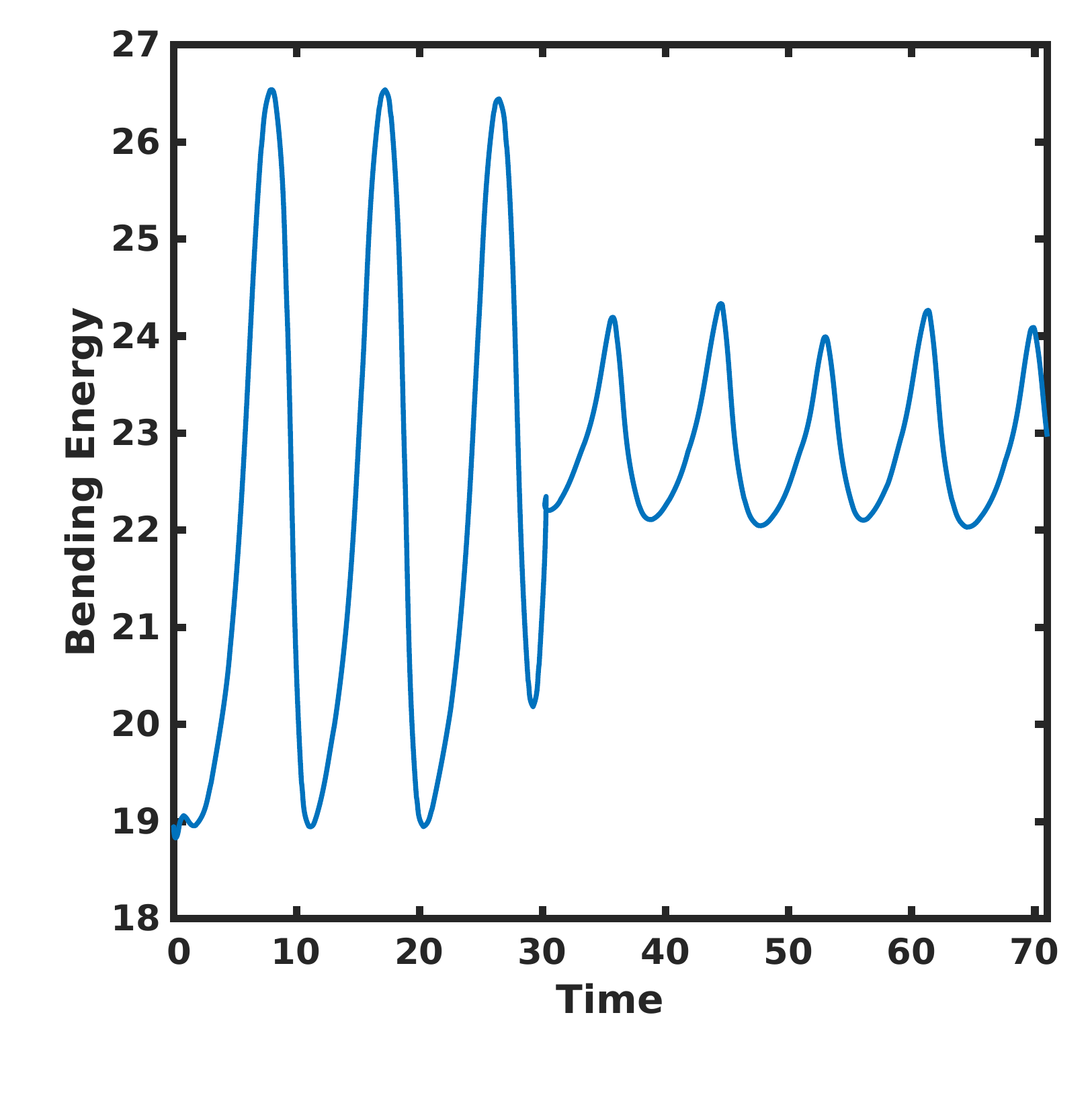}
		} 
		\qquad	
		\subfigure[Domain Boundary Energy]{			
			\includegraphics[height=5.5cm]{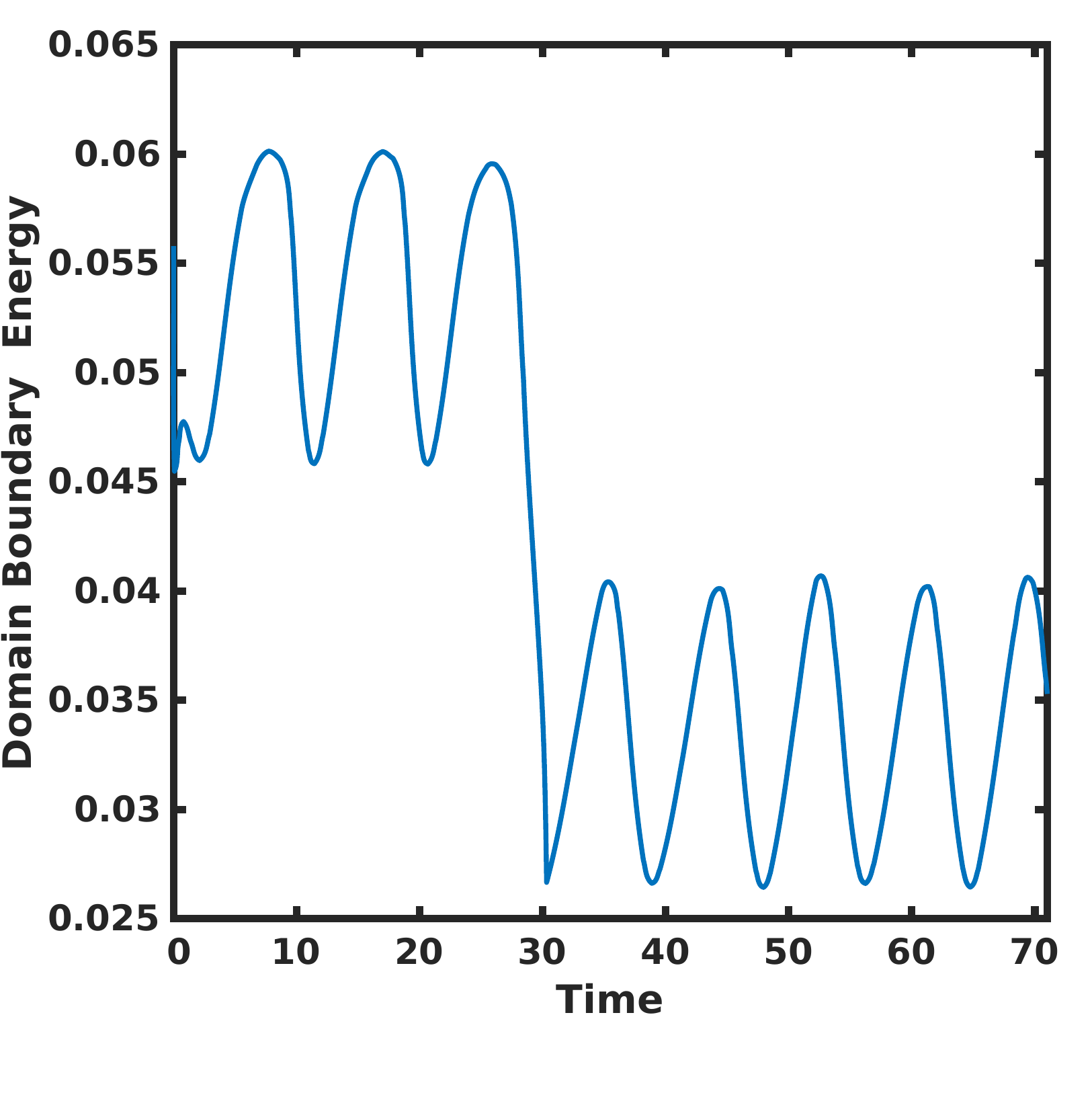}	
		}		
	\end{center}
		\caption{The bending and domain boundary energy
			for the results shown in Fig.~\ref{fig:Tread_Pep3Alphap5} and \ref{fig:Tread_Pep3Alphap5_XY_boxed}.
			The parameters are $\bar{c}=0.4$, $\alpha=0.5$, $\kappa_c^B =0.4$, and $\Pe=0.3$.}		 
		\label{fig:Tread_Pep3Alphap5_Energy}
	
\end{figure}

\begin{figure}[!hp]
	\begin{center}		
		\subfigure[Inclination Angle]{
			\includegraphics[height=5.5cm]{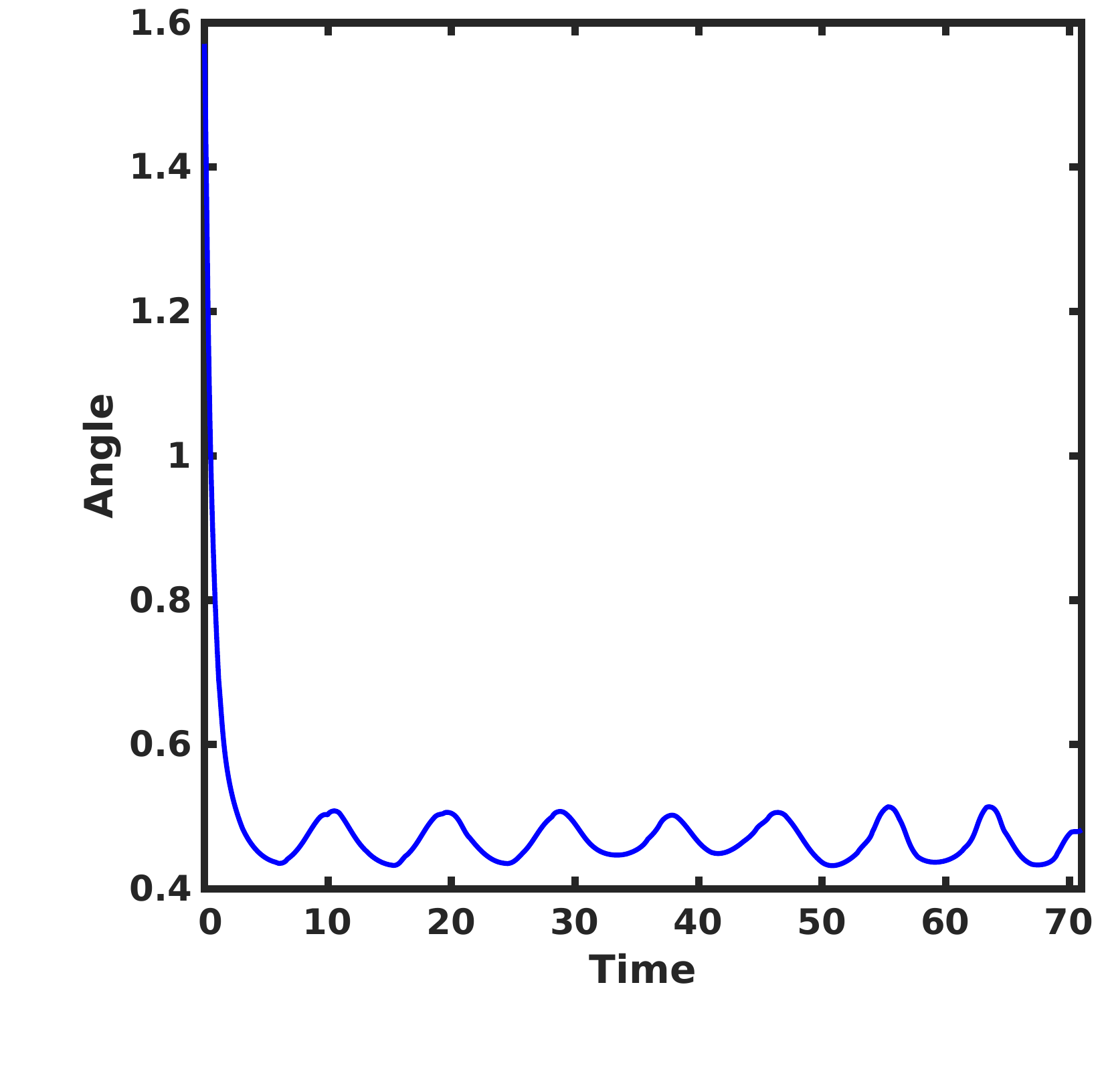}
		} 
		\qquad
		\subfigure[Vesicle Center]{			
			\includegraphics[height=5.5cm]{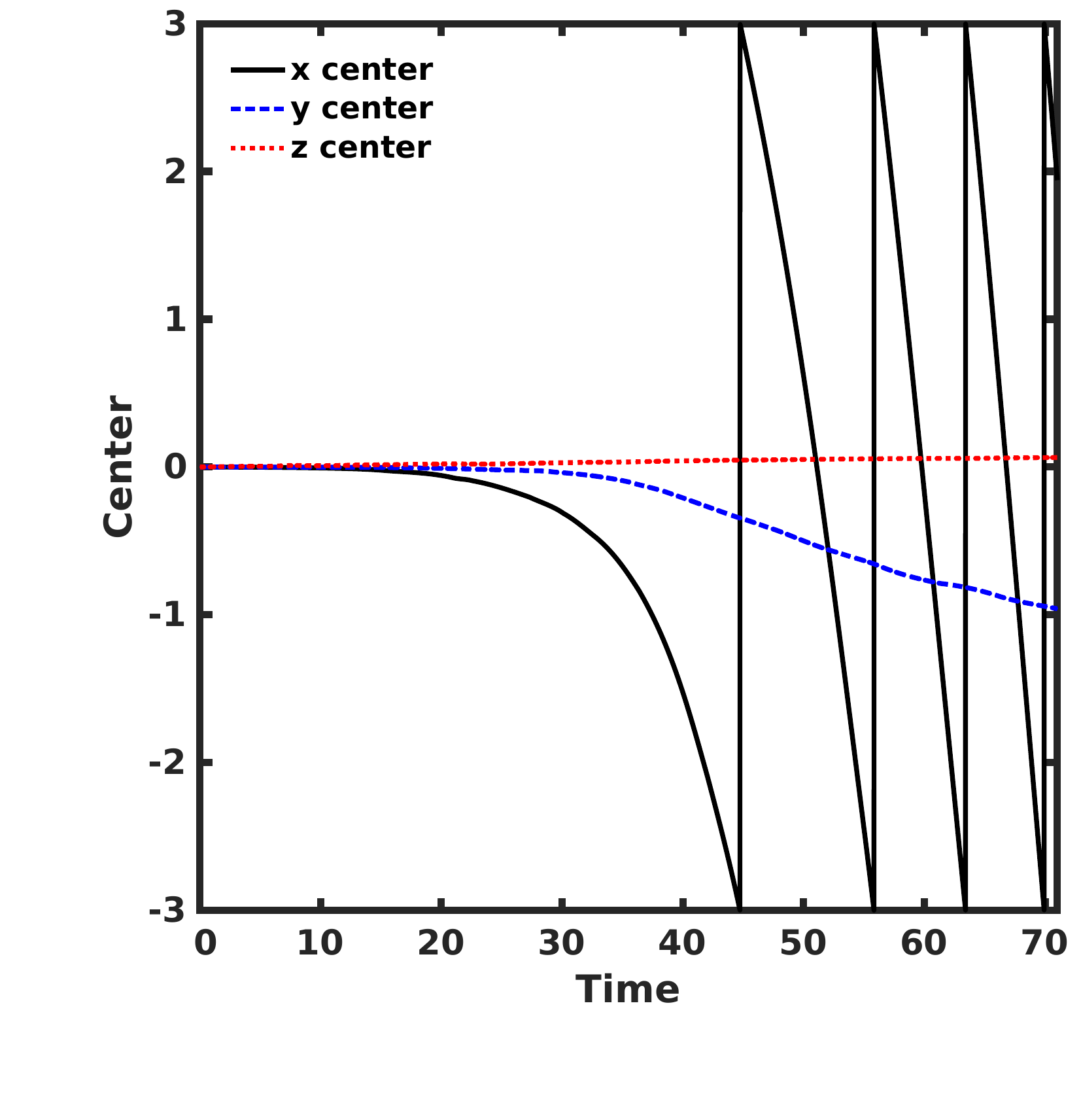}	
		}			
	\end{center}
		\caption{The vesicle inclination angle and center location
			for the results shown in Fig.~\ref{fig:Tread_Pep3Alphap5} and \ref{fig:Tread_Pep3Alphap5_XY_boxed}.
			The parameters are $\bar{c}=0.4$, $\alpha=0.5$, $\kappa_c^B =0.4$, and $\Pe=0.3$.}		 
		\label{fig:Tread_Pep3Alphap5_AngleCenter}
	
\end{figure}

A sample result with $\alpha=0.5$, $\kappa_c^B=0.4$ and $\Pe=0.3$ is seen in 
Figs.~\ref{fig:Tread_Pep3Alphap5}-\ref{fig:Tread_Pep3Alphap5_AngleCenter}.
The results of Fig.~\ref{fig:Tread_Pep3Alphap5} are centered on the vesicle,
while those in Fig.~\ref{fig:Tread_Pep3Alphap5_XY_boxed} demonstrate the motion
induced after symmetry breaking.


Up to a time of $t=30$, there are two domains phase treading on the membrane with a period
of 9.23. At $t=30$, a large drop in the domain boundary energy occurs, which indicates
that there is now a single domain on the membrane. After merging, the single domain continues
to phase tread, with a lower period of 8.54. 
See Fig.~\ref{fig:Tread_Pep3Alphap5_Energy}
for the bending and domain boundary energy over time.
After merging, the vesicle
moves slightly downwards, exposing it to a higher fluid velocity.
It is suspected that this is the cause for the decrease in the treading period.

Due to the phase treading of the surface phases, there 
are periodic fluctuations of the inclination angle.
Additionally, after merging the symmetry of the membrane domains is lost, which results in
a non-symmetric velocity field. This results in the movement of the vesicle center.
See Fig.~\ref{fig:Tread_Pep3Alphap5_AngleCenter} for the vesicle inclination angle
and center over time.

\section{Discussion}

\begin{figure}[!ht]
	\begin{center}		
		\subfigure[Time to Merge]{
			\includegraphics[height=5.5cm]{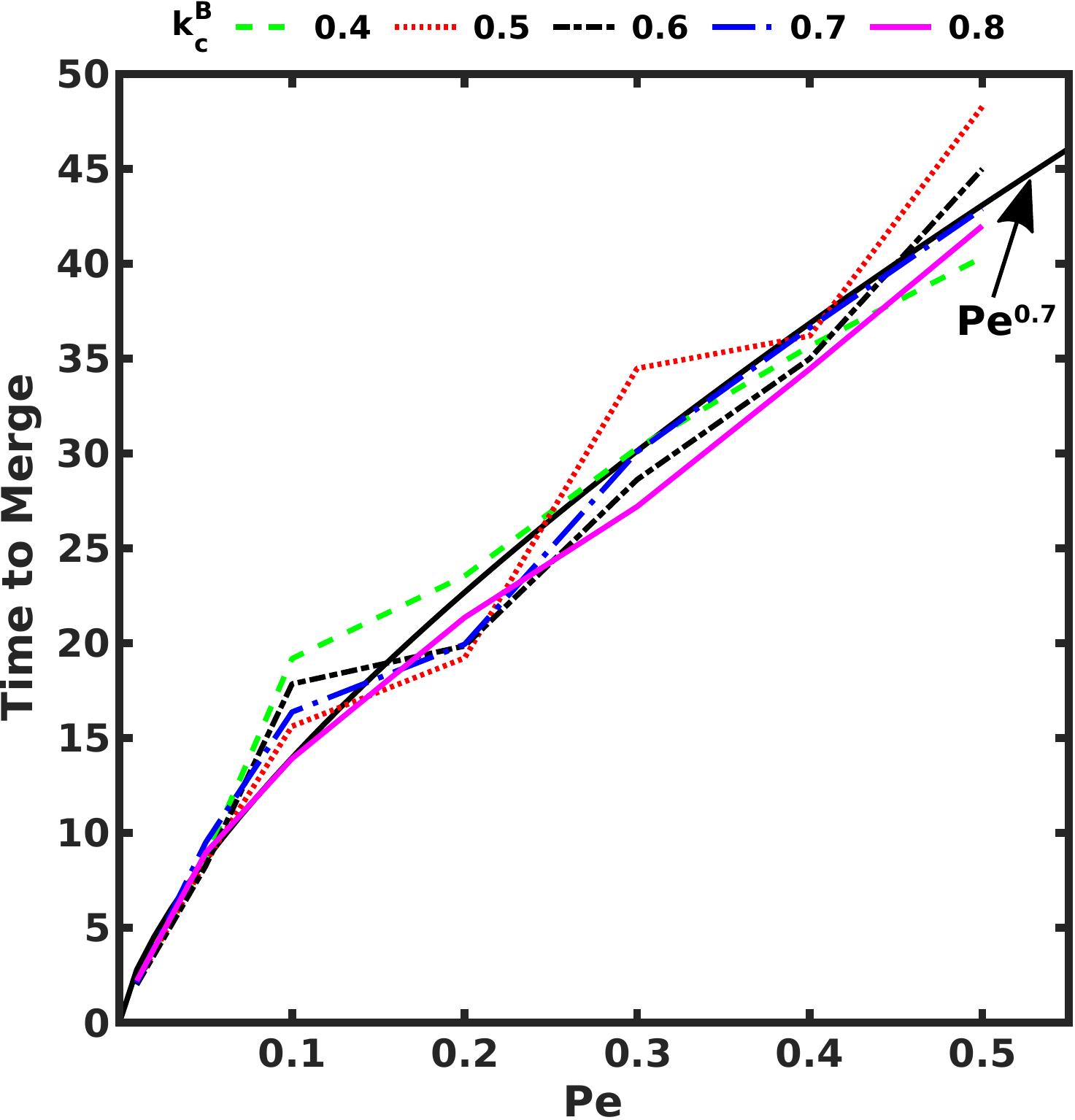}		
		} 
		\qquad
		\subfigure[Phase Period Before Merging]{
			\includegraphics[height=5.5cm]{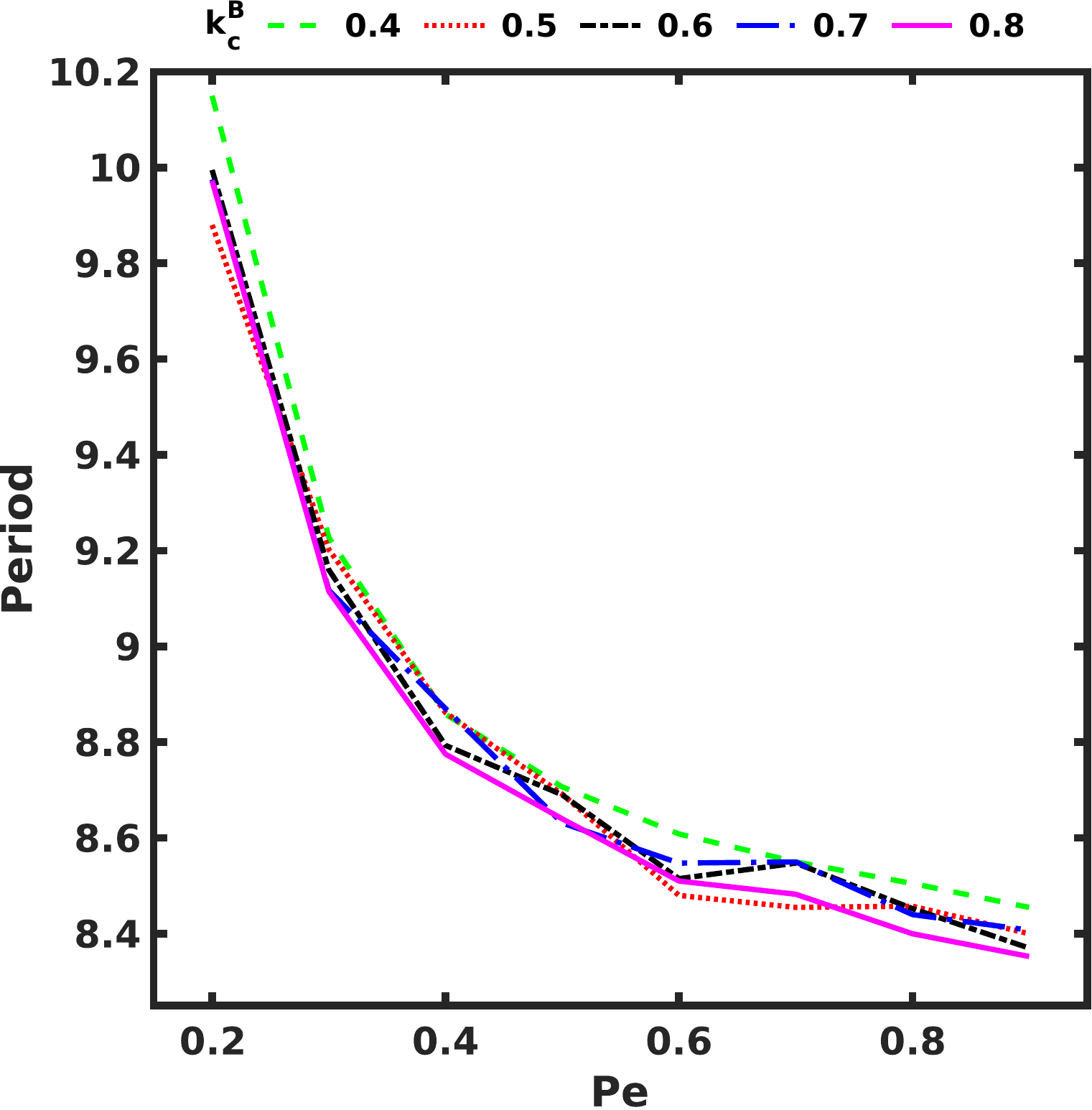}
		} 
		\qquad
		\subfigure[Phase Period After Merging]{			
			\includegraphics[height=5.5cm]{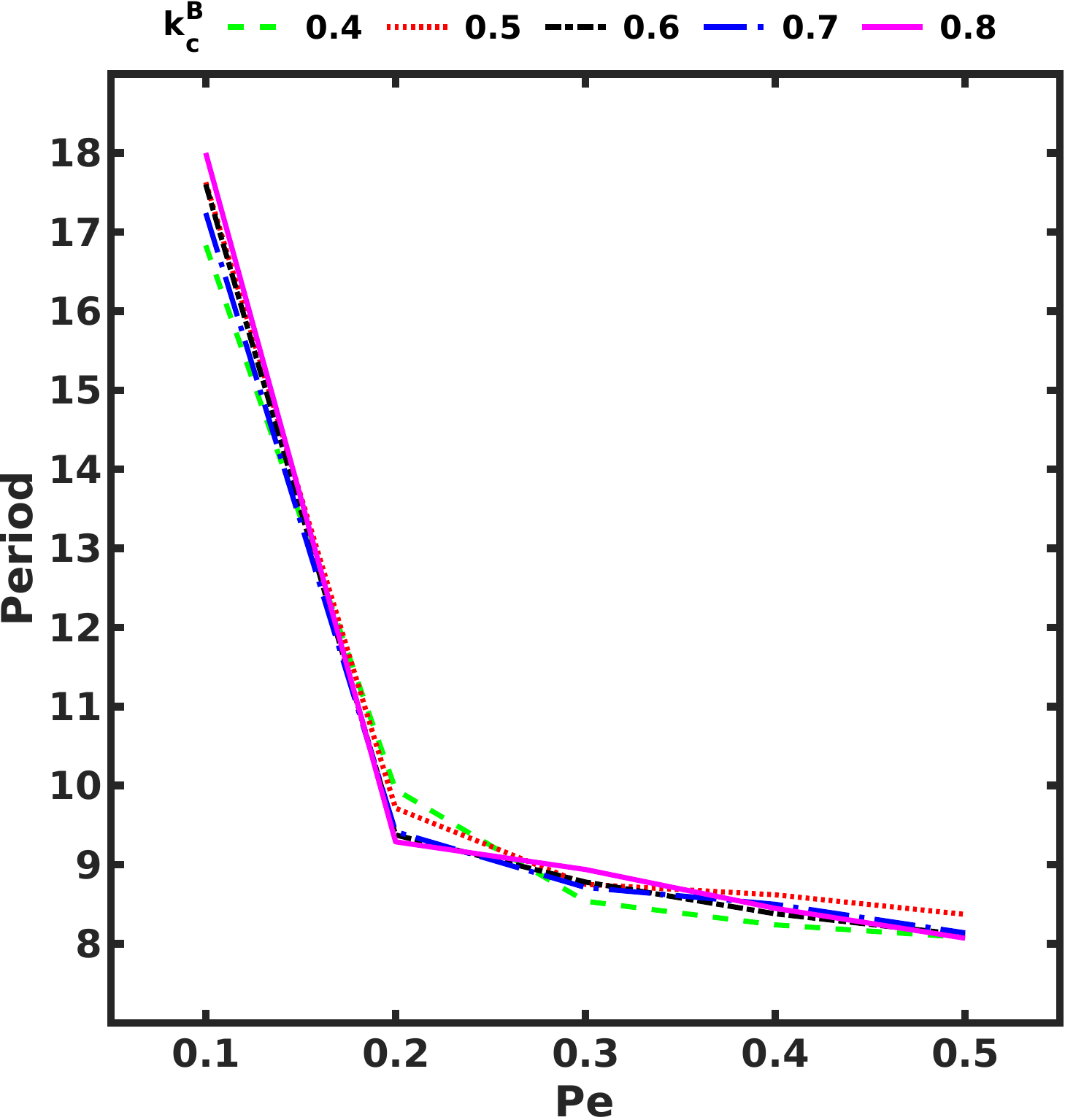}	
		}	
	\end{center}
	\caption{The time to merge and the phase period as a function of Peclet number for a vesicle with 
       $\bar{c}=0.4$ and $\alpha=0.5$.}
	\label{fig:Alphap5_TimeToMerge_Period}
\end{figure}

It is clear from the prior results that the properties of the membrane components play a crucial role
in determining the dynamics of the system. To further explore this, consider the amount of time 
needed for merging, the phase treading period
before and after merging, in addition to the 
phase treading period both before and after merging 
for a vesicle with $\alpha=0.5$, Fig.~\ref{fig:Alphap5_TimeToMerge_Period}.
The phase treading period is defined as the time between peaks of the bending energy,
while the time to merge is calculated as when a large drop in the domain boundary energy occurs
and is verified by visually observing the domains on the membrane.
Note that 
the $\alpha=20$ case has too much variation in the domain shapes to provide useful information.

As can be seen, the amount of time required for the merging of the two domains into the single domain
depends strongly on the surface Peclet number, with the bending rigidity difference between the two phases
playing a little-to-no role. Assuming that the limit of the time to merge is zero as $\Pe$ approaches zero,
the time to merge scales as $\Pe^{0.7}$ for $\alpha=0.5$ and this particular initial condition.

The phase treading period, both before and after merging, depends much more on the surface Peclet 
number than the bending rigidity difference. Note that the phase periods after merging 
for $\Pe$ values larger than 0.5 are not reported due to the time required to obtain results.
Additionally, systems with $\Pe=0.1$ do not under go phase treading before merging, and thus
only the period after merging is reported.

In both situations, the phase period appears to decay exponentially with the Peclet number. 
Following from the prior results, small values of the surface Peclet number result in long
phase treading periods, as the soft domains remain at the high curvature tips longer.
It is interesting to note that after merging, the phase treading period decreases slightly
for all values of $\Pe$.

\begin{figure*}
	\begin{center}
		\subfigure[$\kappa_c^B =0.5$: Bending Energy]{
			\includegraphics[height=5.5cm]{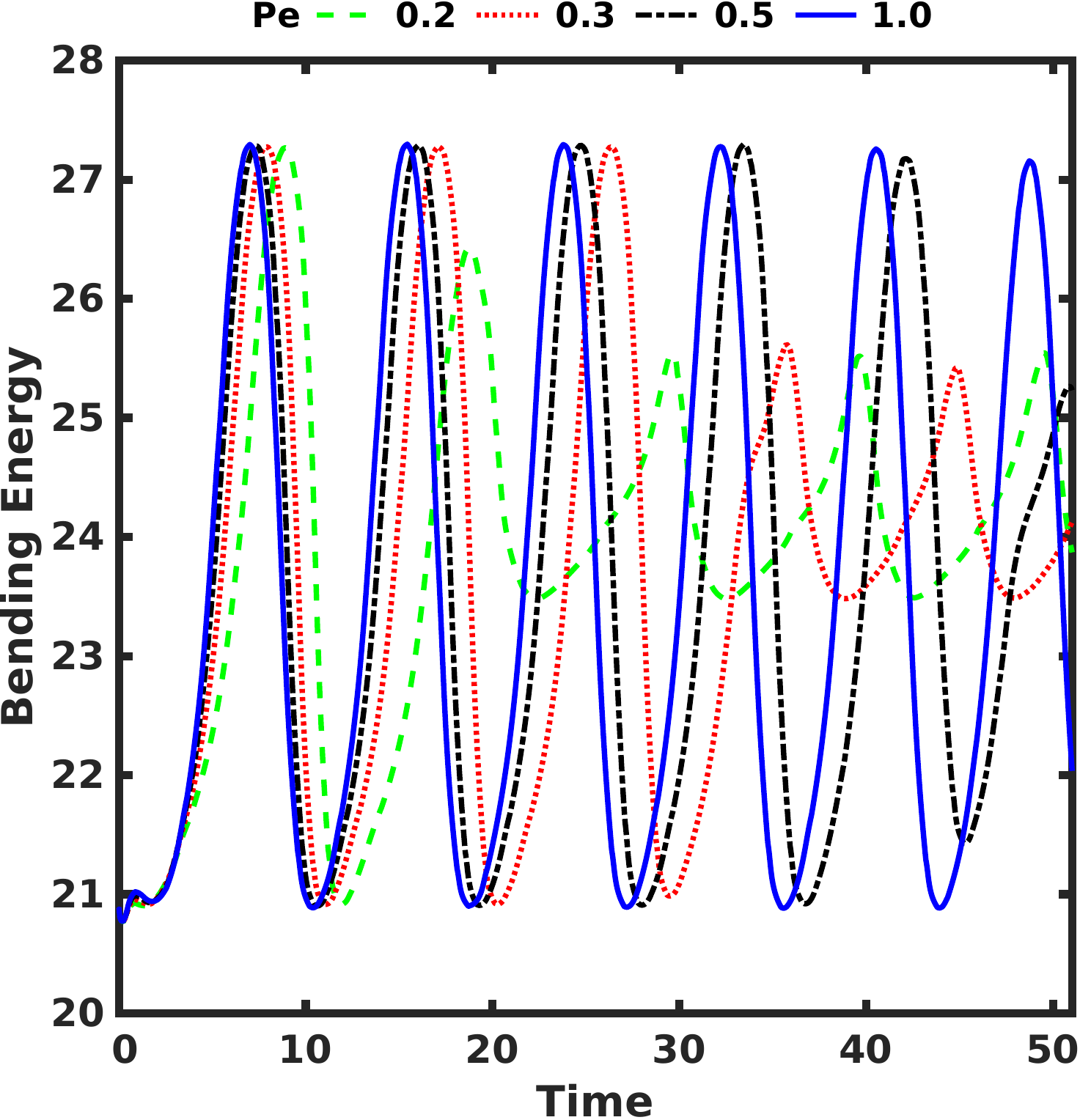}
		} 
		\qquad
		\subfigure[$\kappa_c^B =0.5$: Domain Boundary Energy]{			
			\includegraphics[height=5.5cm]{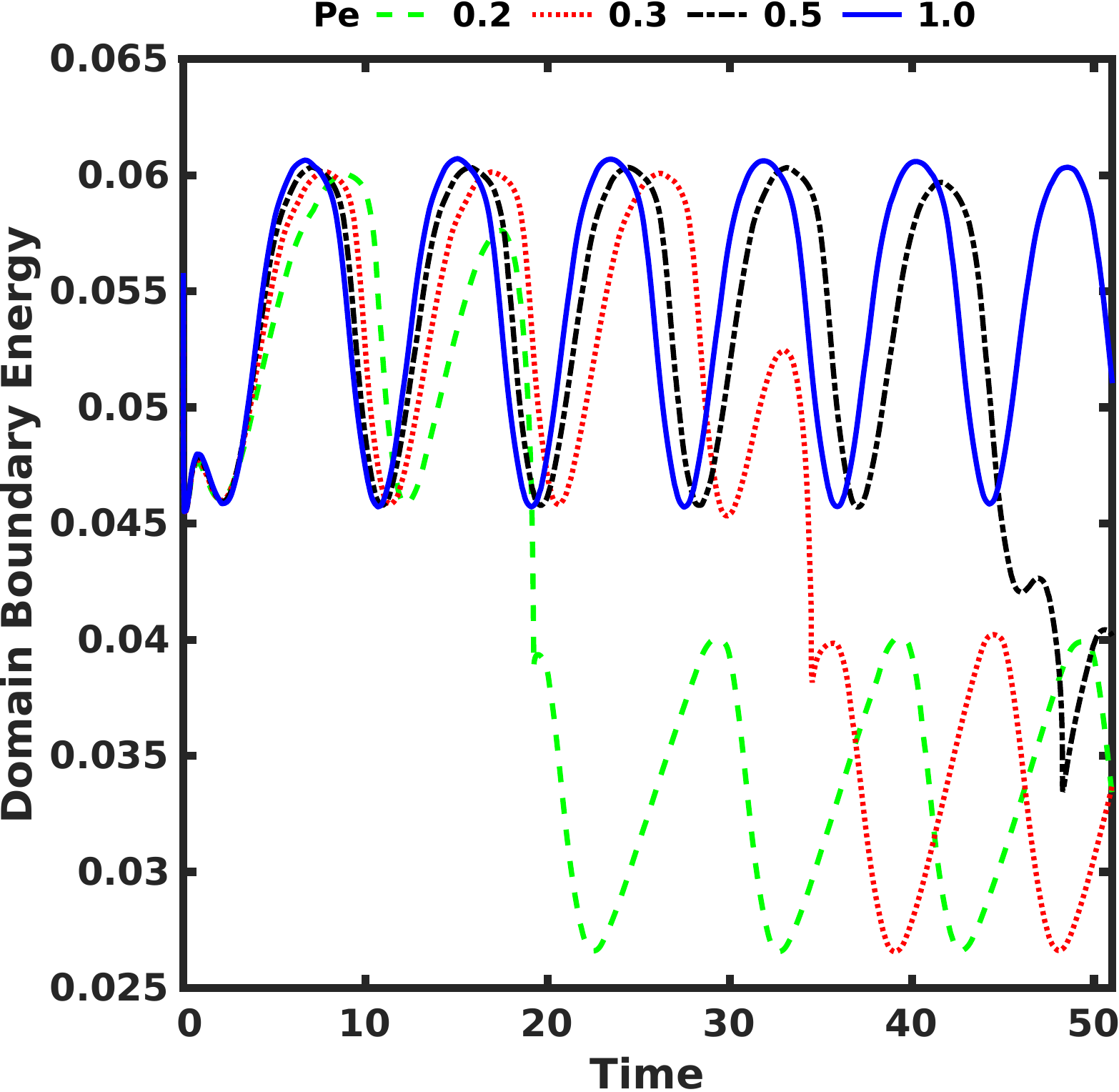}	
		}
		\\
		\subfigure[$\Pe=0.5$: Bending Energy]{
			\includegraphics[height=5.5cm]{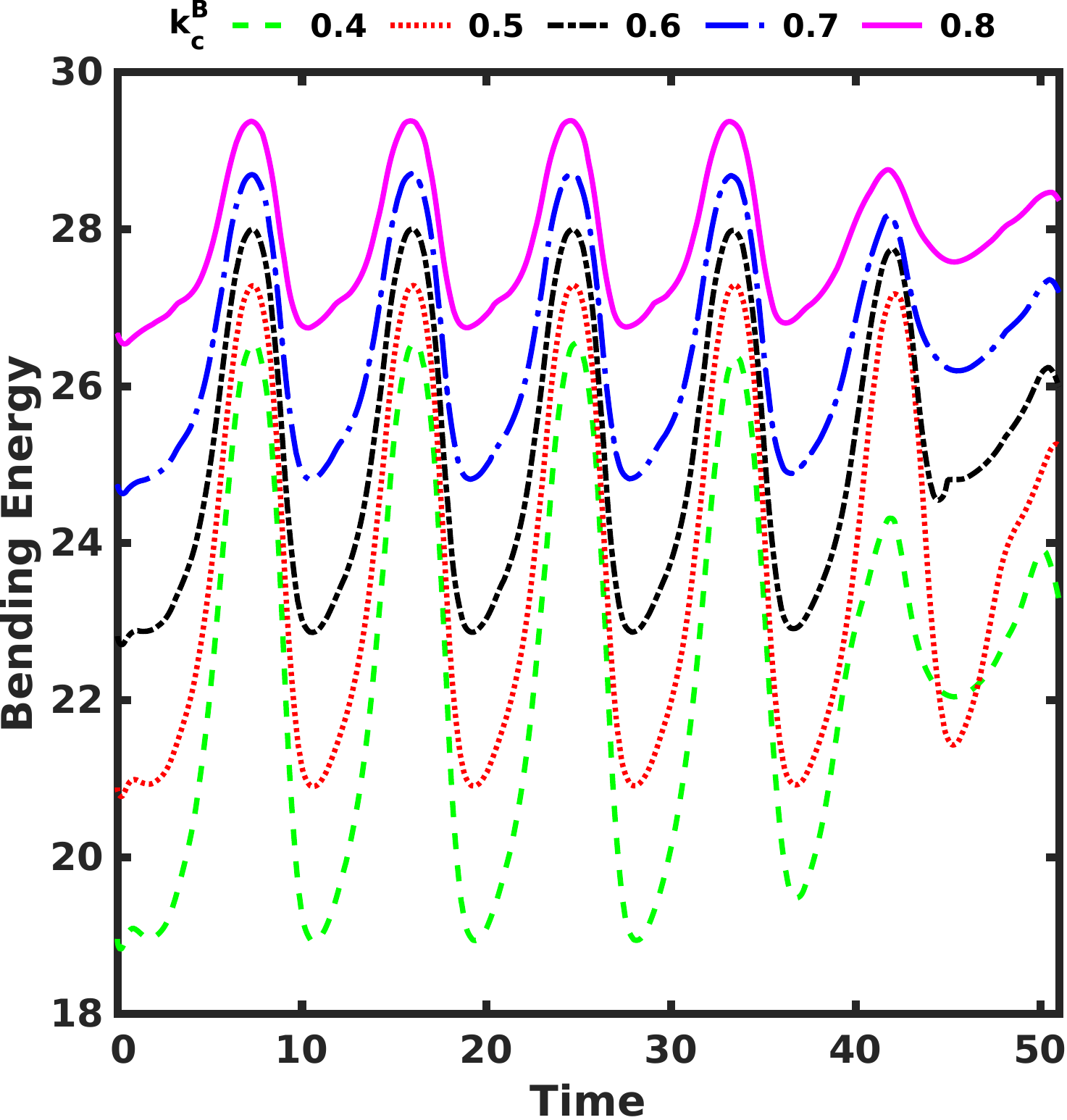}
		} 
		\qquad
		\subfigure[$\Pe=0.5$: Domain Boundary Energy]{			
			\includegraphics[height=5.5cm]{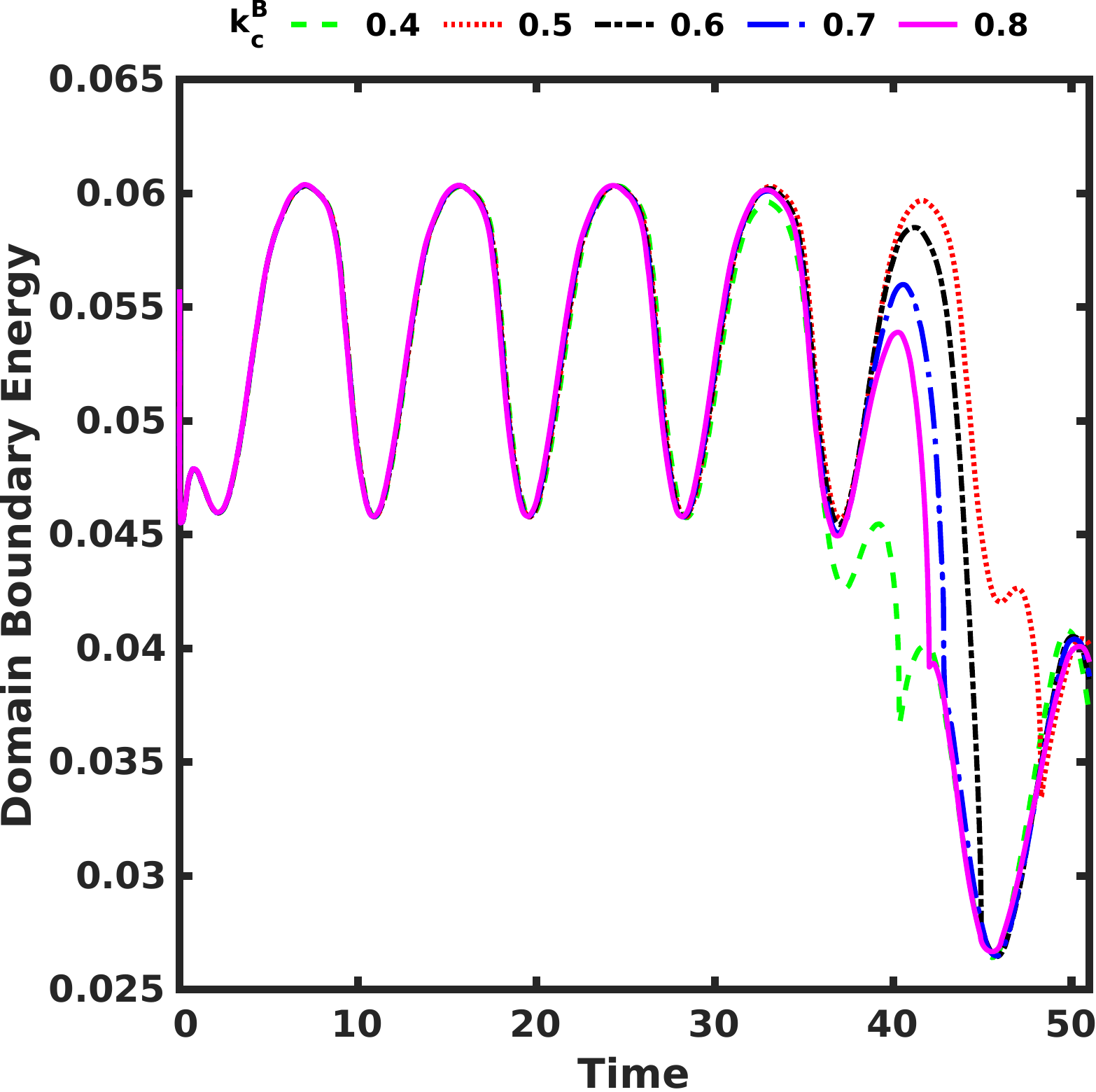}	
		}	
	\end{center}
		\caption{The bending and domain boundary energies for varying Peclet and bending rigidity 
		for a vesicle with $\bar{c}=0.4$ and $\alpha=0.5$.}
		\label{fig:Alphap5_EnergyPlots}
	
\end{figure*}

\begin{figure*}
	\begin{center}
		\subfigure[$\kappa_c^B =0.4$: Bending Energy]{
			\includegraphics[height=5.5cm]{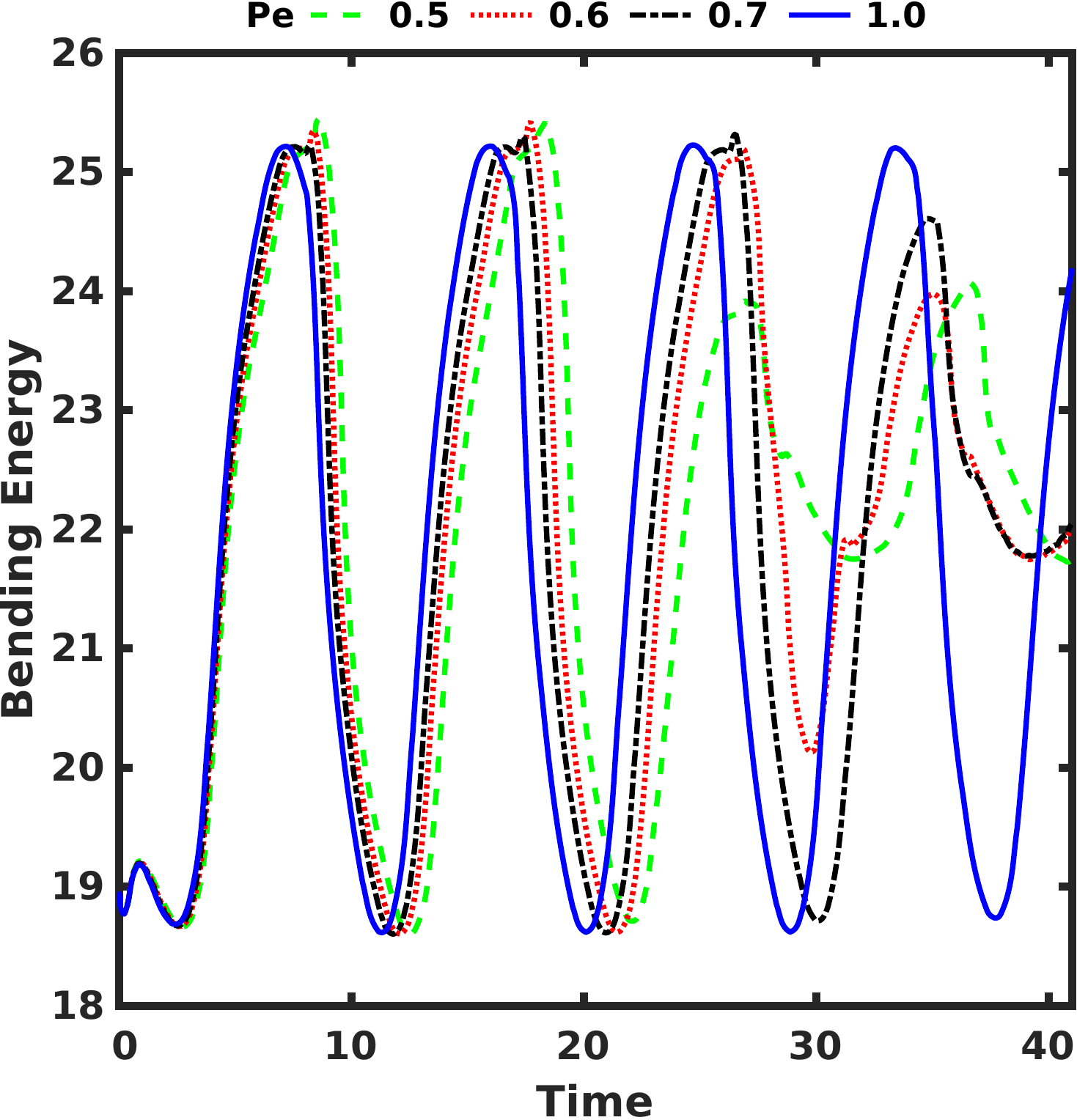}
		} 
		\qquad
		\subfigure[$\kappa_c^B =0.4$: Domain Boundary Energy]{			
			\includegraphics[height=5.5cm]{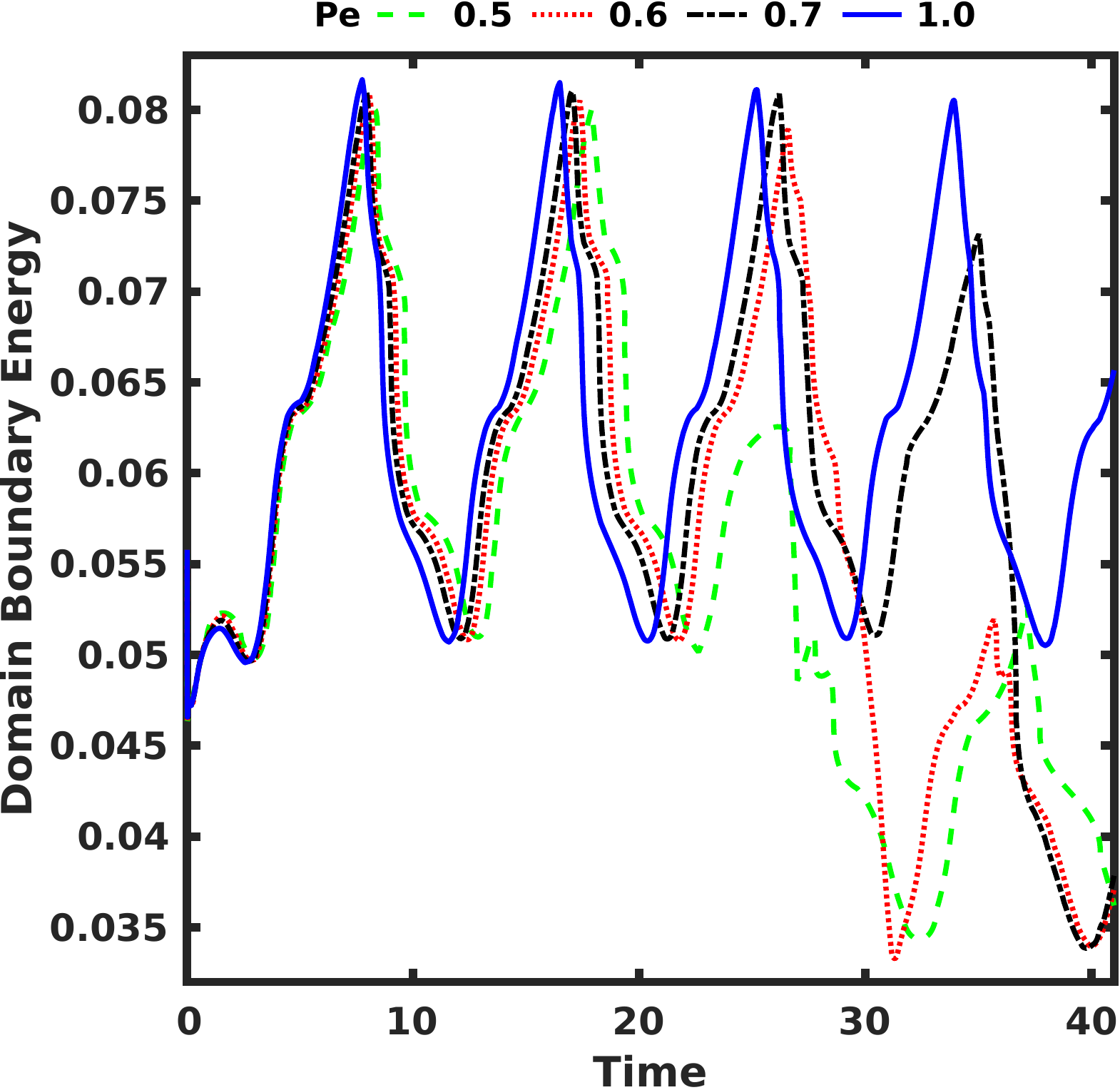}	
		}
		\\
		\subfigure[$\Pe=1$: Bending Energy]{
			\includegraphics[height=5.5cm]{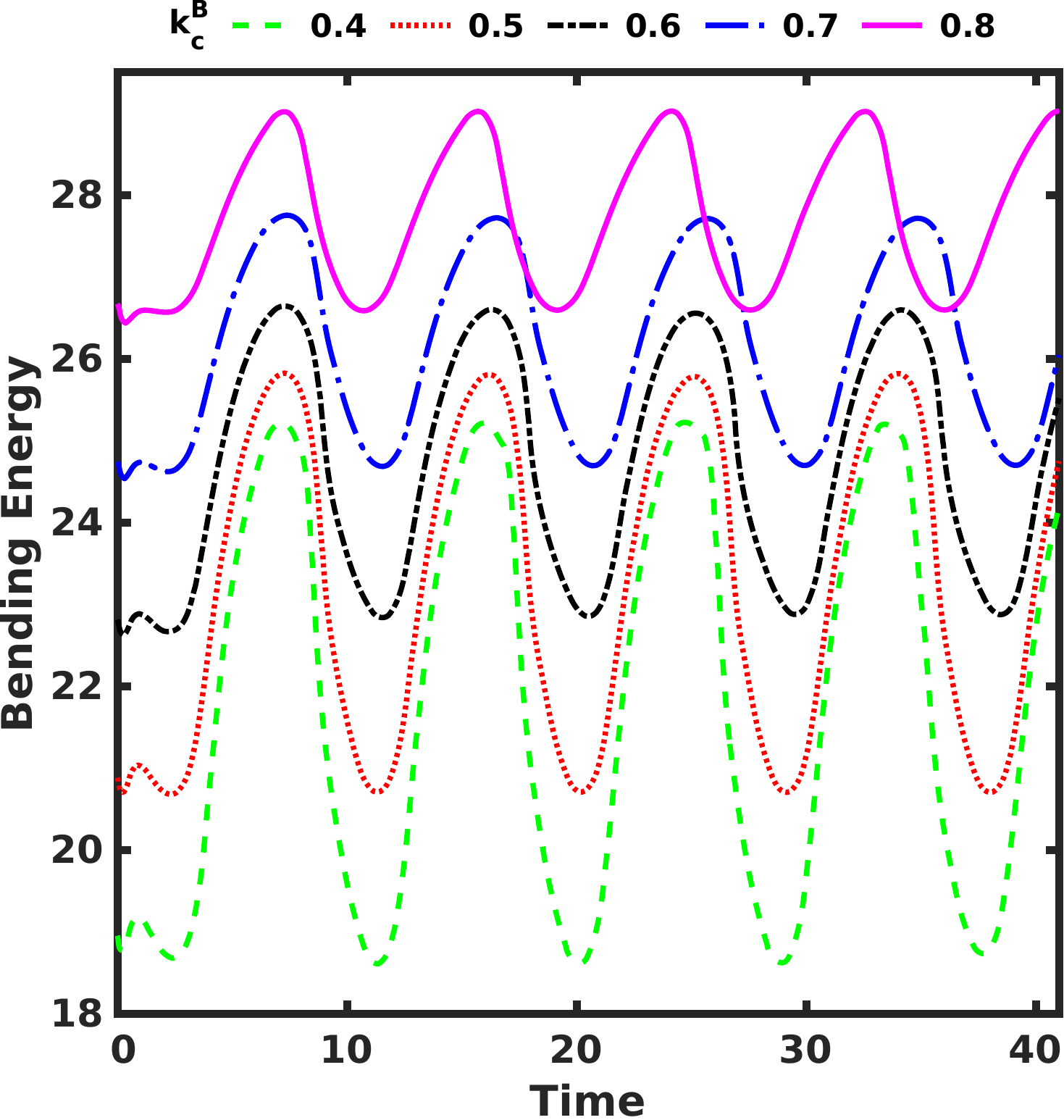}
		} 
		\qquad
		\subfigure[$\Pe=1$: Domain Boundary Energy]{			
			\includegraphics[height=5.5cm]{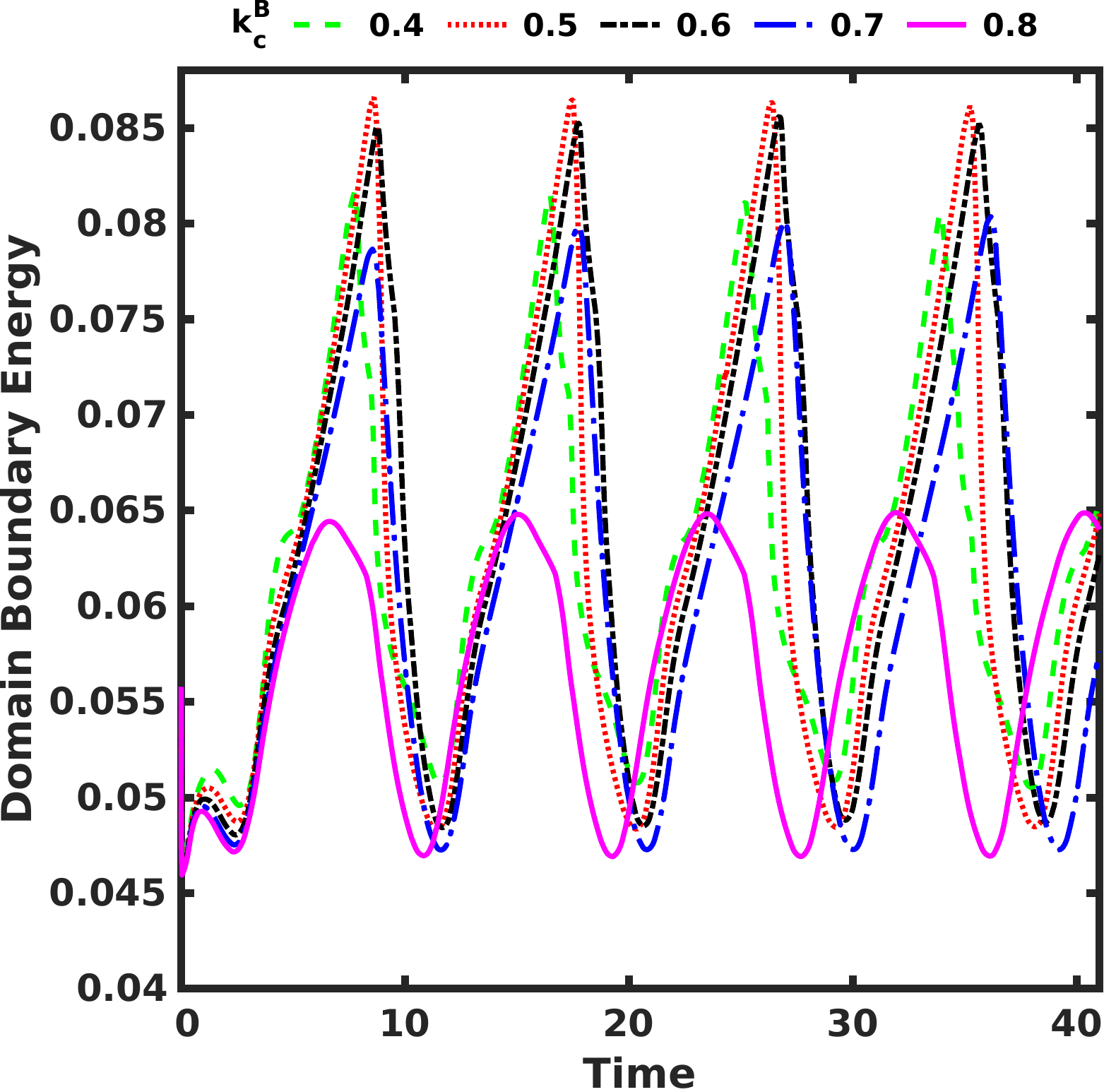}	
		}		
	\end{center}
		\caption{The bending and domain boundary energies for varying Peclet and bending rigidity 
		for a vesicle with $\bar{c}=0.4$ and $\alpha=20$.}
		\label{fig:Alpha20_EnergyPlots}
	
\end{figure*}

\begin{figure}
	\begin{center}
		\includegraphics[height=5.5cm]{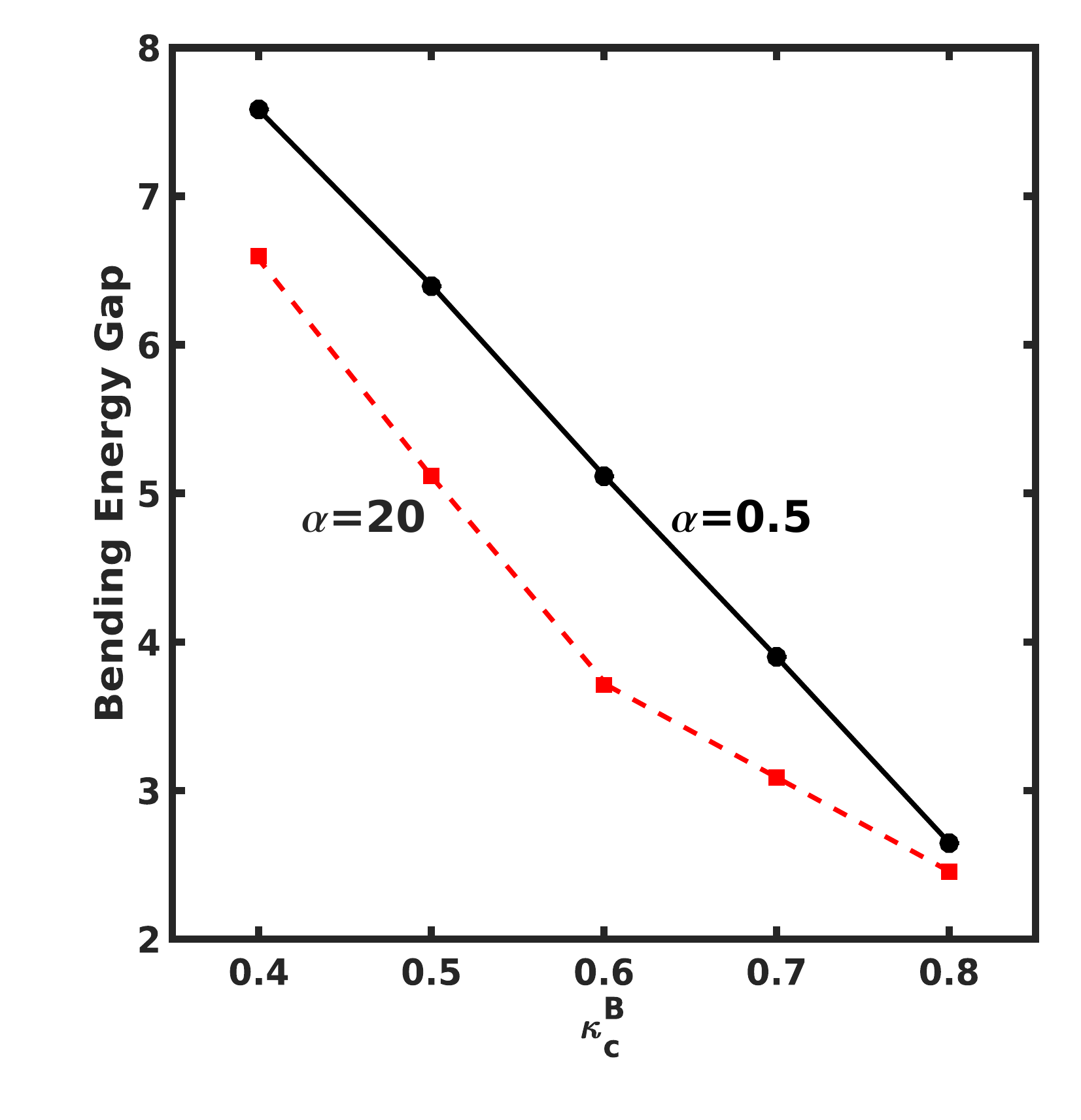}
	\end{center}
		\caption{The difference between the maximum and minimum bending energy before domain
	       merging for a vesicle with $\bar{c}=0.4$ with $\alpha=0.5$ or $\alpha=20$.
	       The bending energy gap is not sensitive to variations of Peclet number.}
		\label{fig:BendingEnergyGap}
	
\end{figure}

Finally, consider the bending and domain
boundary energy as a function of time for various Peclet numbers and bending rigidities. 
The results for $\alpha=0.5$ are shown in Fig.~\ref{fig:Alphap5_EnergyPlots} while those for
$\alpha=20$ are shown in Fig.~\ref{fig:Alpha20_EnergyPlots}.
From these results, several items become apparent. First, the difference between the highest and lowest
bending energy during phase treading, denoted as the bending energy gap, depends only on the difference between
the bending rigidity of the two phases and not the surface Peclet number. This should be expected, as the 
bending energy primarily depends on the bending rigidity difference. The surface Peclet number influences the 
phase treading period, as demonstrated earlier. Additionally, the bending energy and bending
energy gap is larger for 
systems with $\alpha=0.5$ than for systems with $\alpha=20$, see Fig.~\ref{fig:BendingEnergyGap}.
This is due to the fact that the surface domains deform more for $\alpha=20$ than for 
$\alpha=0.5$, as shown in prior sections. This allows for more of the phase with the 
lower bending rigidity to stay in contact with the high curvature tips during phase treading,
lowering the overall bending energy. 

The shape of the energy curves also follows the prior qualitative results. In particular, 
for $\alpha=0.5$ the domains remain relatively circular, and thus the domain boundary energy
has a relatively smooth transition from high energy values to low energy ones, Fig.~\ref{fig:Alphap5_EnergyPlots}.
For $\alpha=20$, the domains become elongated, with a tail remaining at the vesicle tips. When the domains
become too large, the domain tails move very quickly, which is seen as a 
periodic sharp drop in the domain boundary energy. This behavior is confirmed by comparing the 
domain boundary energy for $\kappa_c^B=0.8$ and those for $\kappa_c^B<0.8$. As the bending energy
gap is smaller for $\kappa_c^B=0.8$ than for the other values, only a small tail is formed, see Fig.~\ref{fig:Tread}.
Thus, the system with $\kappa_c^B=0.8$ and $\alpha=20$ behaves more like the $\alpha=0.5$ cases.

\section{Conclusions}

In this work, the dynamics of a three-dimensional multicomponent vesicle in shear flow has been investigated.
The focus of the study was on the influence of the bending rigidity difference, the rate of surface diffusion,
and domain boundary line energy on the dynamics. The system in three
dimensions allows for inclusion of domain line energy/tension, which results in a
more accurate model.

To allow for a systematic study of the influence of properties and parameters, this work focused on initially
pre-segregated and symmetric domains. Three types of dynamics were observed: stationary phase, phase treading,
and a new dynamic called vertical banding. In general, the dynamics are due to the complex interplay
between the difference in domain bending rigidity, the speed of diffusion as measured by the surface Peclet
number, and the relative strength of the domain line energy to the bending energy. When the domain line energy
is weak, as denoted by a large value of $\alpha$, domains can elongate and form vertical bands. When the line 
energy is strong, given by small values of $\alpha$, the influence of bending rigidity difference deceases and the 
dynamics are primarily determined by the surface Peclet number.

Given enough time, all cases considered here will result in a single phase domain. When this occurs 
the symmetry of the system is lost, and the vesicle begins to tremble about an equilibrium
inclination angle. During this time, the vesicle is pulled from the center and begins to migrate laterally.

The results presented here demonstrate that a complete understanding of the dynamics of multicomponent
vesicles can only be obtained via fully three-dimensional models. Due to the nature of two-dimensional models,
they are not able to capture the full influence of the domain line energy, which has been demonstrated plays a 
critical role in the determination of the dynamics. In the future, it will be necessary to relax some assumptions made
in this work. In particular, the use of a variable surface mobility might better match 
the physical system. The inclusion of varying fluid density and more complex bending rigidity models
may also result in additional interesting dynamics.

%\balance 

\section*{Conflicts of interest}
There are no conflicts to declare.

\section*{Acknowledgements}
The authors acknowledge support from the National Science Foundation,
Division of Chemical, Bioengineering, Environmental and Transport Systems, (NSF-CBET)
grant CBET-1253739. The computations shown here were performed at the Center for Computational
Research, University at Buffalo.


%%%REFERENCES%%%
\providecommand*{\mcitethebibliography}{\thebibliography}
\csname @ifundefined\endcsname{endmcitethebibliography}
{\let\endmcitethebibliography\endthebibliography}{}

\bibliographystyle{rsc}

\end{document}